\documentclass[pdftex,10pt,journal]{IEEEtran}

\usepackage[cmex10]{amsmath} 

\usepackage{latexsym,amssymb,bm,cite,ifthen}
\usepackage{afterpage}
\usepackage[pdftex]{graphicx,color}
\addtocounter{dbltopnumber}{1}
\addtocounter{totalnumber}{1}
\newcommand{\iftwocolumn}[2]{\ifthenelse{\boolean{@twocolumn}}{#1}{#2}}

\newcommand{\ignore}[1]{}

\newcommand{\visual}{}

\newcommand{\Fig}[1]{Fig.~\ref{#1}}

\newcommand{\eqdef}{\stackrel{\scriptscriptstyle\bigtriangleup}{=} }
\newcommand{\smpl}[2]{#1^{(#2)}}

\newcommand{\cond}{\hspace{0.02em}|\hspace{0.08em}}

\newcommand{\rangedot}{\cdot}
\newcommand{\T}{\mathsf{T}}
\renewcommand{\H}{\mathsf{H}}

\newcommand{\R}{\mathbb{R}}
\newcommand{\C}{{\mathbb C}}

\newcommand{\FT}{\mathcal{F}}

\newcommand{\calY}{\mathcal{Y}}

\newcommand{\ccj}[1]{\overline{#1}}

\newcommand{\E}{\operatorname{E}}

\newcommand{\tr}{\operatorname{tr}}

\newcommand{\wig}[1]{\mathbf{#1}}

\newcounter{examplecntr}
{\begin{trivlist}\small\item[]\refstepcounter{examplecntr}%
 {\bfseries Example~\theexamplecntr%
  \ifthenelse{\equal{#1}{}}{}{ (#1)}.
}}%
{\end{trivlist}}

\newcounter{definitioncntr}
{\begin{trivlist}\item[]\refstepcounter{definitioncntr}%
{\bfseries Definition~\thedefinitioncntr.}}%
{\hfill$\Box$\end{trivlist}}

\newcounter{theoremcntr}
{\begin{trivlist}\item[]\refstepcounter{theoremcntr}%
{\bfseries Theorem~\thetheoremcntr%
  \ifthenelse{\equal{#1}{}}{}{ (#1)}.
}}%
{\hfill$\Box$\end{trivlist}}

\newcounter{propositioncntr}
\newenvironment{proposition}[1][]%
{\begin{trivlist}\item[]\refstepcounter{propositioncntr}%
{\bfseries Proposition~\thepropositioncntr%
  \ifthenelse{\equal{#1}{}}{}{ (#1)}.
}}%
{\hfill$\Box$\end{trivlist}}

\newenvironment{proofof}[1]{\begin{trivlist}\item[]{\bfseries Proof\ifthenelse{\equal{#1}{}}{}{ #1}:}
 }{\hfill$\blacksquare$\end{trivlist}}

\newcommand{\eproofnegspace}{\\[-1.5\baselineskip]\rule{0em}{0ex}}

\definecolor{gray}{rgb}{0.5, 0.5, 0.5}

\newcommand{\gray}{\color{gray}}

\newcommand{\cent}[1]{\makebox(0,0){#1}}
\newcommand{\pos}[2]{\makebox(0,0)[#1]{#2}}
\newcommand{\connectionDot}{\circle*{1}}
\newcommand{\markerDot}{\circle*{1}}

\newcommand{\knownBox}{\cent{\rule{1.75\unitlength}{1.75\unitlength}}}

\begin{document}
\title{Factor Graphs for Quantum Probabilities%
}

\author{%
Hans-Andrea Loeliger
and Pascal O.\ Vontobel%
\thanks{%
H.-A.~Loeliger is with 
the Department of Information Technology and Electrical Engineering, 
ETH Zurich, Switzerland.
Email: loeliger@isi.ee.ethz.ch.

P.~O.~Vontobel is with 
the Department of Information Engineering, 
The Chinese University of Hong Kong.
Email: pascal.vontobel@ieee.org.

This paper was presented in part at the 2012 IEEE Int.\ Symp.\ on Information Theory \cite{LVo:ISIT2012c}.

To appear in IEEE Transactions on Information Theory, 2017.

}
}

\maketitle

\begin{abstract}
A factor-graph representation of quantum-mechani\-cal probabilities 
(involving any number of measurements) 
is proposed. 
Unlike standard statistical models, 
the proposed representation uses auxiliary variables (state variables)
that are not random variables. 
All joint probability distributions are marginals of
some complex-valued function $q$, 
and it is demonstrated how the basic concepts 
of quantum mechanics relate to factorizations and marginals of $q$.
\end{abstract}


\section{Introduction}
\label{sec:Intro}

Factor graphs \cite{KFL:fg2001,Forney:nr2001,Lg:ifg2004}
and similar graphical notations 
\cite{MeMo:ipc,Jo:gm2004,bi:prml,KoFr:PGMb} are widely used 
to represent statistical models with many variables. 
Factor graphs have become quite standard in coding theory \cite{RiUr:mct}, 
but their applications include also communications \cite{Wym:ird}, 
signal processing \cite{LDHKLK:fgsp2007,LBMWZ:ITA2016c}, 
combinatorics \cite{Vo:bp2013}, and much more.
The graphical notation can be helpful in various ways, 
including the elucidation of the model itself 
and the derivation of algorithms for statistical inference.

In this paper, we show how quantum mechanical probabilities 
(including, in particular, joint distributions over several measurements) 
can be expressed in factor graphs
that are fully compatible with factor graphs of standard statistical models
and error correcting codes. 
This is not trivial: 
despite being a statistical theory, 
quantum mechanics does not fit into standard statistical categories 
and it is not built on the Kolmogorov axioms of probability theory.
Existing graphical representations of quantum mechanics
such as Feynman diagrams \cite{Ve:Diagr1994}, 
tensor diagrams \cite{GuLeWe:prb2008,CiVe:rtp2009,Coe:qp2010,WBC:tngc2015}, 
and quantum circuits \cite[Chap.~4]{NiChuang:QCI} 
do not explicitly represent probabilities, 
and they are not compatible with ``classical'' graphical models.

Therefore, this paper is not just about a graphical notation, 
but it offers a perspective 
of quantum mechanics
that has not (as far as we know) been proposed before.

In order to introduce this perspective, 
recall that statistical models usually contain auxiliary variables 
(also called hidden variables or state variables),
which are essential for factorizing 
the joint probability distribution. 
For example, a hidden Markov model with 
primary variables $Y_1,\ldots,Y_n$ is defined 
by a joint probability mass function of the form
\begin{equation} \label{eqn:HMM}
p(y_1,\ldots,y_n,x_0,\ldots,x_n) = p(x_0) \prod_{k=1}^n p(y_k,x_k\cond x_{k-1}),
\end{equation}
where $X_0,X_1,\ldots, X_n$ are auxiliary variables (hidden variables)
that are essential for the factorization (\ref{eqn:HMM}). 
More generally, the joint distribution $p(y_1,\ldots,y_n)$ 
of some primary variables $Y_1,\ldots,Y_n$
is structured by a factorization of the joint distribution
$p(y_1,\ldots,y_n, x_0,\ldots,x_m)$ with auxiliary variables $X_0,\ldots,X_m$
and
\begin{equation} \label{eqn:PyAsMarginal}
p(y_1,\ldots,y_n) = \sum_{x_0,\ldots,x_m} p(y_1,\ldots,y_n,x_0,\ldots,x_m),
\end{equation}
where the sum is over all possible values of 
$X_0,\ldots,X_m$.
(For the sake of exposition, we assume here that all variables have finite alphabets.) 
However, quantum-mechanical joint probabilities cannot, in general, be structured
in this way.

We now generalize $p(y_1,\ldots,y_n, x_0,\ldots,x_m)$ in (\ref{eqn:PyAsMarginal}) to 
an arbitrary complex-valued function 
$q(y_1,\ldots,y_n,x_0,\ldots,x_m)$ such that
\begin{equation} \label{eqn:pfromq}
p(y_1,\ldots,y_n) = \sum_{x_0,\ldots,x_m} q(y_1,\ldots,y_n,x_0,\ldots,x_m).
\end{equation}
The purpose of $q$ is still to enable a nice factorization,
for which there may now be more opportunities. 
Note that the concept of marginalization carries over to $q$;
in particular, all marginals of $p(y_1,\ldots,y_n)$ (involving any number of variables) 
are also marginals of $q$. 
However, the auxiliary variables $X_0,\ldots,X_m$ are not, in general, random variables, 
and marginals of $q$ involving one or several of these variables are not, in general, 
probability distributions.

We will show that this generalization allows natural
representations
of quantum-mechanical probabilities
involving any number of measurements.
In particular, 
the factor graphs of this paper will represent pertinent factorizations of 
complex-valued functions $q$ as in (\ref{eqn:pfromq}).

This paper is not concerned with physics, 
but only with the peculiar joint probability distributions
that arise in quantum mechanics. 
However, we will show how the basic concepts and terms of quantum mechanics 
relate to factorizations and marginals of suitable functions $q$.
For the sake of clarity, we will restrict ourselves to finite alphabets 
(with some exceptions, especially in Appendix~\ref{sec:WignerWeyl}),
but this restriction is not essential. 
Within this limited scope, this paper may even be used 
as a self-contained introduction to 
the pertinent concepts of quantum mechanics.

To the best of our knowlege, describing quantum probabilities 
(and, indeed, any probabilities)
by explicitly using a function $q$ 
as in (\ref{eqn:pfromq}) is new. 
Nonetheless, 
this paper is, 
of course, related 
to much previous work in quantum mechanics and quantum computation. 
For example, quantum circuits as in \cite[Chap.~4]{NiChuang:QCI} 
have natural interpretations in terms of factor graphs 
as will be demonstrated in Sections \ref{sec:QuantumCircuits}
and~\ref{sec:CodesChannels}.
Our factor graphs are also related to tensor diagrams 
\cite{GuLeWe:prb2008,CiVe:rtp2009,Coe:qp2010,WBC:tngc2015,ChiAP:pp2010},
see Sections \ref{sec:FGMatrices} and Appendix~\ref{sec:Related}.
Also related is the very recent work by Mori \cite{Mo:htgp2015c}.
On the other hand, quantum Bayesian networks (see, e.g., \cite{Tu:qi})
and quantum belief propagation (see, e.g., \cite{LePo:qbp2008})
are not immediately related to our approach
since they are not based on (\ref{eqn:pfromq}) 
(and they lack Proposition~\ref{prop:FGMain} in Section~\ref{sec:FG}).
Finally, we mention that 
the factor graphs of this paper 
are used in \cite{CaoVo:ISIT2017} 
for estimating the information rate of certain quantum channels,
and iterative sum-product message passing in such factor graphs 
is considered in \cite{CaoVo:defg2017c}.

The paper is structured as follows. 
Section~\ref{sec:FG} reviews factor graphs and their connection to linear algebra. 
In Section~\ref{sec:ElementaryQM}, we express elementary 
quantum mechanics (with a single projection measurement) in factor graphs;
we also demonstrate how the Schr\"odinger picture, 
the Heisenberg picture, and even an elementary form of Feynman path integrals
are naturally expressed 
in terms of factor graphs.
Multiple and more general measurements are discussed in Section~\ref{sec:QMFG}. 
Section~\ref{sec:Decompositions} addresses partial measurements, 
decompositions of unitary operators (including quantum circuits),
and the emergence of non-unitary operators from unitary interactions.  
In Section~\ref{sec:MeasRecons}, we revisit measurements 
and briefly address 
their realization in terms of unitary interactions,
and in Section~\ref{sec:RVRev}, we comment on the origin of randomness.
In Section~\ref{sec:CodesChannels}, we further illustrate 
the use of factor graphs by an elementary introduction to quantum coding.
Section~\ref{sec:Concl} concludes the main part of the paper.
In Appendix~\ref{sec:Related},
we offer some additional remarks on the prior literature. 
In Appendix~\ref{sec:WignerWeyl}, we briefly discuss the Wigner--Weyl representation,
which leads to 
an alternative factor-graph representation.
In Appendix~\ref{sec:MonteCarlo}, 
we outline
the extension of Monte Carlo methods to the factor graphs of this paper.

This paper contains many figures of factor graphs
that represent some complex function $q$ 
as in (\ref{eqn:pfromq}). 
The main figures are Figs.\ \ref{fig:SingleMeasurement},
\ref{fig:GenQFG}, \ref{fig:PartialMeasurement}, and \ref{fig:MargSubsystem};
in a sense, the whole paper is about explaining and exploring these four figures. 

We will use standard linear algebra notation rather 
than the bra-ket notation of quantum mechanics. 
The Hermitian transpose of a complex matrix $A$ 
will be denoted by $A^\H \eqdef \ccj{A^\T}$,
where $A^\T$ is the transpose of $A$ and $\ccj{A}$ is the componentwise complex conjugate. 
An identity matrix will be denoted by $I$. 
The symbol ``$\propto$'' denotes equality of functions up to a scale factor.

\section{On Factor Graphs}
\label{sec:FG}

\subsection{Basics}

Factor graphs represent factorizations of functions of several variables. 
We will use Forney factor graphs%
\footnote{%
Factor graphs as in \cite{KFL:fg2001}
represent variables not by edges, but by variable nodes.
Adapting Proposition~\ref{prop:FGMain} for such factor graphs is awkward.

Henceforth in this paper, ``factor graph'' means ``Forney factor graph''; 
the qualifier ``Forney'' (or ``normal'') will sometimes be added 
to emphasize that the distinction matters.
}
(also called normal factor graphs)
as in \cite{Forney:nr2001,Lg:ifg2004,LDHKLK:fgsp2007},
where nodes (depicted as boxes) represent factors and edges represent variables.
For example, assume that some function $f(x_1,\ldots,x_5)$
can be written as
\begin{equation}  \label{eqn:ExampleFactorgraph}
f(x_1,\ldots,x_5) = f_1(x_1,x_2,x_5) f_2(x_2,x_3) f_3(x_3,x_4,x_5).
\end{equation}
The corresponding factor graph is shown in \Fig{fig:FactorGraph}. 

In this paper,
all variables in factor graphs take values in finite alphabets
(with some exceptions, especially in Appendix~\ref{sec:WignerWeyl})
and all functions take values in $\C$.

The factor graph of the hidden Markov model (\ref{eqn:HMM}) is shown in \Fig{fig:HMM}.
As in this example, variables in factor graphs are often denoted
by capital letters.

\begin{figure}
\begin{center}
\begin{picture}(55,22.5)(3,7.5)
\put(10,15){\framebox(5,5){}}     \put(9,17.5){\pos{cr}{$f_1$}}
 \put(12.5,15){\line(0,-1){7.5}}    \put(13.5,7.5){\pos{bl}{$x_1$}}
\put(15,17.5){\line(1,0){15}}     \put(22.5,18.5){\pos{cb}{$x_2$}}
\put(30,15){\framebox(5,5){}}     \put(32.5,21){\pos{cb}{$f_2$}}
\put(35,17.5){\line(1,0){15}}     \put(42.5,18.5){\pos{cb}{$x_3$}}
\put(50,15){\framebox(5,5){}}     \put(56,17.5){\pos{cl}{$f_3$}}
 \put(52.5,15){\line(0,-1){7.5}}  \put(53.5,7.5){\pos{bl}{$x_4$}}
\put(12.5,20){\line(0,1){10}}
\put(12.5,30){\line(1,0){40}}     \put(22.5,28.5){\pos{ct}{$x_5$}}
\put(52.5,20){\line(0,1){10}}
\end{picture}
\caption{\label{fig:FactorGraph}%
Factor graph (i.e., Forney factor graph) of (\ref{eqn:ExampleFactorgraph}).}
\end{center}
\vspace{\floatsep}

\begin{center}
\begin{picture}(67.5,17)(0,0)
\put(0,10){\framebox(5,5){}}
\put(5,12.5){\line(1,0){12.5}}     \put(11.25,13.5){\pos{cb}{$X_0$}}
\put(17.5,10){\framebox(5,5){}}
 \put(20,10){\line(0,-1){10}}      \put(21,0){\pos{bl}{$Y_1$}}
\put(22.5,12.5){\line(1,0){12.5}}  \put(28.75,13.5){\pos{cb}{$X_1$}}
\put(35,10){\framebox(5,5){}}
 \put(37.5,10){\line(0,-1){10}}   \put(38.5,0){\pos{bl}{$Y_2$}}
\put(40,12.5){\line(1,0){12.5}}    \put(46.75,13.5){\pos{cb}{$X_2$}}
\put(52.5,10){\framebox(5,5){}}
 \put(55,10){\line(0,-1){10}}     \put(56,0){\pos{bl}{$Y_3$}}
\put(57.5,12.5){\line(1,0){10}}    \put(63.75,13.5){\pos{cb}{$X_3$}}
\end{picture}
\caption{\label{fig:HMM}%
Factor graph of the hidden Markov model~(\ref{eqn:HMM}) for $n=3$.}
\end{center}
\end{figure}

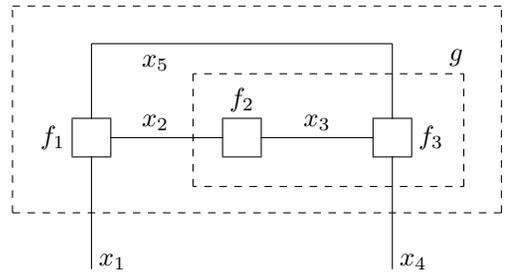
\begin{figure}[t]
\begin{center}
\begin{picture}(65,35)(2,0)
\put(10,15){\framebox(5,5){}}     \put(9,17.5){\pos{cr}{$f_1$}}
 \put(12.5,15){\line(0,-1){15}}    \put(13.5,0){\pos{bl}{$x_1$}}
\put(15,17.5){\line(1,0){15}}     \put(21,18.5){\pos{cb}{$x_2$}}
\put(30,15){\framebox(5,5){}}     \put(32.5,21){\pos{cb}{$f_2$}}
\put(35,17.5){\line(1,0){15}}     \put(42.5,18.5){\pos{cb}{$x_3$}}
\put(50,15){\framebox(5,5){}}     \put(56,17.5){\pos{cl}{$f_3$}}
 \put(52.5,15){\line(0,-1){15}}  \put(53.5,0){\pos{bl}{$x_4$}}
\put(12.5,20){\line(0,1){10}}
\put(12.5,30){\line(1,0){40}}    \put(21,28.5){\pos{ct}{$x_5$}}
\put(52.5,20){\line(0,1){10}}
\put(26,11){\dashbox(36,15){}}     \put(61,27){\pos{bc}{$g$}}  
\put(2,7.5){\dashbox(65,27.5){}}      
\end{picture}
\caption{\label{fig:ClosingBoxes}%
Closing boxes in factor graphs.}
\end{center}
\end{figure}

The Forney factor-graph notation is intimately connected 
with the idea of opening and closing boxes 
\cite{Lg:ifg2004,LDHKLK:fgsp2007,VoLg:FGEN2002}. 
Consider the dashed boxes in \Fig{fig:ClosingBoxes}. 
The \emph{exterior function} of such a box
is defined to be the product of all factors inside the box, 
summed over all its internal variables. 
The exterior function of the inner dashed box in \Fig{fig:ClosingBoxes} is  
\begin{equation} \label{eqn:InnerBoxExt}
g(x_2,x_4,x_5) \eqdef \sum_{x_3} f_2(x_2,x_3) f_3(x_3,x_4,x_5),
\IEEEeqnarraynumspace
\end{equation}
and the exterior function of the outer dashed box is 
\begin{equation} \label{eqn:OuterBoxExt}
f(x_1,x_4) \eqdef \sum_{x_2,x_3,x_5} f(x_1,\ldots,x_5).
\end{equation}
The summations in (\ref{eqn:InnerBoxExt}) and (\ref{eqn:OuterBoxExt}) 
range over all possible values of the corresponding variable(s).

\emph{Closing a box}
means replacing the box with a single node 
that represents the exterior function of the box.
For example, closing the inner dashed box in \Fig{fig:ClosingBoxes}
replaces the two nodes/factors 
$f_2(x_2,x_3)$ and $f_3(x_3,x_4,x_5)$ by 
the single node/factor (\ref{eqn:InnerBoxExt});
closing the outer dashed box in \Fig{fig:ClosingBoxes}
replaces all nodes/factors in (\ref{eqn:ExampleFactorgraph})
by the single node/factor (\ref{eqn:OuterBoxExt});
and closing first the inner dashed box and then the outer dashed box
replaces all nodes/factors in (\ref{eqn:ExampleFactorgraph}) by
\begin{equation} \label{eqn:FactorGraphExampleHierachicalMarginal}
\sum_{x_2,x_5} f_1(x_1,x_2,x_5) g(x_2,x_4,x_5)
= f(x_1,x_4).
\end{equation}
Note the equality between (\ref{eqn:FactorGraphExampleHierachicalMarginal}) 
and (\ref{eqn:OuterBoxExt}), which holds in general:
\begin{proposition} \label{prop:FGMain}
Closing an inner box within some outer box 
(by summing over the internal variables of the inner box) 
does not change the exterior function of the outer box.
\end{proposition}
This simple fact is the pivotal property of Forney factor graphs. 
Closing boxes in factor graphs is thus compatible with marginalization 
both of probability mass functions and of complex-valued functions 
$q$ as in (\ref{eqn:pfromq}), which is the basis of the present paper.

\emph{Opening a box}
in a factor graph means the reverse operation 
of expanding a node/factor into a factor graph of its own.

A \emph{half edge} in a factor graph is an edge that is connected to only one node 
(such as $x_1$ in \Fig{fig:FactorGraph}). 
The \emph{exterior function of a factor graph%
\footnote{What we here call the exterior function of a factor graph, is 
called \emph{partition function} in \cite{FoVo:pfnfg2011c}.
The term ``exterior function'' was first used in \cite{BaMa:nfght2011}.}%
}
is defined to be 
the exterior function of a box that contains all nodes and all full edges, 
but all half edges stick out
(such as the outer box in \Fig{fig:ClosingBoxes}). 
For example, 
the exterior function of \Fig{fig:FactorGraph} is~(\ref{eqn:OuterBoxExt}).
The \emph{partition sum%
\footnote{What we call here the partition sum has often been called 
\emph{partition function}.}}
of a factor graph is the exterior function 
of a box that contains the whole factor graph, including all half edges; 
the partition sum is a constant.

The exterior function of \Fig{fig:HMM} is $p(x_n,y_1,\ldots,y_n)$, 
and its partition sum equals one.

\begin{figure}
\begin{center}
\begin{picture}(25,10)(0,0)
\put(0,0){\framebox(5,5){}}   \put(2.5,6.5){\pos{cb}{$p$}}
\put(5,2.5){\line(1,0){15}}   \put(12.5,4){\pos{cb}{$X$}}
\put(20,0){\framebox(5,5){}}  \put(22.5,6.5){\pos{cb}{$g$}}
\end{picture}
\caption{\label{fig:FGEgX}%
Factor graph of $\E[g(X)]$ according to (\ref{eqn:FGEgX}).}
\end{center}
\end{figure}
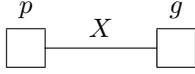

Factor graphs can also express expectations: 
the partition sum (and the exterior function) of \Fig{fig:FGEgX} is
\begin{equation} \label{eqn:FGEgX}
\E[g(X)] = \sum_x p(x) g(x),
\end{equation}
where $p(x)$ is a probability mass function and $g$ is an arbitrary real-valued 
(or complex-valued) function.

The equality constraint function $f_=$ is defined as
\begin{equation} \label{eqn:EqualityConstraint}
f_=(x_1,\ldots,x_n) = \left\{
 \begin{array}{ll}
   1, & \text{if $x_1= \cdots = x_n$} \\
   0, & \text{otherwise.}
 \end{array}
\right.
\end{equation}
The corresponding node (which is denoted by ``$=$'') can serve
as a branching point in a factor graph 
(cf.\ Figs.\ \ref{fig:MultiStepUnitaryModel}--\ref{fig:HeisenbergPicture}):
only configurations with $x_1=\ldots = x_n$ contribute
to the exterior function of any boxes containing these variables.

A variable with a fixed known value will be marked
by a solid square as in Figs.\ \ref{fig:KnownRemovesEqu}
and~\ref{fig:SchroedingerPictureKnownInitial}.

\subsection{Factor Graphs and Matrices}
\label{sec:FGMatrices}

A matrix $A\in \C^{m\times n}$ may be viewed as a function 
\begin{equation}
\{ 1,\ldots, m\} \times \{ 1,\ldots, n\} \rightarrow \C:\, (x,y) \mapsto A(x,y).
\end{equation}
The multiplication of two matrices $A$ and $B$ 
can then be written as
\begin{equation} \label{eqn:MatrixMult}
(AB)(x,z) = \sum_{y} A(x,y) B(y,z),
\end{equation}
which is the exterior function of \Fig{fig:FactorGraphMatrixMult}.
Note that the identity matrix corresponds to an equality constraint 
function $f_=(x,y)$. 

\begin{figure}
\begin{center}
\begin{picture}(55,19)(0,0)
\put(0,7.5){\line(1,0){15}}     \put(5,8.5){\pos{cb}{$x$}}
\put(15,5){\framebox(5,5){}}    \put(17.5,4){\pos{ct}{$A$}}
 \put(15,7.5){\markerDot}
\put(20,7.5){\line(1,0){15}}    \put(27.5,8.5){\pos{cb}{$y$}}
\put(35,5){\framebox(5,5){}}    \put(37.5,4){\pos{ct}{$B$}}
 \put(35,7.5){\markerDot}
\put(40,7.5){\line(1,0){15}}    \put(50,8.5){\pos{cb}{$z$}}
\put(10,0){\dashbox(35,15){}}   \put(27.5,16.5){\pos{cb}{$AB$}}

\end{picture}
\caption{\label{fig:FactorGraphMatrixMult}%
Factor-graph representation of matrix multiplication~(\ref{eqn:MatrixMult}).
The small dot denotes the variable that indexes the rows of the corresponding matrix.
}
\vspace{\floatsep}

\begin{picture}(15,14)(0,-4)
\put(0,10){\line(1,0){15}}
\put(0,10){\line(0,-1){7.5}}
\put(0,2.5){\line(1,0){5}}
\put(5,0){\framebox(5,5){}}     \put(7.5,-4){\pos{cb}{$A$}}
 \put(5,2.5){\markerDot}
\put(10,2.5){\line(1,0){5}}
\put(15,10){\line(0,-1){7.5}}
\end{picture}
\hspace{15mm}
\begin{picture}(30,14)(0,-4)
\put(0,10){\line(1,0){30}}
\put(0,10){\line(0,-1){7.5}}
\put(0,2.5){\line(1,0){5}}
\put(5,0){\framebox(5,5){}}     \put(7.5,-4){\pos{cb}{$A$}}
 \put(5,2.5){\markerDot}
\put(10,2.5){\line(1,0){10}}
\put(20,0){\framebox(5,5){}}     \put(22.5,-4){\pos{cb}{$B$}}
 \put(20,2.5){\markerDot}
\put(25,2.5){\line(1,0){5}}
\put(30,10){\line(0,-1){7.5}}
\end{picture}
\caption{\label{fig:FactorGraphTrace}%
Factor graph of $\tr(A)$ (left) and of $\tr(AB) = \tr(BA)$ (right).}
\end{center}
\end{figure}

In this notation, 
the trace of a square matrix $A$ is 
\begin{equation} \label{eqn:Trace}
\tr(A) = \sum_{x} A(x,x),
\end{equation}
which is the exterior function 
(and the partition sum) 
of the factor graph in \Fig{fig:FactorGraphTrace} (left). 
\Fig{fig:FactorGraphTrace} (right) shows
the graphical proof of the identity $\tr(AB) = \tr(BA)$.

In this way, closing and opening boxes in factor graphs
may thus be viewed as generalizations of matrix multiplication 
and matrix factorization, respectively.

The factor graph of a diagonal matrix with diagonal elements 
from some vector $v$ is shown in \Fig{fig:DiagonalMatrix}.
\Fig{fig:SpectralDecomp} shows 
the decomposition of a Hermitian matrix $A$
according to the spectral theorem into 
\begin{equation} \label{eqn:SpectralDecomp}
A = U \Lambda U^\H,
\end{equation}
where $U$ is unitary and where $\Lambda$ is diagonal and real 
with diagonal elements from some vector $\lambda$.

\begin{figure}
\begin{center}
\begin{picture}(35,27.5)(0,1)
\put(15,17.5){\framebox(5,5){}}   \put(17.5,24){\pos{cb}{$v$}}
\put(17.5,10){\line(0,1){7.5}}
\put(0,7.5){\line(1,0){15}}
\put(15,5){\framebox(5,5){$=$}}
\put(20,7.5){\line(1,0){15}}
\put(10,1){\dashbox(15,27.5){}}
\end{picture}
\caption{\label{fig:DiagonalMatrix}%
Factor graph of a diagonal matrix with diagonal vector $v$.
The node labeled ``$=$'' represents 
the equality constraint function (\ref{eqn:EqualityConstraint}).}
\vspace{\floatsep}

\begin{picture}(52.5,37)(0,-2)
\put(5,12.5){\framebox(5,5){}}  \put(7.5,19){\pos{cb}{$\lambda$}}
\put(10,15){\line(1,0){10}}
\put(20,12.5){\framebox(5,5){$=$}}
\put(22.5,17.5){\line(0,1){5}}
\put(22.5,22.5){\line(1,0){10}}
\put(32.5,20){\framebox(5,5){}}   \put(35,18.5){\pos{ct}{$U$}}
 \put(37.5,22.5){\connectionDot}
\put(37.5,22.5){\line(1,0){15}}
\put(22.5,12.5){\line(0,-1){5}}
\put(22.5,7.5){\line(1,0){10}}
\put(32.5,5){\framebox(5,5){}}    \put(35,3.5){\pos{ct}{$U^\H$}}
 \put(32.5,7.5){\connectionDot}
\put(37.5,7.5){\line(1,0){15}}
\put(0,-2){\dashbox(42.5,32.5){}}  \put(21,32){\pos{cb}{$A$}}
\end{picture}
\caption{\label{fig:SpectralDecomp}%
Factor graph of decomposition (\ref{eqn:SpectralDecomp})
according to the spectral theorem.}
\end{center}
\end{figure}

Factor graphs for linear algebra operations 
such as \Fig{fig:FactorGraphMatrixMult}
and \Fig{fig:FactorGraphTrace} (and the corresponding generalizations to tensors) 
are essentially tensor diagrams (or trace diagrams) 
as in \cite{WBC:tngc2015,Cv:gtbt,Pe:ula2009}. 
This connection between factor graphs and tensor diagrams 
was noted in \cite{BaMa:nfght2011,FoVo:pfnfg2011c,BMV:nfgla2011c}
and will further be discussed in Appendix~\ref{sec:Related}.

\subsection{Reductions}

Reasoning with factor graphs typically involves 
``local'' manipulations of some nodes/factors (such as opening or closing boxes) 
that preserve the exterior function of all surrounding boxes.
Some such reductions are shown in Figs.\ 
\ref{fig:RedTwoVarEqu}--\ref{fig:KnownRemovesEqu};
these (very simple) reductions will be essential for understanding 
the proposed factor graphs for quantum-mechanical probabilities.

\begin{figure}
\centering
\begin{picture}(40,30)(2.5,-5)
\put(2.5,17.5){\line(1,0){12.5}}
 \put(5,11){\cent{$\vdots$}}
\put(2.5,2.5){\line(1,0){12.5}}
\put(15,0){\framebox(10,20){}}
\put(25,17.5){\line(1,0){10}}
 \put(35,17.5){\line(0,-1){5}}
\put(32.5,7.5){\framebox(5,5){$=$}}
 \put(35,7.5){\line(0,-1){5}}
\put(25,2.5){\line(1,0){10}}
\put(10,-5){\dashbox(32.5,30){}}
\end{picture}
\hfill
\begin{picture}(0,30)(0,-5)
\put(0,10){\cent{$=$}}
\end{picture}
\hfill
\begin{picture}(32.5,30)(2.5,-5)
\put(2.5,17.5){\line(1,0){12.5}}
 \put(5,11){\cent{$\vdots$}}
\put(2.5,2.5){\line(1,0){12.5}}
\put(15,0){\framebox(10,20){}}
\put(25,17.5){\line(1,0){5}}
\put(30,17.5){\line(0,-1){15}}
\put(25,2.5){\line(1,0){5}}
\put(10,-5){\dashbox(25,30){}}
\end{picture}
\caption{\label{fig:RedTwoVarEqu}%
A two-variable equality constraint 
(i.e., an identity matrix) can be dropped or addded.}
\vspace{\floatsep}

\begin{picture}(30,25)(2.5,-2.5)
\put(2.5,17.5){\line(1,0){15}}
 \put(17.5,17.5){\line(0,-1){5}}
\put(15,7.5){\framebox(5,5){$=$}}
 \put(20,10){\line(1,0){7.5}}
 \put(17.5,7.5){\line(0,-1){5}}
\put(2.5,2.5){\line(1,0){15}}
\put(10,-2.5){\dashbox(22.5,25){}}
\end{picture}
\hspace{7.5mm}
\begin{picture}(0,25)(0,-2.5)
\put(0,10){\cent{$=$}}
\end{picture}
\hspace{7.5mm}
\begin{picture}(22.5,25)(2.5,-2.5)
\put(2.5,17.5){\line(1,0){15}}
 \put(17.5,17.5){\line(0,-1){5}}
\put(15,7.5){\framebox(5,5){$=$}}
 \put(17.5,7.5){\line(0,-1){5}}
\put(2.5,2.5){\line(1,0){15}}
\put(10,-2.5){\dashbox(15,25){}}
\end{picture}
\caption{\label{fig:RedHalfEdgeEqu}%
A half edge out of an equality constraint node (of any degree) can be dropped or added.}
\vspace{\floatsep}

\begin{picture}(35,30)(2.5,-6)
\put(2.5,17.5){\line(1,0){12.5}}
\put(15,15){\framebox(5,5){}}     \put(17.5,13.5){\pos{ct}{$A^{-1}$}}
 \put(20,17.5){\markerDot}
\put(20,17.5){\line(1,0){10}}
 \put(30,17.5){\line(0,-1){5}}
\put(27.5,7.5){\framebox(5,5){$=$}}
 \put(30,7.5){\line(0,-1){5}}
\put(2.5,2.5){\line(1,0){12.5}}
\put(15,0){\framebox(5,5){}}     \put(17.5,-1.5){\pos{ct}{$A$}}
 \put(15,2.5){\markerDot}
\put(20,2.5){\line(1,0){10}}
\put(10,-6){\dashbox(27.5,30){}}
\end{picture}
\hspace{7.5mm}
\begin{picture}(0,25)(0,-2.5)
\put(0,10){\cent{$=$}}
\end{picture}
\hspace{7.5mm}
\begin{picture}(22.5,25)(2.5,-2.5)
\put(2.5,17.5){\line(1,0){15}}
 \put(17.5,17.5){\line(0,-1){5}}
\put(15,7.5){\framebox(5,5){$=$}}
 \put(17.5,7.5){\line(0,-1){5}}
\put(2.5,2.5){\line(1,0){15}}
\put(10,-2.5){\dashbox(15,25){}}
\end{picture}
\caption{\label{fig:RedMatrixMultEqu}%
A regular square matrix $A$ multiplied by its inverse reduces to an identity matrix 
(i.e., a two-variable equality constraint).}
\vspace{\floatsep}

\begin{picture}(30,25)(2.5,-2.5)
\put(2.5,17.5){\line(1,0){15}}
 \put(17.5,17.5){\line(0,-1){5}}
\put(15,7.5){\framebox(5,5){$=$}}
 \put(20,10){\line(1,0){7.5}}
  \put(27.5,10){\knownBox}       \put(27.5,12){\pos{cb}{$x$}}
 \put(17.5,7.5){\line(0,-1){5}}
\put(2.5,2.5){\line(1,0){15}}
\put(10,-2.5){\dashbox(22.5,25){}}
\end{picture}
\hspace{7.5mm}
\begin{picture}(0,25)(0,-2.5)
\put(0,10){\cent{$=$}}
\end{picture}
\hspace{7.5mm}
\begin{picture}(22.5,25)(2.5,-2.5)
\put(2.5,17.5){\line(1,0){15}}
 \put(17.5,17.5){\knownBox}    \put(17.5,15.5){\pos{ct}{$x$}}
\put(2.5,2.5){\line(1,0){15}}
 \put(17.5,2.5){\knownBox}     \put(17.5,4.5){\pos{cb}{$x$}}
\put(10,-2.5){\dashbox(15,25){}}
\end{picture}
\caption{\label{fig:KnownRemovesEqu}%
A fixed known value (depicted as a small solid square) 
propagates through, and thereby eliminates, an equality constraint.}
\end{figure}

\subsection{Complex Conjugate Pairs}
\label{sec:ConjPairs}

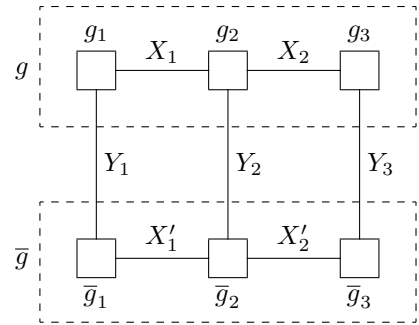
\begin{figure}
\begin{center}
\begin{picture}(50,42)(0,-1)
\put(0,25){\dashbox(50,16){}}        \put(-1.5,32.5){\pos{cr}{$g$}}
\put(5,30){\framebox(5,5){}}         \put(7.5,36.25){\pos{cb}{$g_1$}}
\put(10,32.5){\line(1,0){12.5}}      \put(16.25,33.5){\pos{cb}{$X_1$}}
\put(22.5,30){\framebox(5,5){}}      \put(25,36.25){\pos{cb}{$g_2$}}
\put(27.5,32.5){\line(1,0){12.5}}    \put(33.75,33.5){\pos{cb}{$X_2$}}
\put(40,30){\framebox(5,5){}}        \put(42.5,36.25){\pos{cb}{$g_3$}}
\put(7.5,30){\line(0,-1){20}}        \put(8.5,20){\pos{cl}{$Y_1$}}
\put(25,30){\line(0,-1){20}}         \put(26,20){\pos{cl}{$Y_2$}}
\put(42.5,30){\line(0,-1){20}}       \put(43.5,20){\pos{cl}{$Y_3$}}
\put(0,-1){\dashbox(50,16){}}         \put(-1.5,7.5){\pos{cr}{$\ccj{g}$}}
\put(5,5){\framebox(5,5){}}          \put(7.5,4){\pos{ct}{$\ccj{g}_1$}}
\put(10,7.5){\line(1,0){12.5}}       \put(16.25,8.5){\pos{cb}{$X_1'$}}
\put(22.5,5){\framebox(5,5){}}       \put(25,4){\pos{ct}{$\ccj{g}_2$}}
\put(27.5,7.5){\line(1,0){12.5}}     \put(33.75,8.5){\pos{cb}{$X_2'$}}
\put(40,5){\framebox(5,5){}}         \put(42.5,4){\pos{ct}{$\ccj{g}_3$}}

\end{picture}
\caption{\label{fig:BasicExampleQFG}%
Factor graph with complex factors 
and nonnegative real marginal (\ref{eqn:ComplexConjClosed}).
}
\end{center}
\end{figure}

A general recipe for constructing complex functions $q$ 
with real and nonnegative marginals as in (\ref{eqn:pfromq})
is illustrated in \Fig{fig:BasicExampleQFG},
where all factors are complex valued. 
Note that the lower dashed box in \Fig{fig:BasicExampleQFG}
mirrors the upper dashed box: all factors in the lower box
are the complex conjugates of the corresponding factors in the upper dashed box. 
The exterior function of the upper dashed box is 
\begin{equation}
g(y_1,y_2,y_3) \eqdef \sum_{x_1,x_2} g_1(x_1,y_1) g_2(x_1,x_2,y_2) g_3(x_2,y_3)
\end{equation}
and the exterior function of the lower dashed box is
\begin{equation}
\sum_{x_1',x_2'} \ccj{g_1(x_1',y_1)}\, \ccj{g_2(x_1',x_2',y_2)}\, \ccj{g_3(x_2',y_3)}
=  \ccj{g(y_1,y_2,y_3)}.
\end{equation}
If follows that closing both boxes in \Fig{fig:BasicExampleQFG}
yields
\begin{equation} \label{eqn:ComplexConjClosed}
g(y_1,y_2,y_3) \ccj{g(y_1,y_2,y_3)} = |g(y_1,y_2,y_3)|^2,
\end{equation}
which is real and nonnegative. 

\begin{figure}[p]
\begin{center}
\begin{picture}(80,30)(0,-6.5)

\put(0,7.5){\framebox(5,5){}}  \put(2.5,6.5){\pos{ct}{$p(x)$}}
\put(5,10){\line(1,0){15}}     \put(10,11){\pos{cb}{$X$}}
\put(20,7.5){\framebox(5,5){$=$}}

\put(22.5,12.5){\line(0,1){5}}
\put(22.5,17.5){\line(1,0){10}}
\put(32.5,15){\framebox(5,5){}}  \put(35,13.5){\pos{ct}{$U$}}
 \put(37.5,17.5){\markerDot}
\put(37.5,17.5){\line(1,0){10}}
\put(47.5,15){\framebox(5,5){}}  \put(50,13.5){\pos{ct}{$B^\H$}}
 \put(52.5,17.5){\markerDot}
\put(52.5,17.5){\line(1,0){10}}
\put(62.5,17.5){\line(0,-1){5}}

\put(22.5,7.5){\line(0,-1){5}}
\put(22.5,2.5){\line(1,0){10}}
\put(32.5,0){\framebox(5,5){}}   \put(35,-1.5){\pos{ct}{$U^\H$}}
 \put(32.5,2.5){\markerDot}
\put(37.5,2.5){\line(1,0){10}}
\put(47.5,0){\framebox(5,5){}}   \put(50,-1.5){\pos{ct}{$B$}}
 \put(47.5,2.5){\markerDot}
\put(52.5,2.5){\line(1,0){10}}
\put(62.5,2.5){\line(0,1){5}}

\put(60,7.5){\framebox(5,5){$=$}}
\put(65,10){\line(1,0){15}}      \put(75,11.25){\pos{cb}{$Y$}}

\put(16,-6.5){\dashbox(53,30){}}  \put(70.25,-5){\pos{bl}{$p(y\cond x)$}}

\end{picture}
\caption{\label{fig:SingleMeasurement}%
Factor graph of an elementary quantum system.}
\end{center}
\vspace{\floatsep}

\begin{center}
\setlength{\unitlength}{0.9mm}
\begin{picture}(61,30)(0,-6.5)

\put(5,17.5){\knownBox}       \put(3,17.5){\pos{cr}{$x$}}
\put(5,17.5){\line(1,0){10}}
\put(15,15){\framebox(5,5){}}  \put(17.5,13.5){\pos{ct}{$U$}}
 \put(20,17.5){\markerDot}
\put(20,17.5){\line(1,0){10}}
\put(30,15){\framebox(5,5){}}  \put(32.5,13.5){\pos{ct}{$B^\H$}}
 \put(35,17.5){\markerDot}
\put(35,17.5){\line(1,0){10}}
\put(45,17.5){\line(0,-1){5}}

\put(5,2.5){\knownBox}        \put(3,2.5){\pos{cr}{$x$}}
\put(5,2.5){\line(1,0){10}}
\put(15,0){\framebox(5,5){}}   \put(17.5,-1.5){\pos{ct}{$U^\H$}}
 \put(15,2.5){\markerDot}
\put(20,2.5){\line(1,0){10}}
\put(30,0){\framebox(5,5){}}   \put(32.5,-1.5){\pos{ct}{$B$}}
 \put(30,2.5){\markerDot}
\put(35,2.5){\line(1,0){10}}
\put(45,2.5){\line(0,1){5}}

\put(42.5,7.5){\framebox(5,5){$=$}}
\put(47.5,10){\line(1,0){10}}   \put(55,11.25){\pos{cb}{$Y$}}

\put(11,-6.5){\dashbox(50,30){}}

\end{picture}
\caption{\label{fig:SingleMeasurementLikelihoodFixedX}%
Dashed box of \Fig{fig:SingleMeasurement} for fixed $X=x$. 
The partition sum of this factor graph equals one.
}
\end{center}
\vspace{\floatsep}

\begin{center}
\setlength{\unitlength}{0.9mm}
\begin{picture}(49,35)(-2.5,-6.5)

\put(5,22.5){\knownBox}       \put(3,22.5){\pos{cr}{$x$}}
\put(5,22.5){\line(1,0){7.5}}
\put(12.5,20){\framebox(5,5){}}  \put(15,18.5){\pos{ct}{$U$}}
 \put(17.5,22.5){\markerDot}
\put(17.5,22.5){\line(1,0){12.5}}
\put(30,20){\framebox(5,5){}}  \put(32.5,18.5){\pos{ct}{$B^\H$}}
 \put(35,22.5){\markerDot}
\put(35,22.5){\line(1,0){7.5}} \put(44.5,22.5){\pos{cl}{$y$}}
 \put(42.5,22.5){\knownBox}
\put(-2.5,13.5){\dashbox(24,15){}}  \put(22.5,29.5){\pos{tl}{$\psi$}}

\put(5,2.5){\knownBox}        \put(3,2.5){\pos{cr}{$x$}}
\put(5,2.5){\line(1,0){7.5}}
\put(12.5,0){\framebox(5,5){}}   \put(15,-1.5){\pos{ct}{$U^\H$}}
 \put(12.5,2.5){\markerDot}
\put(17.5,2.5){\line(1,0){12.5}}
\put(30,0){\framebox(5,5){}}   \put(32.5,-1.5){\pos{ct}{$B$}}
 \put(30,2.5){\markerDot}
\put(35,2.5){\line(1,0){7.5}}
 \put(42.5,2.5){\knownBox}   \put(44.5,2.5){\pos{cl}{$y$}}
\put(-2.5,-6.5){\dashbox(24,15){}}  \put(22.5,9.5){\pos{tl}{$\psi^\H$}}

\end{picture}
\caption{\label{fig:SingleMeasurementProbabilityFixedX}%
Derivation of (\ref{eqn:ElementaryProbability}) 
and (\ref{eqn:ElementaryProbabilityQuantumState}).}
\end{center}
\vspace{\floatsep}

\begin{center}
\begin{picture}(82.5,35)(-4,-6.5)

\put(0,7.5){\framebox(5,5){}}  \put(2.5,6.5){\pos{ct}{$p(x)$}}
\put(5,10){\line(1,0){10}}     \put(10,11){\pos{cb}{$X$}}
\put(15,7.5){\framebox(5,5){$=$}}

\put(17.5,12.5){\line(0,1){5}}
\put(17.5,17.5){\line(1,0){10}}
\put(27.5,15){\framebox(5,5){}}  \put(30,13.5){\pos{ct}{$U$}}
 \put(32.5,17.5){\markerDot}
\put(32.5,17.5){\line(1,0){15}}
\put(47.5,15){\framebox(5,5){}}  \put(50,13.5){\pos{ct}{$B^\H$}}
 \put(52.5,17.5){\markerDot}
\put(52.5,17.5){\line(1,0){10}}
\put(62.5,17.5){\line(0,-1){5}}

\put(17.5,7.5){\line(0,-1){5}}
\put(17.5,2.5){\line(1,0){10}}
\put(27.5,0){\framebox(5,5){}}   \put(30,-1.5){\pos{ct}{$U^\H$}}
 \put(27.5,2.5){\markerDot}
\put(32.5,2.5){\line(1,0){15}}
\put(47.5,0){\framebox(5,5){}}   \put(50,-1.5){\pos{ct}{$B$}}
 \put(47.5,2.5){\markerDot}
\put(52.5,2.5){\line(1,0){10}}
\put(62.5,2.5){\line(0,1){5}}

\put(60,7.5){\framebox(5,5){$=$}}
\put(65,10){\line(1,0){7.5}}
\put(71,11){\pos{cb}{$Y$}}

\put(-4,-6.5){\dashbox(40.5,30){}}   \put(16,25){\pos{cb}{$\rho$}}
\put(43.5,-6.5){\dashbox(35,30){}}  \put(62,25){\pos{cb}{$f_=$}}

\end{picture}
\caption{\label{fig:DensityMatrix}%
Regrouping \Fig{fig:SingleMeasurement} 
into a density matrix $\rho$ and an equality constraint.}
\end{center}
\vspace{\floatsep}

\begin{center}
\begin{picture}(82.5,35)(-4,-6.5)

\put(0,7.5){\framebox(5,5){}}  \put(2.5,6.5){\pos{ct}{$p(x)$}}
\put(5,10){\line(1,0){10}}     \put(10,11){\pos{cb}{$X$}}
\put(15,7.5){\framebox(5,5){$=$}}

\put(17.5,12.5){\line(0,1){5}}
\put(17.5,17.5){\line(1,0){10}}
\put(27.5,15){\framebox(5,5){}}  \put(30,13.5){\pos{ct}{$U$}}
 \put(32.5,17.5){\markerDot}
\put(32.5,17.5){\line(1,0){15}}
\put(47.5,15){\framebox(5,5){}}  \put(50,13.5){\pos{ct}{$B^\H$}}
 \put(52.5,17.5){\markerDot}
\put(52.5,17.5){\line(1,0){10}}
\put(62.5,17.5){\line(0,-1){5}}

\put(17.5,7.5){\line(0,-1){5}}
\put(17.5,2.5){\line(1,0){10}}
\put(27.5,0){\framebox(5,5){}}   \put(30,-1.5){\pos{ct}{$U^\H$}}
 \put(27.5,2.5){\markerDot}
\put(32.5,2.5){\line(1,0){15}}
\put(47.5,0){\framebox(5,5){}}   \put(50,-1.5){\pos{ct}{$B$}}
 \put(47.5,2.5){\markerDot}
\put(52.5,2.5){\line(1,0){10}}
\put(62.5,2.5){\line(0,1){5}}

\put(60,7.5){\framebox(5,5){$=$}}
\put(65,10){\line(1,0){7.5}}
\put(72.5,10){\knownBox}    \put(72.5,12){\pos{cb}{$y$}}

\put(-4,-6.5){\dashbox(40.5,30){}}   \put(16,25){\pos{cb}{$\rho$}}
\put(43.5,-6.5){\dashbox(35,30){}}  \put(62,25){\pos{cb}{$B(\rangedot,y)B(\rangedot,y)^\H$}}

\end{picture}
\caption{\label{fig:ObsProbTrace}%
\Fig{fig:DensityMatrix} for fixed \mbox{$Y=y$}.}
\end{center}
\end{figure}

All factor graphs for quantum-mechanical probabilities 
that will be proposed in this paper (except in Appendix~\ref{sec:WignerWeyl}) are 
special cases of this general form.
With two parts that are complex conjugates of each other, 
such representations might seem redundant. 
Indeed, one of the two parts could certainly be \emph{depicted} in some abbreviated form;
however, \emph{as mathematical objects} 
subject to Proposition~\ref{prop:FGMain}, 
our factor graphs must contain both parts.
(Also, the Monte Carlo methods of Appendix~\ref{sec:MonteCarlo}
work with samples where $x_k'\neq x_k$.)

\section{Elementary Quantum Mechanics\\ in Factor Graphics}
\label{sec:ElementaryQM}

\subsection{Born's Rule}
\label{sec:BornsRule}

We begin with an elementary situation with a single measurement 
as shown in \Fig{fig:SingleMeasurement}. 
In this factor graph, 
$p(x)$ is a probability mass function, 
$U$ and $B$ are complex-valued unitary $M\times M$ matrices,
and all variables take values in the set $\{ 1,\ldots, M\}$. 
The matrix $U$ describes the unitary evolution of the initial state $X$. 
The matrix $B$ defines the basis for the projection measurement whose outcome is $Y$
(as will further be discussed below).
The exterior function of the dashed box is $p(y \cond x)$, 
which we will examine below;
with that box closed, the factor graph represents the joint 
distribution 
\begin{equation} \label{eqn:ElementaryQM}
p(x, y) = p(x) p(y \cond x).
\end{equation}

We next verify that the dashed box in \Fig{fig:SingleMeasurement} 
can indeed represent a conditional probability distribution $p(y \cond x)$. 
For fixed $X=x$, this dashed box 
turns into (all of) \Fig{fig:SingleMeasurementLikelihoodFixedX}
(see \Fig{fig:KnownRemovesEqu}).
By the reductions from Figs.\ \ref{fig:RedHalfEdgeEqu} and~\ref{fig:RedMatrixMultEqu},
closing the dashed box in \Fig{fig:SingleMeasurementLikelihoodFixedX}
turns it into an identity matrix. 
It follows that the partition sum of \Fig{fig:SingleMeasurementLikelihoodFixedX} 
is $I(x,x)=1$ 
(i.e., the element in row $x$ and column $x$ of an identity matrix),
thus complying with the requirement $\sum_y p(y \cond x)=1$.

It is then clear from (\ref{eqn:ElementaryQM}) 
that the partition sum of \Fig{fig:SingleMeasurement} equals~1.

For fixed $X=x$ and $Y=y$, the dashed box in \Fig{fig:SingleMeasurement}
turns into \Fig{fig:SingleMeasurementProbabilityFixedX}, 
and $p(y\cond x)$ is the partition sum of that factor graph.
The partition sum of the upper part 
of \Fig{fig:SingleMeasurementProbabilityFixedX} 
is $B(\rangedot,y)^\H U(\rangedot,x)$, 
where $U(\rangedot,x)$ is column $x$ of $U$ and $B(\rangedot,y)$ is column $y$ of $B$.  
The partition sum of the lower part 
of \Fig{fig:SingleMeasurementProbabilityFixedX}
is $U(\rangedot,x)^\H B(\rangedot,y)$. 
Therefore, the partition sum of \Fig{fig:SingleMeasurementProbabilityFixedX} 
is the product of these two terms, i.e., 
\begin{IEEEeqnarray}{rCl}
p(y\cond x) & = & \left| B(\rangedot,y)^\H U(\rangedot,x) \right|^2 \IEEEeqnarraynumspace
  \label{eqn:ElementaryProbability}\\
 & = & \left| B(\rangedot,y)^\H \psi \right|^2,
   \label{eqn:ElementaryProbabilityQuantumState}
\end{IEEEeqnarray}
where $\psi \eqdef U(\rangedot,x)$ is the quantum state (or the wave function).

With a little practice, 
the auxiliary Figs.~\ref{fig:SingleMeasurementLikelihoodFixedX}
and \ref{fig:SingleMeasurementProbabilityFixedX} need not actually 
be drawn and (\ref{eqn:ElementaryProbability}) 
can be directly read off \Fig{fig:SingleMeasurement}.

\subsection{Density Matrix}
\label{sec:DensityMatrix}

Consider Figs.\ \ref{fig:DensityMatrix} and~\ref{fig:ObsProbTrace},
which are regroupings of \Fig{fig:SingleMeasurement}.
The exterior function of the left-hand dashed box in these figures 
is the density matrix $\rho$ of quantum mechanics,
which can be decomposed into
\begin{equation} \label{eqn:DensityMatrixDecomp}
\rho = \sum_{x} p(x) U(\rangedot,x) U(\rangedot,x)^\H
\end{equation}
(cf.\ \Fig{fig:SpectralDecomp})
and which satisfies
\begin{IEEEeqnarray}{rCl}
\tr(\rho) & = & \sum_x p(x) \tr\!\left(U(\rangedot,x) U(\rangedot,x)^\H\right) \IEEEeqnarraynumspace\\
 & = & \sum_x p(x) \tr\!\left(U(\rangedot,x)^\H U(\rangedot,x)\right) \\
 & = & \sum_x p(x)\, \|(U(\rangedot,x)\|^2 \\
 & = & \sum_x p(x) \\
 & = & 1.  \label{eqn:DensityMatrixTrace}
\end{IEEEeqnarray}

The exterior function of the right-hand dashed box in \Fig{fig:DensityMatrix}
is an identity matrix (i.e., an equality constraint function), 
as is obvious from the reductions of 
Figs.\ \ref{fig:RedHalfEdgeEqu} and \ref{fig:RedMatrixMultEqu}. 
It is then obvious (cf.\ \Fig{fig:FactorGraphTrace}) 
that the partition sum of \Fig{fig:DensityMatrix}
is $\tr(\rho)$, which equals~1 by (\ref{eqn:DensityMatrixTrace}). 
(But we already established in Section~\ref{sec:BornsRule} 
that the partition sum of Figs.\ \ref{fig:SingleMeasurement} 
and \ref{fig:DensityMatrix} is~1.)

The exterior function of the right-hand dashed box 
in \Fig{fig:ObsProbTrace} (with fixed $Y=y$) is 
the matrix $B(\rangedot,y)B(\rangedot,y)^\H$.
From \Fig{fig:SingleMeasurement}, we know that the partition sum 
of \Fig{fig:ObsProbTrace} is $\sum_x p(x,y) = p(y)$. 
Using \Fig{fig:FactorGraphTrace}, this partition sum can be expressed as
\begin{IEEEeqnarray}{rCl}
p(y) & = & \tr\!\left( \rho B(\rangedot,y)B(\rangedot,y)^\H \right) \\
& = & \tr\!\left( B(\rangedot,y)^\H \rho B(\rangedot,y) \right) \IEEEeqnarraynumspace\\
& = & B(\rangedot,y)^\H \rho B(\rangedot,y). \label{eqn:ProbFromDensity}
\end{IEEEeqnarray}
Plugging (\ref{eqn:DensityMatrixDecomp})
into (\ref{eqn:ProbFromDensity}) is, of course, 
consistent with (\ref{eqn:ElementaryProbability}).

\subsection{Observables}

In most standard formulations of quantum mechanics, 
the outcome of a physical experiment is not $Y$ 
as in \Fig{fig:SingleMeasurement}, but some 
(essentially arbitrary) real-valued function $g(Y)$.

In \Fig{fig:ObsExp}, we have augmented \Fig{fig:SingleMeasurement} by 
a corresponding factor $g(Y)$. The partition sum of \Fig{fig:ObsExp} is thus
\begin{equation} \label{eqn:ObsExp}
\E[g(Y)] = \sum_{y} p(y) g(y),
\end{equation}
cf.\ \Fig{fig:FGEgX}.
Regrouping \Fig{fig:ObsExp} as in \Fig{fig:ObsProbTrace} yields 
\Fig{fig:ObsTrace}, the partition sum of which is
\begin{equation} \label{eqn:ObsTrace}
\E[g(Y)] = \tr\!\left( \rho O \right),
\end{equation}
where the matrix $O$ is the right-hand dashed box in \Fig{fig:ObsProbTrace}. 
Note that, by the spectral theorem,
every Hermitian matrix $O$ can be represented as in \Fig{fig:ObsTrace}
(cf.\ \Fig{fig:SpectralDecomp})
and $g(1), \ldots, g(M)$ are the eigenvalues of $O$.

In this paper, however, we will focus on probabilities 
and we will not further use such expectations.

\begin{figure}[t]
\begin{center}
\begin{picture}(84,30)(-4,-6.5)

\put(0,7.5){\framebox(5,5){}}  \put(2.5,6.5){\pos{ct}{$p(x)$}}
\put(5,10){\line(1,0){10}}     \put(10,11){\pos{cb}{$X$}}
\put(15,7.5){\framebox(5,5){$=$}}

\put(17.5,12.5){\line(0,1){5}}
\put(17.5,17.5){\line(1,0){10}}
\put(27.5,15){\framebox(5,5){}}  \put(30,13.5){\pos{ct}{$U$}}
 \put(32.5,17.5){\markerDot}
\put(32.5,17.5){\line(1,0){10}}
\put(42.5,15){\framebox(5,5){}}  \put(45,13.5){\pos{ct}{$B^\H$}}
 \put(47.5,17.5){\markerDot}
\put(47.5,17.5){\line(1,0){10}}
\put(57.5,17.5){\line(0,-1){5}}

\put(17.5,7.5){\line(0,-1){5}}
\put(17.5,2.5){\line(1,0){10}}
\put(27.5,0){\framebox(5,5){}}   \put(30,-1.5){\pos{ct}{$U^\H$}}
 \put(27.5,2.5){\markerDot}
\put(32.5,2.5){\line(1,0){10}}
\put(42.5,0){\framebox(5,5){}}   \put(45,-1.5){\pos{ct}{$B$}}
 \put(42.5,2.5){\markerDot}
\put(47.5,2.5){\line(1,0){10}}
\put(57.5,2.5){\line(0,1){5}}

\put(55,7.5){\framebox(5,5){$=$}}
\put(60,10){\line(1,0){15}}      \put(69.5,11.25){\pos{cb}{$Y$}}
\put(75,7.5){\framebox(5,5){}}   \put(77.5,6.5){\pos{ct}{$g(y)$}}

\put(-4,-6.5){\dashbox(68,30){}} \put(65.5,-5){\pos{bl}{$p(y)$}}
\end{picture}
\caption{\label{fig:ObsExp}%
Factor graph of expectation (\ref{eqn:ObsExp}).}
\end{center}
\vspace{\floatsep}

\begin{center}
\begin{picture}(88,34)(-4,-6.5)

\put(0,7.5){\framebox(5,5){}}  \put(2.5,6.5){\pos{ct}{$p(x)$}}
\put(5,10){\line(1,0){10}}     \put(10,11){\pos{cb}{$X$}}
\put(15,7.5){\framebox(5,5){$=$}}

\put(17.5,12.5){\line(0,1){5}}
\put(17.5,17.5){\line(1,0){10}}
\put(27.5,15){\framebox(5,5){}}  \put(30,13.5){\pos{ct}{$U$}}
 \put(32.5,17.5){\markerDot}
\put(32.5,17.5){\line(1,0){15}}
\put(47.5,15){\framebox(5,5){}}  \put(50,13.5){\pos{ct}{$B^\H$}}
 \put(52.5,17.5){\markerDot}
\put(52.5,17.5){\line(1,0){10}}
\put(62.5,17.5){\line(0,-1){5}}

\put(17.5,7.5){\line(0,-1){5}}
\put(17.5,2.5){\line(1,0){10}}
\put(27.5,0){\framebox(5,5){}}   \put(30,-1.5){\pos{ct}{$U^\H$}}
 \put(27.5,2.5){\markerDot}
\put(32.5,2.5){\line(1,0){15}}
\put(47.5,0){\framebox(5,5){}}   \put(50,-1.5){\pos{ct}{$B$}}
 \put(47.5,2.5){\markerDot}
\put(52.5,2.5){\line(1,0){10}}
\put(62.5,2.5){\line(0,1){5}}

\put(60,7.5){\framebox(5,5){$=$}}
\put(65,10){\line(1,0){10}}      \put(70,11.25){\pos{cb}{$Y$}}
\put(75,7.5){\framebox(5,5){}}   \put(77.5,6.5){\pos{ct}{$g(y)$}}

\put(-4,-6.5){\dashbox(40.5,30){}}   \put(16,25){\pos{cb}{$\rho$}}
\put(43.5,-6.5){\dashbox(40.5,30){}}  \put(63.5,25){\pos{cb}{$O$}}

\end{picture}
\caption{\label{fig:ObsTrace}%
Factor graph of expectation (\ref{eqn:ObsTrace}) with general 
Hermitian matrix $O$.}
\end{center}
\end{figure}

\subsection{Evolution over Time: Schr\"odinger, Heisenberg, Feynman}

Consider the factor graph of \Fig{fig:MultiStepUnitaryModel}, 
which agrees with \Fig{fig:SingleMeasurement} except 
that the matrix $U$ is expanded into the product $U=U_n\cdots U_1$.
One interpretation of this factor graph is that 
the initial state $X$ 
evolves unitarily over $n$ discrete time steps 
until it is measured by a projection measurement 
as in \Fig{fig:SingleMeasurement}.
Note that a continuous-time picture may be obtained, if desired, 
by a suitable limit with $n\rightarrow \infty$.

In this setting, the so-called Schr\"odinger and Heisenberg pictures 
correspond to sequentially closing boxes 
(from the innermost dashed box to the outermost dashed box) 
as in Figs.~\ref{fig:SchroedingerPicture} and \ref{fig:HeisenbergPicture}, 
respectively; the former propagates the quantum state $\psi$ 
(or the density matrix $\rho$) 
forward in time while the latter propagates the measurement backwards in time.
The resulting probability distribution over $Y$ is identical 
by Proposition~\ref{prop:FGMain}.

\begin{figure*}
\setlength{\unitlength}{0.9mm}
\begin{center}
\begin{picture}(110,36)(0,-7.5)
\put(0,10){\framebox(5,5){}}      \put(2.5,8.5){\pos{ct}{$p(x)$}}
\put(5,12.5){\line(1,0){10}}         \put(10,14){\pos{cb}{$X$}}
\put(15,10){\framebox(5,5){$=$}}
\put(17.5,15){\line(0,1){7.5}}
\put(17.5,22.5){\line(1,0){10}}
\put(27.5,20){\framebox(5,5){}}    \put(30,18.5){\pos{ct}{$U_1$}}
 \put(32.5,22.5){\markerDot}
\put(32.5,22.5){\line(1,0){10}}
\put(42.5,20){\framebox(5,5){}}    \put(45,18.5){\pos{ct}{$U_2$}}
 \put(47.5,22.5){\markerDot}
\put(47.5,22.5){\line(1,0){5}}
\put(57.5,22.5){\cent{\ldots}}
\put(62.5,22.5){\line(1,0){5}}
\put(67.5,20){\framebox(5,5){}}    \put(70,18.5){\pos{ct}{$U_n$}}
 \put(72.5,22.5){\markerDot}
\put(72.5,22.5){\line(1,0){10}}
\put(82.5,20){\framebox(5,5){}}    \put(85,18.5){\pos{ct}{$B^\H$}}
 \put(87.5,22.5){\markerDot}
\put(87.5,22.5){\line(1,0){10}}
\put(97.5,22.5){\line(0,-1){7.5}}
\put(17.5,10){\line(0,-1){7.5}}
\put(17.5,2.5){\line(1,0){10}}
\put(27.5,0){\framebox(5,5){}}     \put(30,-1.5){\pos{ct}{$U_1^\H$}}
 \put(27.5,2.5){\markerDot}
\put(32.5,2.5){\line(1,0){10}}
\put(42.5,0){\framebox(5,5){}}     \put(45,-1.5){\pos{ct}{$U_2^\H$}}
 \put(42.5,2.5){\markerDot}
\put(47.5,2.5){\line(1,0){5}}
\put(57.5,2.5){\cent{\ldots}}
\put(62.5,2.5){\line(1,0){5}}
\put(67.5,0){\framebox(5,5){}}     \put(70,-1.5){\pos{ct}{$U_n^\H$}}
 \put(67.5,2.5){\markerDot}
\put(72.5,2.5){\line(1,0){10}}
\put(82.5,0){\framebox(5,5){}}     \put(85,-1.5){\pos{ct}{$B$}}
 \put(82.5,2.5){\markerDot}
\put(87.5,2.5){\line(1,0){10}}
\put(97.5,2.5){\line(0,1){7.5}}
\put(95,10){\framebox(5,5){$=$}}
\put(100,12.5){\line(1,0){10}}        \put(107.5,14){\pos{cb}{$Y$}}
\put(23.5,13){\dashbox(53,15.5){}}
\put(23.5,-7.5){\dashbox(53,16){}}
\end{picture}
\vspace{1ex}
\caption{\label{fig:MultiStepUnitaryModel}%
Elementary quantum mechanics: 
unitary evolution over time in $n$ steps followed by a single projection measurement.}
\end{center}
\vspace{\dblfloatsep}

\begin{center}
\begin{picture}(130,50)(-15,-16.5)
\put(0,7.5){\framebox(5,5){}}      \put(2.5,6){\pos{ct}{$p(x)$}}
\put(5,10){\line(1,0){10}}         \put(10,11.5){\pos{cb}{$X$}}
\put(15,7.5){\framebox(5,5){$=$}}
\put(17.5,12.5){\line(0,1){5}}
\put(17.5,17.5){\line(1,0){10}}
\put(27.5,15){\framebox(5,5){}}    \put(30,13.5){\pos{ct}{$U_1$}}
 \put(32.5,17.5){\markerDot}
\put(32.5,17.5){\line(1,0){10}}
\put(42.5,15){\framebox(5,5){}}    \put(45,13.5){\pos{ct}{$U_2$}}
 \put(47.5,17.5){\markerDot}
\put(47.5,17.5){\line(1,0){8.625}}
\put(61.25,17.5){\cent{\ldots}}
\put(65,17.5){\line(1,0){5}}
\put(70,15){\framebox(5,5){}}    \put(72.5,13.5){\pos{ct}{$U_n$}}
 \put(75,17.5){\markerDot}
\put(75,17.5){\line(1,0){10}}
\put(85,15){\framebox(5,5){}}    \put(87.5,13.5){\pos{ct}{$B^\H$}}
 \put(90,17.5){\markerDot}
\put(90,17.5){\line(1,0){12.5}}
\put(102.5,17.5){\line(0,-1){5}}
\put(17.5,7.5){\line(0,-1){5}}
\put(17.5,2.5){\line(1,0){10}}
\put(27.5,0){\framebox(5,5){}}     \put(30,-1.5){\pos{ct}{$U_1^\H$}}
 \put(27.5,2.5){\markerDot}
\put(32.5,2.5){\line(1,0){10}}
\put(42.5,0){\framebox(5,5){}}     \put(45,-1.5){\pos{ct}{$U_2^\H$}}
 \put(42.5,2.5){\markerDot}
\put(47.5,2.5){\line(1,0){8.625}}
\put(61.25,2.5){\cent{\ldots}}
\put(65,2.5){\line(1,0){5}}
\put(70,0){\framebox(5,5){}}     \put(72.5,-1.5){\pos{ct}{$U_n^\H$}}
 \put(70,2.5){\markerDot}
\put(75,2.5){\line(1,0){10}}
\put(85,0){\framebox(5,5){}}     \put(87.5,-1.5){\pos{ct}{$B$}}
 \put(85,2.5){\markerDot}
\put(90,2.5){\line(1,0){12.5}}
\put(102.5,2.5){\line(0,1){5}}
\put(100,7.5){\framebox(5,5){$=$}}
\put(105,10){\line(1,0){10}}        \put(112.5,11.5){\pos{cb}{$Y$}}
\put(-5,-7.5){\dashbox(41.5,32){}}    \put(37.5,24.5){\pos{tl}{$\rho_1$}}
\put(-10,-11.5){\dashbox(61.5,40){}}  \put(52.5,28.5){\pos{tl}{$\rho_2$}}
\put(-15,-16.5){\dashbox(94,50){}}    \put(80,33.5){\pos{tl}{$\rho_n$}}
\end{picture}
\vspace{1ex}
\caption{\label{fig:SchroedingerPicture}%
Schr\"{o}dinger picture.}
\end{center}
\vspace{\dblfloatsep}

\begin{center}
\visual\vspace{-5mm}
\begin{picture}(110,64)(2.5,-13)
\put(20,37.5){\knownBox}           \put(18,37.5){\pos{cr}{$x$}}
\put(20,37.5){\line(1,0){7.5}}
\put(27.5,35){\framebox(5,5){}}    \put(30,33.5){\pos{ct}{$U_1$}}
 \put(32.5,37.5){\markerDot}
\put(32.5,37.5){\line(1,0){10}}
\put(42.5,35){\framebox(5,5){}}    \put(45,33.5){\pos{ct}{$U_2$}}
 \put(47.5,37.5){\markerDot}
\put(47.5,37.5){\line(1,0){8.75}}
\put(61,37.5){\cent{\ldots}}
\put(65,37.5){\line(1,0){5}}
\put(70,35){\framebox(5,5){}}    \put(72.5,33.5){\pos{ct}{$U_n$}}
 \put(75,37.5){\markerDot}
\put(75,37.5){\line(1,0){10}}
\put(85,35){\framebox(5,5){}}    \put(87.5,33.5){\pos{ct}{$B^\H$}}
 \put(90,37.5){\markerDot}
\put(90,37.5){\line(1,0){12.5}}
\put(102.5,37.5){\line(0,-1){15}}
\put(12.5,28){\dashbox(24,17){}}   \put(37.5,45){\pos{tl}{$\psi_1$}}
\put(7.5,25.5){\dashbox(44,22){}}  \put(52.5,47.5){\pos{tl}{$\psi_2$}}
\put(2.5,23){\dashbox(76.5,28){}}  \put(80,51){\pos{tl}{$\psi_n$}}
\put(20,2.5){\knownBox}            \put(18,2.5){\pos{cr}{$x$}}
\put(20,2.5){\line(1,0){7.5}}
\put(27.5,0){\framebox(5,5){}}     \put(30,-1.5){\pos{ct}{$U_1^\H$}}
 \put(27.5,2.5){\markerDot}
\put(32.5,2.5){\line(1,0){10}}
\put(42.5,0){\framebox(5,5){}}     \put(45,-1.5){\pos{ct}{$U_2^\H$}}
 \put(42.5,2.5){\markerDot}
\put(47.5,2.5){\line(1,0){8.75}}
\put(61,2.5){\cent{\ldots}}
\put(65,2.5){\line(1,0){5}}
\put(70,0){\framebox(5,5){}}     \put(72.5,-1.5){\pos{ct}{$U_n^\H$}}
 \put(70,2.5){\markerDot}
\put(75,2.5){\line(1,0){10}}
\put(85,0){\framebox(5,5){}}     \put(87.5,-1.5){\pos{ct}{$B$}}
 \put(85,2.5){\markerDot}
\put(90,2.5){\line(1,0){12.5}}
\put(102.5,2.5){\line(0,1){15}}
\put(12.5,-7.5){\dashbox(24,17){}}  \put(37.5,9.5){\pos{tl}{$\psi_1^\H$}}
\put(7.5,-10){\dashbox(44,22){}}    \put(52.5,12){\pos{tl}{$\psi_2^\H$}}
\put(2.5,-13){\dashbox(76.5,28){}}  \put(80,15){\pos{tl}{$\psi_n^\H$}}
\put(100,17.5){\framebox(5,5){$=$}}
\put(105,20){\line(1,0){10}}        \put(112.5,21.5){\pos{cb}{$Y$}}
\end{picture}
\vspace{1ex}
\caption{\label{fig:SchroedingerPictureKnownInitial}%
Schr\"{o}dinger picture with known initial state $X=x$ 
and unitarily evolving quantum state (or wave function) $\psi$.}
\end{center}
\vspace{\dblfloatsep}

\begin{center}
\begin{picture}(110,50)(-2.5,-16)
\put(-2.5,7.5){\framebox(5,5){}}   \put(0,6){\pos{ct}{$p(x)$}}
\put(2.5,10){\line(1,0){10}}         \put(7.5,11.5){\pos{cb}{$X$}}
\put(12.5,7.5){\framebox(5,5){$=$}}
\put(15,12.5){\line(0,1){5}}
\put(15,17.5){\line(1,0){12.5}}
\put(27.5,15){\framebox(5,5){}}    \put(30,13.5){\pos{ct}{$U_1$}}
 \put(32.5,17.5){\markerDot}
\put(32.5,17.5){\line(1,0){10}}
\put(42.5,15){\framebox(5,5){}}    \put(45,13.5){\pos{ct}{$U_2$}}
 \put(47.5,17.5){\markerDot}
\put(47.5,17.5){\line(1,0){5}}
\put(57.5,17.5){\cent{\ldots}}
\put(62.5,17.5){\line(1,0){10}}
\put(72.5,15){\framebox(5,5){}}    \put(75,13.5){\pos{ct}{$B^\H$}}
 \put(77.5,17.5){\markerDot}
\put(77.5,17.5){\line(1,0){10}}
\put(87.5,17.5){\line(0,-1){5}}
\put(15,7.5){\line(0,-1){5}}
\put(15,2.5){\line(1,0){12.5}}
\put(27.5,0){\framebox(5,5){}}     \put(30,-1.5){\pos{ct}{$U_1^\H$}}
 \put(27.5,2.5){\markerDot}
\put(32.5,2.5){\line(1,0){10}}
\put(42.5,0){\framebox(5,5){}}     \put(45,-1.5){\pos{ct}{$U_2^\H$}}
 \put(42.5,2.5){\markerDot}
\put(47.5,2.5){\line(1,0){5}}
\put(57.5,2.5){\cent{\ldots}}
\put(62.5,2.5){\line(1,0){10}}
\put(72.5,0){\framebox(5,5){}}     \put(75,-1.5){\pos{ct}{$B$}}
 \put(72.5,2.5){\markerDot}
\put(77.5,2.5){\line(1,0){10}}
\put(87.5,2.5){\line(0,1){5}}
\put(85,7.5){\framebox(5,5){$=$}}
\put(90,10){\line(1,0){17.5}}        \put(105,11.5){\pos{cb}{$Y$}}
\put(67.5,-6){\dashbox(27.5,30){}}
\put(37.5,-11){\dashbox(60,40){}}
\put(22.5,-16){\dashbox(77.5,50){}}
\end{picture}
\vspace{1ex}
\caption{\label{fig:HeisenbergPicture}%
Heisenberg picture.}
\end{center}
\end{figure*}

Both the Schr\"odinger picture and the Heisenberg picture 
can be reduced to sum-product message passing 
in a cycle-free graph as follows. 
In the Schr\"odinger picture, assume first that the initial state $X$ is known. 
In this case, we obtain the cycle-free factor graph of 
\Fig{fig:SchroedingerPictureKnownInitial}, in which 
$p(y \cond x)$ is easily computed by left-to-right sum-product message passing 
(cf.\ \cite{KFL:fg2001,Lg:ifg2004}), which amounts 
to a sequence of matrix-times-vector multiplications
\begin{equation}
\psi_k = U_k \psi_{k-1}
\end{equation}
with $\psi_1 \eqdef U_1(\rangedot,x)$ (= column $x$ of $U_1$). 
The quantities $\psi_1,\ldots,\psi_n$ 
in \Fig{fig:SchroedingerPictureKnownInitial} 
are the wave functions 
propagated up to the corresponding time.
Since \Fig{fig:SchroedingerPictureKnownInitial} 
consists of two complex conjugate parts, 
it suffices to carry out these computations
for one of the two parts.

If the initial state $X$ is not known, we write
\begin{equation} 
p(y) = \sum_{x} p(x) p(y \cond x),
\end{equation}
and each term $p(y \cond x)$ can be computed 
as in \Fig{fig:SchroedingerPictureKnownInitial}. 
This decomposition carries over to the relation 
\begin{IEEEeqnarray}{rCl}
\rho_k(x',x'') & = & \sum_x p(x) \psi_k(x') \psi_k^\H(x'') \\
 & = & \sum_x p(x) \psi_k(x') \ccj{\psi_k(x'')}
\end{IEEEeqnarray}
between the wave function $\psi_k$
and the density matrix $\rho_k$ 
(see Figs.~\ref{fig:SchroedingerPicture} and~\ref{fig:SchroedingerPictureKnownInitial})
for $k=1,\ldots,n$.

In the Heisenberg picture (\Fig{fig:HeisenbergPicture}), 
we can proceed analogously. For any fixed $Y=y$, 
this value can be plugged into the factors/matrices $B$ and $B^\H$,
which turns \Fig{fig:HeisenbergPicture} into a cycle-free factor graph 
that looks almost like a time-reversed version of 
\Fig{fig:SchroedingerPictureKnownInitial}.
In consequence, $p(y)$ can be computed by right-to-left sum-product message passing, 
which again amounts to a sequence of matrix-times-vector multiplications.

Finally, we note that the dashed boxes in \Fig{fig:MultiStepUnitaryModel} 
encode Feynman's path integral in its most elementary embodiment. 
Each internal configuration (i.e., an assignment of values to all variables)
in such a box may be viewed as a ``path'', and the corresponding 
product of all factors inside the box may be viewed as the (complex) weight 
of the path. 
The exterior function of the box is (by definition) the sum, 
over all internal configurations/paths,
of the weight of each configuration/path.

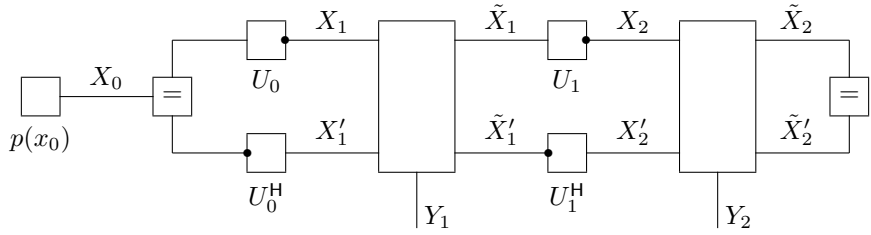
\begin{figure*}[p]
\centering
\begin{picture}(112.5,30)(0,-7.5)
\put(0,7.5){\framebox(5,5){}}      \put(2.5,6){\pos{ct}{$p(x_0)$}}
\put(5,10){\line(1,0){12.5}}       \put(11.25,11){\pos{cb}{$X_0$}}
\put(20,17.5){\line(1,0){10}}
 \put(20,17.5){\line(0,-1){5}}
\put(17.5,7.5){\framebox(5,5){$=$}}
 \put(20,7.5){\line(0,-1){5}}
\put(20,2.5){\line(1,0){10}}
\put(30,15){\framebox(5,5){}}     \put(32.5,13.5){\pos{ct}{$U_0$}}
 \put(35,17.5){\markerDot}
\put(35,17.5){\line(1,0){12.5}}   \put(41.25,18.5){\pos{cb}{$X_1$}}
\put(30,0){\framebox(5,5){}}      \put(32.5,-1.5){\pos{ct}{$U_0^\H$}}
 \put(30,2.5){\markerDot}
\put(35,2.5){\line(1,0){12.5}}    \put(41.25,3.5){\pos{cb}{$X_1'$}}
\put(47.5,0){\framebox(10,20){}}
\put(52.5,0){\line(0,-1){7.5}}    \put(53.5,-7.5){\pos{bl}{$Y_1$}}
\put(57.5,17.5){\line(1,0){12.5}} \put(63.75,18.5){\pos{cb}{$\tilde{X}_1$}}
\put(70,15){\framebox(5,5){}}     \put(72.5,13.5){\pos{ct}{$U_1$}}
 \put(75,17.5){\markerDot}
\put(75,17.5){\line(1,0){12.5}}   \put(81.25,18.5){\pos{cb}{$X_2$}}
\put(57.5,2.5){\line(1,0){12.5}}  \put(63.75,3.5){\pos{cb}{$\tilde{X}_1'$}}
\put(70,0){\framebox(5,5){}}      \put(72.5,-1.5){\pos{ct}{$U_1^\H$}}
 \put(70,2.5){\markerDot}
\put(75,2.5){\line(1,0){12.5}}    \put(81.25,3.5){\pos{cb}{$X_2'$}}
\put(87.5,0){\framebox(10,20){}}
\put(92.5,0){\line(0,-1){7.5}}    \put(93.5,-7.5){\pos{bl}{$Y_2$}}
\put(97.5,17.5){\line(1,0){12.5}}  \put(103,18.5){\pos{cb}{$\tilde{X}_2$}}
\put(110,17.5){\line(0,-1){5}}
\put(107.5,7.5){\framebox(5,5){$=$}}
\put(110,7.5){\line(0,-1){5}}
\put(97.5,2.5){\line(1,0){12.5}}   \put(103,3.5){\pos{cb}{$\tilde{X}_2'$}}
\end{picture}
\caption{\label{fig:GenQFG}%
Factor graph of a quantum system with two measurements and the corresponding observations $Y_1$ and $Y_2$.
}
\end{figure*}

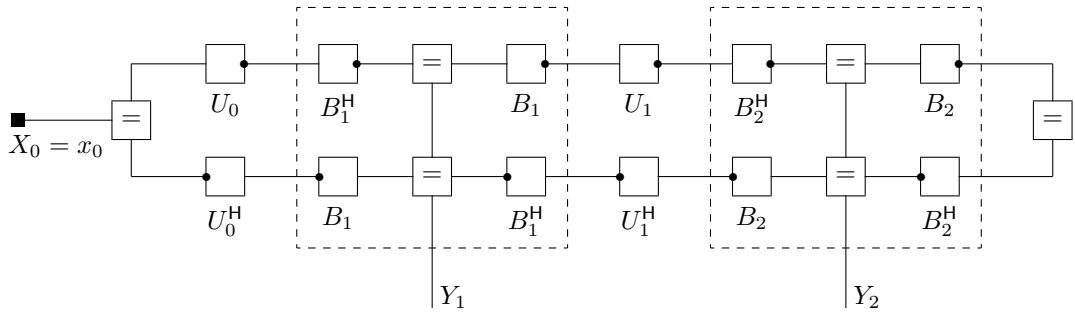
\begin{figure*}
\centering
\begin{picture}(140,40)(-10,-15)
\put(-10,10){\knownBox}
\put(-10,10){\line(1,0){12.5}}
 \put(-5,8){\pos{tc}{$X_0=x_0$}}
\put(5,17.5){\line(1,0){10}}
 \put(5,17.5){\line(0,-1){5}}
\put(2.5,7.5){\framebox(5,5){$=$}}
 \put(5,7.5){\line(0,-1){5}}
\put(5,2.5){\line(1,0){10}}
\put(15,15){\framebox(5,5){}}        \put(17.5,13.5){\pos{ct}{$U_0$}}
 \put(20,17.5){\markerDot}
\put(20,17.5){\line(1,0){10}}
\put(30,15){\framebox(5,5){}}        \put(32.5,13.5){\pos{ct}{$B_1^\H$}}
 \put(35,17.5){\markerDot}
\put(35,17.5){\line(1,0){7.5}}
\put(42.5,15){\framebox(5,5){$=$}}
 \put(45,15){\line(0,-1){10}}
\put(47.5,17.5){\line(1,0){7.5}}
\put(55,15){\framebox(5,5){}}        \put(57.5,13.5){\pos{ct}{$B_1$}}
 \put(60,17.5){\markerDot}
\put(60,17.5){\line(1,0){10}}
\put(70,15){\framebox(5,5){}}        \put(72.5,13.5){\pos{ct}{$U_1$}}
 \put(75,17.5){\markerDot}
\put(75,17.5){\line(1,0){10}}
\put(85,15){\framebox(5,5){}}        \put(87.5,13.5){\pos{ct}{$B_2^\H$}}
 \put(90,17.5){\markerDot}
\put(90,17.5){\line(1,0){7.5}}
\put(97.5,15){\framebox(5,5){$=$}}
 \put(100,15){\line(0,-1){10}}
\put(102.5,17.5){\line(1,0){7.5}}
\put(110,15){\framebox(5,5){}}       \put(112.5,13.5){\pos{ct}{$B_2$}}
 \put(115,17.5){\markerDot}
\put(115,17.5){\line(1,0){12.5}}
\put(127.5,17.5){\line(0,-1){5}}
\put(125,7.5){\framebox(5,5){$=$}}
\put(127.5,7.5){\line(0,-1){5}}
\put(115,2.5){\line(1,0){12.5}}
\put(27,-7){\dashbox(36,32){}}
\put(82,-7){\dashbox(36,32){}}
\put(15,0){\framebox(5,5){}}         \put(17.5,-1.5){\pos{ct}{$U_0^\H$}}
 \put(15,2.5){\markerDot}
\put(20,2.5){\line(1,0){10}}
\put(30,0){\framebox(5,5){}}         \put(32.5,-1.5){\pos{ct}{$B_1$}}
 \put(30,2.5){\markerDot}
\put(35,2.5){\line(1,0){7.5}}
\put(42.5,0){\framebox(5,5){$=$}}
 \put(45,0){\line(0,-1){15}}         \put(46,-15){\pos{bl}{$Y_1$}}
\put(47.5,2.5){\line(1,0){7.5}}
\put(55,0){\framebox(5,5){}}         \put(57.5,-1.5){\pos{ct}{$B_1^\H$}}
 \put(55,2.5){\markerDot}
\put(60,2.5){\line(1,0){10}}
\put(70,0){\framebox(5,5){}}         \put(72.5,-1.5){\pos{ct}{$U_1^\H$}}
 \put(70,2.5){\markerDot}
\put(75,2.5){\line(1,0){10}}
\put(85,0){\framebox(5,5){}}         \put(87.5,-1.5){\pos{ct}{$B_2$}}
 \put(85,2.5){\markerDot}
\put(90,2.5){\line(1,0){7.5}}
\put(97.5,0){\framebox(5,5){$=$}}
 \put(100,0){\line(0,-1){15}}        \put(101,-15){\pos{bl}{$Y_2$}}
\put(102.5,2.5){\line(1,0){7.5}}
\put(110,0){\framebox(5,5){}}        \put(112.5,-1.5){\pos{ct}{$B_2^\H$}}
 \put(110,2.5){\markerDot}
\end{picture}
\caption{\label{fig:BasicQMFG}%
Important special case of \Fig{fig:GenQFG}:
all matrices are unitary and the initial state $X_0=x_0$ is known. 
In quantum-mechanical terms, 
such measurements are projection measurements with one-dimensional eigenspaces. 
}
\end{figure*}

\begin{figure*}[p]
\centering
\begin{picture}(120,40)(-5,-10)
\put(0,7.5){\framebox(5,5){}}      \put(2.5,6){\pos{ct}{$p(x_0)$}}
\put(5,10){\line(1,0){12.5}}       \put(11.25,11){\pos{cb}{$X_0$}}
\put(20,17.5){\line(1,0){10}}
 \put(20,17.5){\line(0,-1){5}}
\put(17.5,7.5){\framebox(5,5){$=$}}
 \put(20,7.5){\line(0,-1){5}}
\put(20,2.5){\line(1,0){10}}
\put(30,15){\framebox(5,5){}}     \put(32.5,13.5){\pos{ct}{$U_0$}}
 \put(35,17.5){\markerDot}
\put(35,17.5){\line(1,0){12.5}}   \put(42,18.5){\pos{cb}{$X_1$}}
\put(30,0){\framebox(5,5){}}      \put(32.5,-1.5){\pos{ct}{$U_0^\H$}}
 \put(30,2.5){\markerDot}
\put(35,2.5){\line(1,0){12.5}}    \put(42,3.5){\pos{cb}{$X_1'$}}
\put(47.5,0){\framebox(10,20){}}
\put(52.5,0){\line(0,-1){7.5}}    \put(53.5,-7.5){\pos{bl}{$Y_1$}}
\put(57.5,17.5){\line(1,0){12.5}} \put(63,18.5){\pos{cb}{$\tilde{X}_1$}}
\put(70,15){\framebox(5,5){}}     \put(72.5,13.5){\pos{ct}{$U_1$}}
 \put(75,17.5){\markerDot}
\put(75,17.5){\line(1,0){12.5}}   \put(81.25,18.5){\pos{cb}{$X_2$}}
\put(57.5,2.5){\line(1,0){12.5}}  \put(63,3.5){\pos{cb}{$\tilde{X}_1'$}}
\put(70,0){\framebox(5,5){}}      \put(72.5,-1.5){\pos{ct}{$U_1^\H$}}
 \put(70,2.5){\markerDot}
\put(75,2.5){\line(1,0){12.5}}    \put(81.25,3.5){\pos{cb}{$X_2'$}}
\put(87.5,0){\framebox(10,20){}}
\put(92.5,0){\line(0,-1){7.5}}    \put(93.5,-7.5){\pos{bl}{$Y_2$}}
\put(97.5,17.5){\line(1,0){12.5}}  \put(103,18.5){\pos{cb}{$\tilde{X}_2$}}
\put(110,17.5){\line(0,-1){5}}
\put(107.5,7.5){\framebox(5,5){$=$}}
\put(110,7.5){\line(0,-1){5}}
\put(97.5,2.5){\line(1,0){12.5}}   \put(103,3.5){\pos{cb}{$\tilde{X}_2'$}}
\put(-5,-10){\dashbox(43,35){}}    \put(16.5,26.5){\pos{cb}{$\rho_1$}}
\put(67,-10){\dashbox(48,35){}}    \put(91,26.5){\pos{cb}{$f_=$}}
\end{picture}
\caption{\label{fig:BackwardMesg}%
The exterior function of the dashed box on the left is the density matrix $\rho_1(x_1,x_1')$.
The exterior function of the dashed box on the right 
is $f_=(\tilde{x}_1, \tilde{x}_1')$ (assuming that $Y_2$ is unknown).
}
\end{figure*}
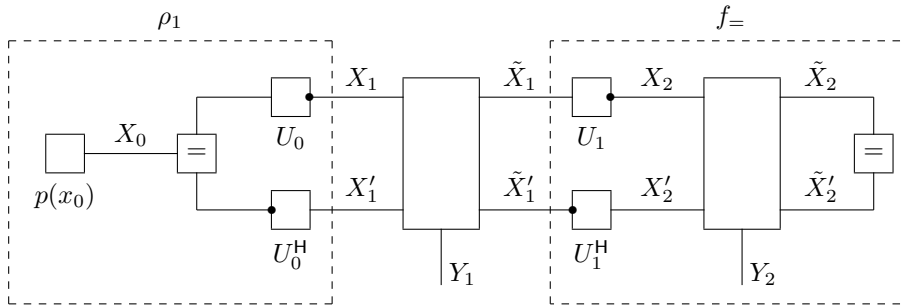

\begin{figure*}[p]
\centering
\begin{picture}(117.5,40)(-5,-10)
\put(0,7.5){\framebox(5,5){}}      \put(2.5,6){\pos{ct}{$p(x_0)$}}
\put(5,10){\line(1,0){12.5}}       \put(11.25,11){\pos{cb}{$X_0$}}
\put(20,17.5){\line(1,0){10}}
 \put(20,17.5){\line(0,-1){5}}
\put(17.5,7.5){\framebox(5,5){$=$}}
 \put(20,7.5){\line(0,-1){5}}
\put(20,2.5){\line(1,0){10}}
\put(30,15){\framebox(5,5){}}     \put(32.5,13.5){\pos{ct}{$U_0$}}
 \put(35,17.5){\markerDot}
\put(35,17.5){\line(1,0){12.5}}   \put(41.25,18.5){\pos{cb}{$X_1$}}
\put(30,0){\framebox(5,5){}}      \put(32.5,-1.5){\pos{ct}{$U_0^\H$}}
 \put(30,2.5){\markerDot}
\put(35,2.5){\line(1,0){12.5}}    \put(41.25,3.5){\pos{cb}{$X_1'$}}
\put(47.5,0){\framebox(10,20){}}
\put(52.5,0){\line(0,-1){7.5}}    
 \put(52.5,-7.5){\knownBox}       \put(51.5,-6){\pos{br}{$Y_1=y_1$}}
\put(57.5,17.5){\line(1,0){12.5}} \put(64.5,18.5){\pos{cb}{$\tilde{X}_1$}}
\put(70,15){\framebox(5,5){}}     \put(72.5,13.5){\pos{ct}{$U_1$}}
 \put(75,17.5){\markerDot}
\put(75,17.5){\line(1,0){12.5}}   \put(81.25,18.5){\pos{cb}{$X_2$}}
\put(57.5,2.5){\line(1,0){12.5}}  \put(64.5,3.5){\pos{cb}{$\tilde{X}_1'$}}
\put(70,0){\framebox(5,5){}}      \put(72.5,-1.5){\pos{ct}{$U_1^\H$}}
 \put(70,2.5){\markerDot}
\put(75,2.5){\line(1,0){12.5}}    \put(81.25,3.5){\pos{cb}{$X_2'$}}
\put(87.5,0){\framebox(10,20){}}
\put(92.5,0){\line(0,-1){7.5}}    \put(93.5,-7.5){\pos{bl}{$Y_2$}}
\put(97.5,17.5){\line(1,0){12.5}}  \put(103,18.5){\pos{cb}{$\tilde{X}_2$}}
\put(110,17.5){\line(0,-1){5}}
\put(107.5,7.5){\framebox(5,5){$=$}}
\put(110,7.5){\line(0,-1){5}}
\put(97.5,2.5){\line(1,0){12.5}}   \put(103,3.5){\pos{cb}{$\tilde{X}_2'$}}
\put(-5,-10){\dashbox(65.5,35){}}    \put(27.75,26.5){\pos{cb}{$\breve\rho_1\propto\tilde{\rho}_1$}}
\end{picture}
\caption{\label{fig:PostMeasurementDensityMatrix}%
The exterior function of the dashed box $\breve\rho_1$
equals the density matrix $\tilde{\rho}_1$, up to a scale factor,
after measuring $Y_1=y_1$, cf.\ (\ref{eqn:PostMeasurementDensityMatrix}).
}
\end{figure*}
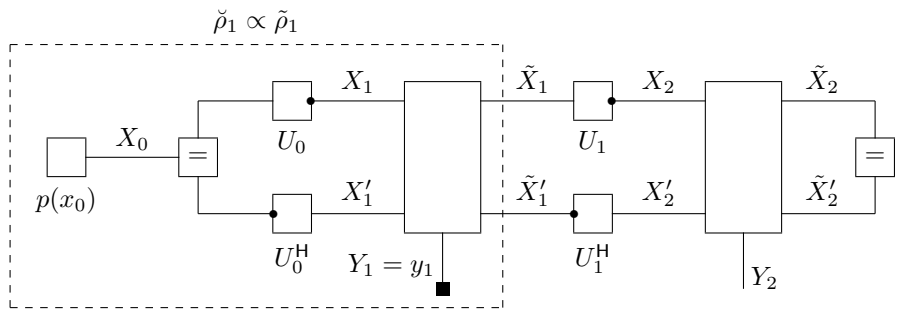

\section{Multiple and More General Measurements}
\label{sec:QMFG}

We now turn to multiple and more general measurements. 
Consider the factor graph of \Fig{fig:GenQFG}. 
In this figure, $U_0$ and $U_1$ are $M\times M$ unitary matrices, 
and all variables except $Y_1$ and $Y_2$ take values in the set $\{ 1,\ldots, M\}$. 
The two large boxes in the figure represent measurements, as will be detailed below. 
The factor/box $p(x_0)$ is a probability mass function over the initial state $X_0$.
We will see that this factor graph 
(with suitable modeling of the measurements) 
represents the joint probability mass function $p(y_1,y_2)$
of a general $M$-dimensional quantum system with two observations $Y_1$ and $Y_2$.
The generalization to more observed variables $Y_1, Y_2, \ldots$ is obvious.

The unitary matrix $U_0$ in \Fig{fig:GenQFG} represents the development 
of the system between the initial state and the first measurement
according to the Schr\"odinger equation; 
the unitary matrix $U_1$ in \Fig{fig:GenQFG} represents the development 
of the system between the two measurements.

In the most basic case, 
the initial state $X_0=x_0$ is known 
and the measurements look as shown in \Fig{fig:BasicQMFG},
where the matrices $B_1$ and $B_2$ are also unitary (cf.\ \Fig{fig:SingleMeasurement}).
In this case, the observed variables $Y_1$ and $Y_2$ 
take values in $\{ 1,\ldots, M\}$ as well. 
Note that the lower part of this factor graph is the complex-conjugate mirror image of the upper part 
(as in \Fig{fig:BasicExampleQFG}).

In quantum-mechanical terms, measurements 
as in \Fig{fig:BasicQMFG} are projection measurements with one-dimensional eigenspaces
(as in Section~\ref{sec:ElementaryQM}). 

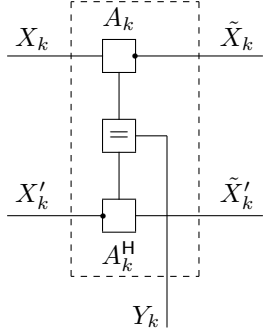
\begin{figure}
\setlength{\unitlength}{0.85mm}
\begin{center}
\begin{picture}(40,51)(0,-10)
\put(0,32.5){\line(1,0){15}}        \put(4,33.5){\pos{cb}{$X_k$}}
\put(15,30){\framebox(5,5){}}       \put(17.5,36.5){\pos{cb}{$A_k$}}
 \put(20,32.5){\markerDot}
\put(20,32.5){\line(1,0){20}}       \put(36,33.5){\pos{cb}{$\tilde{X}_k$}}
\put(17.5,30){\line(0,-1){7.5}}     
\put(15,17.5){\framebox(5,5){$=$}}
 \put(20,20){\line(1,0){5}}
 \put(25,20){\line(0,-1){30}}      \put(24,-10){\pos{br}{$Y_k$}}
\put(17.5,17.5){\line(0,-1){7.5}}
\put(0,7.5){\line(1,0){15}}         \put(4,8.5){\pos{cb}{$X_k'$}}
\put(15,5){\framebox(5,5){}}        \put(17.5,3.5){\pos{ct}{$A_k^\H$}}
 \put(15,7.5){\markerDot}
\put(20,7.5){\line(1,0){20}}        \put(36,8.5){\pos{cb}{$\tilde{X}_k'$}}
\put(10,-2){\dashbox(20,43){}}
\end{picture}
\caption{\label{fig:GenMeasurement}%
General measurement as in \cite[Chap.~2]{NiChuang:QCI}. 
Condition (\ref{eqn:GenMeasurementCond}) must be satisfied.
}
\end{center}
\end{figure}

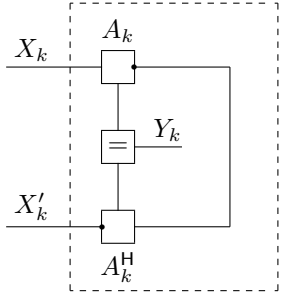
\begin{figure}
\setlength{\unitlength}{0.85mm}
\centering
\begin{picture}(42.5,45)(0,-2.5)
\put(0,32.5){\line(1,0){15}}        \put(4,33.5){\pos{cb}{$X_k$}}
\put(15,30){\framebox(5,5){}}       \put(17.5,36.5){\pos{cb}{$A_k$}}
 \put(20,32.5){\markerDot}
\put(20,32.5){\line(1,0){15}}       
 \put(35,32.5){\line(0,-1){25}}
\put(17.5,30){\line(0,-1){7.5}}     
\put(15,17.5){\framebox(5,5){$=$}}
 \put(20,20){\line(1,0){7.5}}       \put(27.5,21){\pos{br}{$Y_k$}}
\put(17.5,17.5){\line(0,-1){7.5}}
\put(0,7.5){\line(1,0){15}}         \put(4,8.5){\pos{cb}{$X_k'$}}
\put(15,5){\framebox(5,5){}}        \put(17.5,3.5){\pos{ct}{$A_k^\H$}}
 \put(15,7.5){\markerDot}
\put(20,7.5){\line(1,0){15}}        
\put(10,-2.5){\dashbox(32.5,45){}}
\end{picture}
\caption{\label{fig:QMBackwardBoxesGenMeas}%
The dashed box reduces to an equality constraint (i.e., an identity matrix) 
if and only if (\ref{eqn:GenMeasurementCond}) holds.
}
\end{figure}

A very general form of measurement is shown in \Fig{fig:GenMeasurement}. 
In this case, the range of $Y_k$ is a finite set $\calY_k$, 
and for each $y_k\in \calY_k$, 
the factor $A_k(\tilde{x}_k,x_k,y_k)$ corresponds to a complex square matrix $A_k(y_k)$ 
(with row index $\tilde{x}_k$ and column index $x_k$) such that 
\begin{equation} \label{eqn:GenMeasurementCond}
\sum_{y_k\in J_k} A_k(y_k)^\H A_k(y_k) = I,
\end{equation}
cf.\ \cite[Chap.~2]{NiChuang:QCI}.
A factor-graphic interpretation of (\ref{eqn:GenMeasurementCond})
is given in \Fig{fig:QMBackwardBoxesGenMeas}.
Condition (\ref{eqn:GenMeasurementCond}) is both necessary 
and sufficient for Proposition~\ref{prop:QMBackwardBoxes} (below) to hold.
Measurements as in \Fig{fig:BasicQMFG}
are included as a special case with $\calY_k = \{1,\ldots, M\}$ and 
\begin{equation} \label{eqn:BasA}
A_k(y_k) = A_k(y_k)^\H = B_k(\rangedot,y_k) B_k(\rangedot,y_k)^\H,
\end{equation}
where $B_k(\rangedot,y_k)$ denotes the $y_k$-th column of $B_k$.
Note that, for fixed $y_k$, (\ref{eqn:BasA}) is a projection matrix.

Measurements will further be discussed 
in sections \ref{sec:Partial} and~\ref{sec:MeasRecons}.

It is clear from Section~\ref{sec:ConjPairs} that 
the exterior function of \Fig{fig:GenQFG} 
(with measurements as in \Fig{fig:BasicQMFG} or as in \Fig{fig:GenMeasurement})
is real and nonnegative. 
We now proceed to analyze these factor graphs and to verify that
they yield the correct quantum-mechanical probabilities $p(y_1,y_2)$
for the respective class of measurements.
To this end, 
we need to understand the exterior functions 
of the dashed boxes in \Fig{fig:BackwardMesg}.
We begin with the dashed box on the right-hand side of \Fig{fig:BackwardMesg}.

\begin{proposition}[Don't Mind the Future] \label{prop:QMBackwardBoxes}
Closing the dashed box on the right-hand side in \Fig{fig:BackwardMesg}
(with a measurement as in \Fig{fig:BasicQMFG} or as in \Fig{fig:GenMeasurement},
but with unknown result $Y_2$ of the measurement)
reduces it to an equality constraint function.
\end{proposition}

\begin{proofof}{}
For measurements as in \Fig{fig:BasicQMFG}, 
the proof amounts to a sequence of reductions according to 
Figs.\ \ref{fig:RedHalfEdgeEqu} and~\ref{fig:RedMatrixMultEqu},
as illustrated in \Fig{fig:QMBackwardBoxes}.

For measurements as in \Fig{fig:GenMeasurement}, 
the key step is the reduction of \Fig{fig:QMBackwardBoxesGenMeas}
to an equality constraint, which is equivalent to the condition~(\ref{eqn:GenMeasurementCond}).
\end{proofof}

Proposition~\ref{prop:QMBackwardBoxes} 
guarantees, in particular, that a future measurement 
(with unknown result)
does not influence present or past observations.
The proposition clearly holds also for the extension of 
\Fig{fig:GenQFG} to any finite number of measurements $Y_1, Y_2, \ldots$ and can 
then be applied recursively from right to left. 

We pause here for a moment to emphasize this point:
it is obvious from Figs.\ \ref{fig:GenQFG} and~\ref{fig:BasicQMFG} 
(generalized to $n$ measurements $Y_1,\ldots,Y_n$) that, in general, 
a measurement resulting in some variable $Y_k$ affects the joint distribution 
of all other variables $Y_1,\ldots,Y_{k-1}, Y_{k+1},\ldots,Y_n$ (both past and future)
\emph{even if the result $Y_k$ of the measurement is not known.}
By Proposition~\ref{prop:QMBackwardBoxes}, however, 
the joint distribution of $Y_1,\ldots,Y_{k-1}$ is not 
affected by the measurement of $Y_k,\ldots,Y_n$ provided 
that no measurement results are known.

\begin{proposition}[Proper Normalization] \label{prop:ProperlyNormalized}
The factor graph of \Fig{fig:GenQFG}
(with measurements as in \Fig{fig:BasicQMFG} or as in \Fig{fig:GenMeasurement})
represents a properly normalized probability mass function, i.e., 
the exterior function $p(y_1,y_2)$ is real and nonnegative 
and $\sum_{y_1,y_2} p(y_1,y_2) = 1$.
\end{proposition}
In particular, the partition sum of \Fig{fig:GenQFG} equals~1.
Again, the proposition clearly holds also for the extension of 
\Fig{fig:GenQFG} to any finite number of measurements $Y_1, Y_2, \ldots$
\begin{proofof}{of Proposition~\ref{prop:ProperlyNormalized}}
Apply reductions according to Proposition~\ref{prop:QMBackwardBoxes}
recursively from right to left in \Fig{fig:GenQFG}, 
followed by the final reduction $\sum_{x_0} p(x_0)=1$.
\end{proofof}

Consider now the dashed boxes on the left in Figs.\ \ref{fig:BackwardMesg} 
and~\ref{fig:PostMeasurementDensityMatrix}, 
which correspond to the density matrix before and after measuring $Y_1$,
respectively. 
A density matrix 
$\rho$ is defined to be properly normalized if
\begin{equation} \label{eqn:normalizedRho}
\tr(\rho) = 1.
\end{equation}
The dashed box left in \Fig{fig:BackwardMesg} is 
properly normalized ($\tr(\rho_1)=1$) by (\ref{eqn:DensityMatrixTrace}).
Proper normalization of $\rho_k$ for $k>1$ follows from 
Propositions \ref{prop:Unitary}--\ref{prop:GeneralMeasurement} below.

Consider next the dashed box in \Fig{fig:PostMeasurementDensityMatrix},
which we will call $\breve\rho_1$; it is not a properly normalized 
density matrix: 
\begin{proposition}[Trace of the Past] 
\label{prop:PostMeasurementDensityMatrix}
\begin{equation}
\tr(\breve\rho_1) = p(y_1);
\end{equation}
more generally, with $k$ measurements $Y_1=y_1,\ldots,Y_k=y_k$ 
inside the dashed box, we have
\begin{equation}
\tr(\breve\rho_k) = p(y_1,\ldots,y_k).
\end{equation}
\eproofnegspace
\end{proposition}
The proof is immediate from Propositions 
\ref{prop:QMBackwardBoxes} and~\ref{prop:ProperlyNormalized} 
(generalized to an arbitrary number of measurements).
The properly normalized post-measurement density matrix is then
\begin{equation} \label{eqn:PostMeasurementDensityMatrix}
\tilde\rho_k \eqdef \breve\rho_k / p(y_1,\ldots,y_k).
\end{equation}

Between measurements, these functions/matrices evolve as follows.
\begin{proposition}[Unitary Evolution Between Measurements] \label{prop:Unitary}
The matrix $\rho_{k+1}$ is obtained from the matrix $\tilde{\rho}_k$ as
\begin{equation}
\rho_{k+1} = U_k \tilde{\rho}_k U_k^\H.
\end{equation}
\eproofnegspace
\end{proposition}
The proof is immediate from \Fig{fig:FactorGraphMatrixMult}.
Note that $\rho_{k+1}$ is properly normalized (provided that $\tilde{\rho}_k$ is so).

\begin{figure}
\centering
\begin{picture}(87.5,37.5)(0,-10)
\put(0,17.5){\line(1,0){15}}
\put(15,15){\framebox(5,5){}}     \put(17.5,13.5){\pos{ct}{$U_1$}}
 \put(20,17.5){\markerDot}
\put(20,17.5){\line(1,0){10}}
\put(30,15){\framebox(5,5){}}     \put(32.5,13.5){\pos{ct}{$B_2^\H$}}
 \put(35,17.5){\markerDot}
\put(35,17.5){\line(1,0){10}}
\put(45,15){\framebox(5,5){$=$}}
 \put(47.5,15){\line(0,-1){10}}
\put(50,17.5){\line(1,0){10}}
\put(60,15){\framebox(5,5){}}     \put(62.5,13.5){\pos{ct}{$B_2$}}
 \put(65,17.5){\markerDot}
\put(65,17.5){\line(1,0){10}}
 \put(75,17.5){\line(0,-1){5}}
\put(72.5,7.5){\framebox(5,5){$=$}}
 \put(75,7.5){\line(0,-1){5}}
\put(0,2.5){\line(1,0){15}}
\put(15,0){\framebox(5,5){}}     \put(17.5,-1.5){\pos{ct}{$U_1^\H$}}
 \put(15,2.5){\markerDot}
\put(20,2.5){\line(1,0){10}}
\put(30,0){\framebox(5,5){}}     \put(32.5,-1.5){\pos{ct}{$B_2$}}
 \put(30,2.5){\markerDot}
\put(35,2.5){\line(1,0){10}}
\put(45,0){\framebox(5,5){$=$}}
 \put(47.5,0){\line(0,-1){7.5}}
\put(50,2.5){\line(1,0){10}}
\put(60,0){\framebox(5,5){}}     \put(62.5,-1.5){\pos{ct}{$B_2^\H$}}
 \put(60,2.5){\markerDot}
\put(65,2.5){\line(1,0){10}}
\put(55,-7){\dashbox(27.5,31){}}
\put(10,-10){\dashbox(77.5,37.5){}}
\end{picture}
\vspace{\floatsep}

\begin{picture}(87.5,37.5)(0,-10)
\put(0,17.5){\line(1,0){15}}
\put(15,15){\framebox(5,5){}}     \put(17.5,13.5){\pos{ct}{$U_1$}}
 \put(20,17.5){\markerDot}
\put(20,17.5){\line(1,0){10}}
\put(30,15){\framebox(5,5){}}     \put(32.5,13.5){\pos{ct}{$B_2^\H$}}
 \put(35,17.5){\markerDot}
\put(35,17.5){\line(1,0){10}}
\put(45,15){\framebox(5,5){$=$}}
 \put(47.5,15){\line(0,-1){10}}
\put(50,17.5){\line(1,0){10}}
 \put(60,17.5){\line(0,-1){5}}
\put(57.5,7.5){\framebox(5,5){$=$}}
 \put(60,7.5){\line(0,-1){5}}
\put(0,2.5){\line(1,0){15}}
\put(15,0){\framebox(5,5){}}     \put(17.5,-1.5){\pos{ct}{$U_1^\H$}}
 \put(15,2.5){\markerDot}
\put(20,2.5){\line(1,0){10}}
\put(30,0){\framebox(5,5){}}     \put(32.5,-1.5){\pos{ct}{$B_2$}}
 \put(30,2.5){\markerDot}
\put(35,2.5){\line(1,0){10}}
\put(45,0){\framebox(5,5){$=$}}
 \put(47.5,0){\line(0,-1){5}}
\put(50,2.5){\line(1,0){10}}
\put(40,-7.5){\dashbox(27.5,32){}}
\put(10,-10){\dashbox(62.5,37.5){}}
\end{picture}
\vspace{\floatsep}

\begin{picture}(87.5,37.5)(0,-10)
\put(0,17.5){\line(1,0){15}}
\put(15,15){\framebox(5,5){}}     \put(17.5,13.5){\pos{ct}{$U_1$}}
 \put(20,17.5){\markerDot}
\put(20,17.5){\line(1,0){10}}
\put(30,15){\framebox(5,5){}}     \put(32.5,13.5){\pos{ct}{$B_2^\H$}}
 \put(35,17.5){\markerDot}
\put(35,17.5){\line(1,0){10}}
 \put(45,17.5){\line(0,-1){5}}
\put(42.5,7.5){\framebox(5,5){$=$}}
 \put(45,7.5){\line(0,-1){5}}
\put(0,2.5){\line(1,0){15}}
\put(15,0){\framebox(5,5){}}     \put(17.5,-1.5){\pos{ct}{$U_1^\H$}}
 \put(15,2.5){\markerDot}
\put(20,2.5){\line(1,0){10}}
\put(30,0){\framebox(5,5){}}     \put(32.5,-1.5){\pos{ct}{$B_2$}}
 \put(30,2.5){\markerDot}
\put(35,2.5){\line(1,0){10}}
\put(25,-7){\dashbox(27.5,31){}}
\put(10,-10){\dashbox(47.5,37.5){}}
\end{picture}
\vspace{\floatsep}

\begin{picture}(87.5,25)(0,-2.5)
\put(0,17.5){\line(1,0){17.5}}
 \put(17.5,17.5){\line(0,-1){5}}
\put(15,7.5){\framebox(5,5){$=$}}
 \put(17.5,7.5){\line(0,-1){5}}
\put(0,2.5){\line(1,0){17.5}}
\put(10,-2.5){\dashbox(15,25){}}
\end{picture}
\caption{\label{fig:QMBackwardBoxes}%
Proof of Proposition~\ref{prop:QMBackwardBoxes} 
for measurements as in \Fig{fig:BasicQMFG}
by a sequence of reductions 
as in Figs.\ \ref{fig:RedMatrixMultEqu} and~\ref{fig:RedHalfEdgeEqu}.}
\end{figure}

\begin{proposition}[Basic Projection Measurement]
\label{prop:BasicProjMeas}
In \Fig{fig:GenQFG} (generalized to any number of observations), 
if $Y_k$ is measured as in \Fig{fig:BasicQMFG}, then
\begin{IEEEeqnarray}{rCl}
\IEEEeqnarraymulticol{3}{l}{
P(Y_k \!=\! y_k \mid Y_{k-1} \!=\! y_{k-1},\ldots,Y_1 \!=\! y_1) 
}\nonumber\\ \quad
 & = &  B_k(\rangedot,y_k)^\H \rho_k B_k(\rangedot,y_k)  \label{eqn:BasicProjectionProbability1} \\
 & = &  \tr\!\left( \rho_k B_k(\rangedot,y_k) B_k(\rangedot,y_k)^\H \right).
        \label{eqn:BasicProjectionProbability2}
\end{IEEEeqnarray}
After measuring/observing $Y_k=y_k$, 
the density matrix is 
\begin{equation} \label{eqn:BasicProjectionPostMeasurementRho}
\tilde{\rho}_k = B_k(\rangedot,y_k) B_k(\rangedot,y_k)^\H.
\end{equation}
\eproofnegspace
\end{proposition}
Note that (\ref{eqn:BasicProjectionPostMeasurementRho}) 
is properly normalized because 
\begin{IEEEeqnarray}{rCl}
\tr(B_k(\rangedot,y_k) B_k(\rangedot,y_k)^\H) 
 & = & \tr(B_k(\rangedot,y_k)^\H B_k(\rangedot,y_k)) 
       \IEEEeqnarraynumspace\\
 & = & \|B_k(\rangedot,y_k)\|^2 = 1.
\end{IEEEeqnarray}

\begin{proofof}{of Proposition~\ref{prop:BasicProjMeas}}
For fixed $y_1,\ldots,y_{k-1}$, we have 
\begin{IEEEeqnarray}{rCl}
\IEEEeqnarraymulticol{3}{l}{
P(Y_k \!=\! y_k \mid Y_{k-1} \!=\! y_{k-1},\ldots,Y_1 \!=\! y_1) \hspace{3em}
}\nonumber\\\quad
& \propto &
p(y_k, y_{k-1},\ldots, y_1),  \label{eqn:BasicProjMeasProb}
\IEEEeqnarraynumspace
\end{IEEEeqnarray}
where $p$ is the exterior function of \Fig{fig:GenQFG} 
(generalized to any number of observations and with measurements 
as in \Fig{fig:BasicQMFG}). 
We now reduce \Fig{fig:GenQFG} to \Fig{fig:ProofBasicProjMeas} as follows:
everything to the right of $Y_k$ reduces to an equality constraint 
according to Proposition~\ref{prop:QMBackwardBoxes} (see also \Fig{fig:QMBackwardBoxes}),
while everything before the measurement of $Y_k$ 
(with $Y_{k-1}=y_{k-1},\ldots,Y_1=y_1$ plugged in) is subsumed by $\rho_k$. 
Note that the partition sum of \Fig{fig:ProofBasicProjMeas} is 
$\tr(\rho_k)=1$ (cf.\ \Fig{fig:DensityMatrix}), 
which means that the exterior function of \Fig{fig:ProofBasicProjMeas} 
equals $p(y_k \cond y_{k-1},\ldots,y_1)$,
i.e., the missing scale factor in (\ref{eqn:BasicProjMeasProb}) 
has been compensated by the normalization of $\rho_k$. 

For any fixed $Y_k=y_k$, 
we can then read (\ref{eqn:BasicProjectionProbability1}) 
and (\ref{eqn:BasicProjectionProbability2}) from \Fig{fig:ProofBasicProjMeas} 
(cf.\ \Fig{fig:ObsProbTrace}).

We now turn to the post-measurement density matrix $\tilde\rho_k$. 
For a measurement $Y_k=y_k$ as in \Fig{fig:BasicQMFG}, 
the dashed box in \Fig{fig:PostMeasurementDensityMatrix}
looks as in \Fig{fig:ProofBasicProjPostMeas},
which decomposes into two unconnected parts
as indicated by the two inner dashed boxes. 
The exterior function of the left-hand inner dashed box in \Fig{fig:ProofBasicProjPostMeas} 
is the constant (\ref{eqn:BasicProjectionProbability1}); 
the right-hand inner dashed box equals (\ref{eqn:BasicProjectionPostMeasurementRho}).
\end{proofof}

\begin{figure}
\centering
\begin{picture}(47.5,30)(0,-7.5)
\put(0,0){\framebox(12.5,20){}}     \put(6.25,21.5){\pos{cb}{$\rho_k$}}
\put(12.5,17.5){\line(1,0){12.5}}   \put(18.75,18.5){\pos{cb}{$X_k$}}
\put(25,15){\framebox(5,5){}}       \put(27.5,13.5){\pos{$ct$}{$B_k^\H$}}
 \put(30,17.5){\markerDot}
\put(30,17.5){\line(1,0){10}}
\put(40,15){\framebox(5,5){$=$}}
\put(12.5,2.5){\line(1,0){12.5}}    \put(18.75,3.5){\pos{cb}{$X_k'$}}
\put(25,0){\framebox(5,5){}}        \put(27.5,-1.5){\pos{$ct$}{$B_k$}}
 \put(25,2.5){\markerDot}
\put(30,2.5){\line(1,0){10}}
\put(40,0){\framebox(5,5){$=$}}
\put(42.5,15){\line(0,-1){10}}
\put(42.5,0){\line(0,-1){7.5}}       \put(43.5,-7.5){\pos{bl}{$Y_k$}}
\end{picture}
\caption{\label{fig:ProofBasicProjMeas}%
Proof of Proposition~\ref{prop:BasicProjMeas}: the exterior function 
equals (\ref{eqn:BasicProjectionProbability1}) and (\ref{eqn:BasicProjectionProbability2}).}
\vspace{\floatsep}

\centering
\setlength{\unitlength}{0.9mm}
\begin{picture}(95,45)(-10,-10)
\put(0,0){\framebox(10,20){}}       \put(5,21.5){\pos{cb}{$\rho_k$}}
\put(10,17.5){\line(1,0){10}}       \put(15,18.5){\pos{cb}{$X_k$}}
\put(20,15){\framebox(5,5){}}       \put(22.5,13.5){\pos{$ct$}{$B_k^\H$}}
 \put(25,17.5){\markerDot}
\put(25,17.5){\line(1,0){7.5}}
 \put(32.5,17.5){\knownBox}         \put(34.5,17.5){\pos{cl}{$y_k$}}
\put(10,2.5){\line(1,0){10}}        \put(15,3.5){\pos{cb}{$X_k'$}}
\put(20,0){\framebox(5,5){}}        \put(22.5,-1.5){\pos{$ct$}{$B_k$}}
 \put(20,2.5){\markerDot}
\put(25,2.5){\line(1,0){7.5}}
 \put(32.5,2.5){\knownBox}          \put(34.5,2.5){\pos{cl}{$y_k$}}
\put(-5,-6.5){\dashbox(45,33){}}
\put(45,-6.5){\dashbox(25,31){}}    \put(57.5,26){\pos{cb}{$\tilde\rho_k$}}
\put(52.5,17.5){\line(1,0){7.5}}
 \put(52.5,17.5){\knownBox}         \put(50.5,17.5){\pos{cr}{$y_k$}}
\put(60,15){\framebox(5,5){}}       \put(62.5,13.5){\pos{$ct$}{$B_k$}}
 \put(65,17.5){\markerDot}
\put(65,17.5){\line(1,0){20}}       \put(80,18.5){\pos{cb}{$\tilde{X}_k$}}
\put(52.5,2.5){\line(1,0){7.5}}
 \put(52.5,2.5){\knownBox}          \put(50.5,2.5){\pos{cr}{$y_k$}}
\put(60,0){\framebox(5,5){}}        \put(62.5,-1){\pos{$ct$}{$B_k^\H$}}
 \put(60,2.5){\markerDot}
\put(65,2.5){\line(1,0){20}}        \put(80,3.5){\pos{cb}{$\tilde{X}_k'$}}
\put(-10,-10){\dashbox(85,41){}}    \put(32.5,32.5){\pos{cb}{$\propto\tilde\rho_k$}}
\end{picture}
\caption{\label{fig:ProofBasicProjPostMeas}%
Proof of Proposition~\ref{prop:BasicProjMeas}: post-measurement density matrix $\tilde\rho_k$.}
\vspace{\floatsep}

\centering
\begin{picture}(47,45)(-12,-10)
\put(2.5,22.5){\knownBox}           \put(0.5,22.5){\pos{cr}{$x_0$}}
\put(2.5,22.5){\line(1,0){7.5}}
\put(10,20){\framebox(5,5){}}       \put(12.5,18.5){\pos{ct}{$U_0$}}
 \put(15,22.5){\markerDot}
\put(15,22.5){\line(1,0){20}}       \put(30,23.5){\pos{cb}{$X_1$}}
\put(-5,14){\dashbox(25,14){}}      \put(-6,16){\pos{br}{$\psi_1$}}
\put(2.5,2.5){\knownBox}            \put(0.5,2.5){\pos{cr}{$x_0$}}
\put(2.5,2.5){\line(1,0){7.5}}
\put(10,0){\framebox(5,5){}}        \put(12.5,-1){\pos{ct}{$U_0^\H$}}
 \put(10,2.5){\markerDot}
\put(15,2.5){\line(1,0){20}}        \put(30,3.5){\pos{cb}{$X_1'$}}
\put(-5,-6){\dashbox(25,14){}}      \put(-6.25,-4){\pos{br}{$\psi_1^\H$}}
\put(-12,-10){\dashbox(37,42){}}    \put(6.5,33.5){\pos{cb}{$\rho_1$}}
\end{picture}
\caption{\label{fig:QuantumState}%
Quantum state $\psi_1$.}
\end{figure}

In the special case of \Fig{fig:BasicQMFG}, 
with known initial state \mbox{$X_0=x_0$}, 
the matrix $\rho_k$ factors as 
\begin{equation}
\rho_k(x_k,x_k') = \psi_k(x_k) \ccj{\psi_k(x_k')},
\end{equation}
or, in matrix notation,
\begin{equation}
\rho_k = \psi_k \psi_k^\H,
\end{equation}
where $\psi_k$ is a column vector of norm~1. 
For $k=1$, we have $\psi_1(x_1) = U_0(x_1,x_0)$,
as shown in \Fig{fig:QuantumState}. 
The post-measurement density matrix $\tilde{\rho}_k$ factors analoguously,
as is obvious from (\ref{eqn:BasicProjectionPostMeasurementRho})
or from \Fig{fig:ProofBasicProjPostMeas}.
In quantum-mechanical terms, $\psi_k$ is the quantum state
(cf.\ Section~\ref{sec:ElementaryQM}). 
The probability (\ref{eqn:BasicProjectionProbability1})
can then be expressed as 
\begin{IEEEeqnarray}{rCl}
\IEEEeqnarraymulticol{3}{l}{
P(Y_k = y \mid Y_{k-1}=y_{k-1},\ldots,Y_1=y_1) 
}\nonumber\\ \quad
 & = &  B_k(\rangedot,y)^\H \psi_k \psi_k^\H B_k(\rangedot,y) \\
 & = &  \| B_k(\rangedot,y)^\H \psi_k \|^2.
\end{IEEEeqnarray}

\begin{figure}
\centering
\begin{picture}(50,37.5)(0,-7.5)
\put(0,0){\framebox(15,25){}}     \put(7.5,26.5){\pos{cb}{$\rho_k$}}
\put(15,22.5){\line(1,0){12.5}}   \put(21.25,23.5){\pos{cb}{$X_k$}}
\put(27.5,20){\framebox(5,5){}}   \put(30,26.5){\pos{cb}{$A_k$}}
 \put(32.5,22.5){\markerDot}
 \put(30,20){\line(0,-1){5}}
\put(27.5,10){\framebox(5,5){$=$}}
 \put(32.5,12.5){\line(1,0){5}}
 \put(37.5,12.5){\line(0,-1){20}}  \put(38.5,-7.5){\pos{bl}{$Y_k$}}
 
\put(32.5,22.5){\line(1,0){15}}   \put(38.75,23.5){\pos{cb}{$\tilde X_k$}}
\put(15,2.5){\line(1,0){12.5}}    \put(21.25,3.5){\pos{cb}{$X_k'$}}
\put(27.5,0){\framebox(5,5){}}    \put(30,-1.5){\pos{ct}{$A_k^\H$}}
 \put(27.5,2.5){\markerDot}
 \put(30,5){\line(0,1){5}}
\put(32.5,2.5){\line(1,0){15}}    
\put(47.5,22.5){\line(0,-1){7.5}}
\put(45,10){\framebox(5,5){$=$}}
\put(47.5,10){\line(0,-1){7.5}}
\end{picture}
\caption{\label{fig:GeneralMeasurementNorm}%
Proof of Proposition~\ref{prop:GeneralMeasurement}: normalization.}
\vspace{\floatsep}

\centering
\begin{picture}(50,35)(0,-5)
\put(0,0){\framebox(15,25){}}     \put(7.5,26.5){\pos{cb}{$\rho_k$}}
\put(15,22.5){\line(1,0){12.5}}   \put(21.25,23.5){\pos{cb}{$X_k$}}
\put(27.5,20){\framebox(5,5){}}   \put(30,26.5){\pos{cb}{$A_k$}}
 \put(32.5,22.5){\markerDot}
 \put(30,20){\line(0,-1){5}}
 \put(30,15){\knownBox}           \put(32,15){\pos{cl}{$y_k$}}
\put(32.5,22.5){\line(1,0){15}}   \put(38.75,23.5){\pos{cb}{$\tilde X_k$}}
\put(15,2.5){\line(1,0){12.5}}    \put(21.25,3.5){\pos{cb}{$X_k'$}}
\put(27.5,0){\framebox(5,5){}}    \put(30,-1.5){\pos{ct}{$A_k^\H$}}
 \put(27.5,2.5){\markerDot}
 \put(30,5){\line(0,1){5}}
 \put(30,10){\knownBox}           \put(32,10){\pos{cl}{$y_k$}}
\put(32.5,2.5){\line(1,0){15}}    \put(38.75,3.5){\pos{cb}{$\tilde X_k'$}}
\put(47.5,22.5){\line(0,-1){7.5}}
\put(45,10){\framebox(5,5){$=$}}
\put(47.5,10){\line(0,-1){7.5}}
\end{picture}
\caption{\label{fig:GeneralMeasurementProb}%
Proof of Proposition~\ref{prop:GeneralMeasurement}: 
probability (\ref{eqn:GenMeasurementProb}).}
\vspace{\floatsep}

\centering
\begin{picture}(52.5,43)(-5,-8)
\put(0,0){\framebox(15,25){}}     \put(7.5,26.5){\pos{cb}{$\rho_k$}}
\put(15,22.5){\line(1,0){12.5}}   \put(21.25,23.5){\pos{cb}{$X_k$}}
\put(27.5,20){\framebox(5,5){}}   \put(30,26.5){\pos{cb}{$A_k$}}
 \put(32.5,22.5){\markerDot}
 \put(30,20){\line(0,-1){5}}
 \put(30,15){\knownBox}           \put(32,15){\pos{cl}{$y_k$}}
\put(32.5,22.5){\line(1,0){15}}   \put(42.5,23.5){\pos{cb}{$\tilde X_k$}}
\put(15,2.5){\line(1,0){12.5}}    \put(21.25,3.5){\pos{cb}{$X_k'$}}
\put(27.5,0){\framebox(5,5){}}    \put(30,-1.5){\pos{ct}{$A_k^\H$}}
 \put(27.5,2.5){\markerDot}
 \put(30,5){\line(0,1){5}}
 \put(30,10){\knownBox}           \put(32,10){\pos{cl}{$y_k$}}
\put(32.5,2.5){\line(1,0){15}}    \put(42.5,3.5){\pos{cb}{$\tilde X_k'$}}
\put(-5,-8){\dashbox(42.5,40){}}  \put(16.25,33){\pos{cb}{$\propto\tilde\rho_k$}}
\end{picture}
\caption{\label{fig:GeneralMeasurementState}%
Proof of Proposition~\ref{prop:GeneralMeasurement}: 
post-measurement density matrix $\tilde\rho_k$.}
\end{figure}

\begin{figure*}
\centering
\setlength{\unitlength}{0.95mm}
\begin{picture}(171,57.5)(-13.5,-7.5)
\put(-12.5,22.5){\framebox(5,5){}}    \put(-10,21){\pos{ct}{$p(x_0)$}}
\put(-7.5,25){\line(1,0){10}}         \put(-2.5,26){\pos{cb}{$X_0$}}
\put(5,40){\line(1,0){10}}
\put(5,40){\line(0,-1){12.5}}
\put(2.5,22.5){\framebox(5,5){$=$}}
\put(5,10){\line(0,1){12.5}}
\put(5,10){\line(1,0){10}}
\put(15,30){\framebox(10,20){}}      \put(20,28.5){\pos{ct}{$U_0$}}
 \put(25,45){\markerDot}
 \put(25,35){\markerDot}
\put(25,45){\line(1,0){50}}
\put(25,35){\line(1,0){10}}
\put(35,32.5){\framebox(5,5){}}      \put(37.5,31){\pos{ct}{$B_1^\H$}}
 \put(40,35){\markerDot}
\put(40,35){\line(1,0){7.5}}
\put(47.5,32.5){\framebox(5,5){$=$}}
 \put(50,32.5){\line(0,-1){15}}
\put(52.5,35){\line(1,0){7.5}}
\put(60,32.5){\framebox(5,5){}}      \put(62.5,31){\pos{ct}{$B_1$}}
 \put(65,35){\markerDot}
\put(65,35){\line(1,0){10}}
\put(15,0){\framebox(10,20){}}       \put(20,-1.5){\pos{ct}{$U_0^\H$}}
 \put(15,10){\markerDot}
\put(25,5){\line(1,0){50}}
\put(25,15){\line(1,0){10}}
\put(35,12.5){\framebox(5,5){}}      \put(37.5,11){\pos{ct}{$B_1$}}
 \put(35,15){\markerDot}
\put(40,15){\line(1,0){7.5}}
\put(47.5,12.5){\framebox(5,5){$=$}}
 \put(50,12.5){\line(0,-1){20}}      \put(51.5,-7.5){\pos{bl}{$Y_1$}}
\put(52.5,15){\line(1,0){7.5}}
\put(60,12.5){\framebox(5,5){}}      \put(62.5,11){\pos{ct}{$B_1^\H$}}
 \put(60,15){\markerDot}
\put(65,15){\line(1,0){10}}
\put(75,30){\framebox(10,20){}}      \put(80,28.5){\pos{ct}{$U_1$}}
 \put(85,45){\markerDot}
 \put(85,35){\markerDot}
\put(85,45){\line(1,0){50}}
\put(85,35){\line(1,0){10}}
\put(95,32.5){\framebox(5,5){}}      \put(97.5,31){\pos{ct}{$B_2^\H$}}
 \put(100,35){\markerDot}
\put(100,35){\line(1,0){7.5}}
\put(107.5,32.5){\framebox(5,5){$=$}}
 \put(110,32.5){\line(0,-1){15}}
\put(112.5,35){\line(1,0){7.5}}
\put(120,32.5){\framebox(5,5){}}     \put(122.5,31){\pos{ct}{$B_2$}}
 \put(125,35){\markerDot}
\put(125,35){\line(1,0){10}}
\put(75,0){\framebox(10,20){}}       \put(80,-1.5){\pos{ct}{$U_1^\H$}}
 \put(75,15){\markerDot}
 \put(75,5){\markerDot}
\put(85,5){\line(1,0){50}}
\put(85,15){\line(1,0){10}}
\put(95,12.5){\framebox(5,5){}}      \put(97.5,11){\pos{ct}{$B_2$}}
 \put(95,15){\markerDot}
\put(100,15){\line(1,0){7.5}}
\put(107.5,12.5){\framebox(5,5){$=$}}
 \put(110,12.5){\line(0,-1){20}}      \put(111.5,-7.5){\pos{bl}{$Y_2$}}
\put(112.5,15){\line(1,0){7.5}}
\put(120,12.5){\framebox(5,5){}}     \put(122.5,11){\pos{ct}{$B_2^\H$}}
 \put(120,15){\markerDot}
\put(125,15){\line(1,0){10}}
\put(135,30){\framebox(10,20){}}     \put(140,28.5){\pos{ct}{$U_2$}}
 \put(145,40){\markerDot}
\put(135,0){\framebox(10,20){}}      \put(140,-1.5){\pos{ct}{$U_2^\H$}}
 \put(135,15){\markerDot}
 \put(135,5){\markerDot}
\put(145,40){\line(1,0){10}}
\put(155,40){\line(0,-1){12.5}}
\put(152.5,22.5){\framebox(5,5){$=$}}
\put(155,10){\line(0,1){12.5}}
\put(145,10){\line(1,0){10}}
\end{picture}
\caption{\label{fig:PartialMeasurement}%
Factor graph of a quantum system with partial measurements.}
\end{figure*}
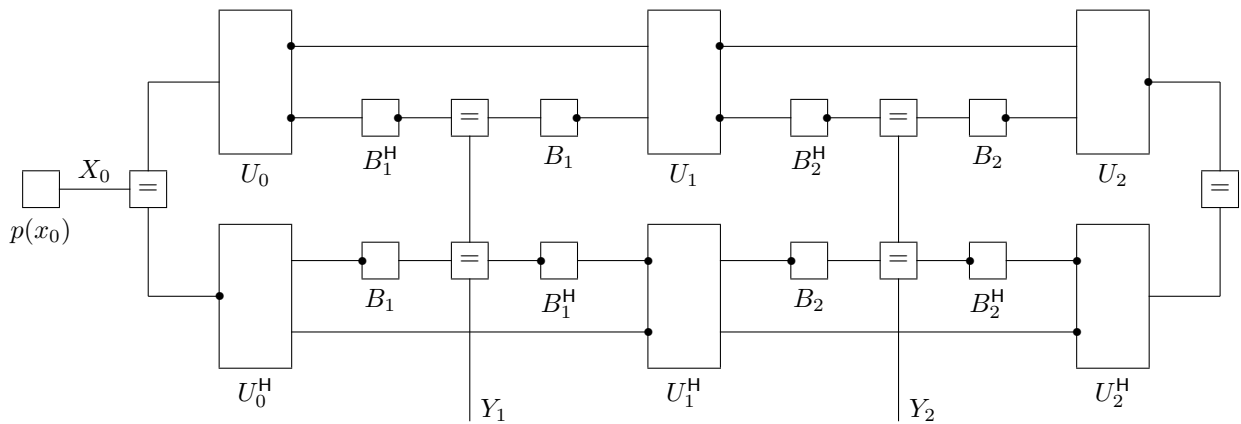

\begin{proposition}[General Measurement] \label{prop:GeneralMeasurement}
In \Fig{fig:GenQFG} (generalized to any number of observations), 
if $Y_k$ is measured as in \Fig{fig:GenMeasurement}, then
\begin{IEEEeqnarray}{rCl}
\IEEEeqnarraymulticol{3}{l}{
P(Y_k \!=\! y_k \mid Y_{k-1} \!=\! y_{k-1},\ldots,Y_1 \!=\! y_1) 
}\nonumber\\ \quad
 & = &  \tr\!\left( A_k(y_k) \rho_k A_k(y_k)^\H \right).
 \label{eqn:GenMeasurementProb}
\end{IEEEeqnarray}
After measuring/observing $Y_k=y_k$, the density matrix is 
\begin{equation} \label{eqn:GenMeasurementState}
\tilde{\rho}_k = \frac{A_k(y_k) \rho_k A_k(y_k)^\H}{\tr\!\left( A_k(y_k) \rho_k A_k(y_k)^\H \right)}
\end{equation}
\eproofnegspace
\end{proposition}

\begin{proofof}{}
The proof is parallel to the proof of Proposition~\ref{prop:BasicProjMeas}.
For fixed $y_{k-1},\ldots,y_1$, we have 
\begin{IEEEeqnarray}{rCl}
\IEEEeqnarraymulticol{3}{l}{
P(Y_k \!=\! y_k \mid Y_{k-1} \!=\! y_{k-1},\ldots,Y_1 \!=\! y_1) \hspace{3em}
}\nonumber\\\quad
& \propto &
p(y_k, y_{k-1},\ldots, y_1),
\IEEEeqnarraynumspace\label{eqn:GeneralMeasurementProb}
\end{IEEEeqnarray}
where $p$ is the exterior function of \Fig{fig:GenQFG} 
(generalized to any number of observations and with measurements 
as in \Fig{fig:GenMeasurement}). 
We now reduce \Fig{fig:GenQFG} to \Fig{fig:GeneralMeasurementNorm}
as follows:
everything to the right of $Y_k$ reduces to an equality constraint 
while everything before the measurement of $Y_k$ 
(with $Y_{k-1}=y_{k-1},\ldots,Y_1=y_1$ plugged in) is subsumed by $\rho_k$. 
From \Fig{fig:QMBackwardBoxesGenMeas}, 
we see that the partition sum of \Fig{fig:GeneralMeasurementNorm} is 
$\tr(\rho_k)=1$, 
which means that the exterior function of \Fig{fig:GeneralMeasurementNorm}
equals $p(y_k \cond y_{k-1},\ldots,y_1)$,
i.e., the missing scale factor in (\ref{eqn:GeneralMeasurementProb}) 
has been compensated by the normalization of $\rho_k$. 

For fixed $Y_k=y_k$, (\ref{eqn:GenMeasurementProb}) 
is then obvious from \Fig{fig:GeneralMeasurementProb}.

Concerning the post-measurement density matrix $\tilde\rho_k$,
for a measurement $Y_k=y_k$ as in \Fig{fig:GenMeasurement}, 
the dashed box in \Fig{fig:PostMeasurementDensityMatrix}
looks as in \Fig{fig:GeneralMeasurementState}.
The numerator of (\ref{eqn:GenMeasurementState}) 
is then obvious from \Fig{fig:GeneralMeasurementState},
and the denominator of (\ref{eqn:GenMeasurementState}) 
is simply the proper normalization (\ref{eqn:normalizedRho}).
\end{proofof}

In summary, 
Propositions \ref{prop:QMBackwardBoxes}--\ref{prop:GeneralMeasurement} 
verify
that the factor graph of \Fig{fig:GenQFG} 
(with measurements as in \Fig{fig:BasicQMFG} or as in \Fig{fig:GenMeasurement})
yields the correct quantum-mechanical probabilities for the respective class of measurements.

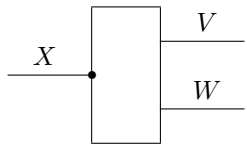
\begin{figure}[p]
\centering
\setlength{\unitlength}{0.9mm}
\begin{picture}(40,20)(0,0)
\put(2.5,10){\line(1,0){12.5}}    \put(8,11.5){\pos{cb}{$X$}}
 \put(15,10){\markerDot}
\put(15,0){\framebox(10,20){}}
\put(25,15){\line(1,0){12.5}}     \put(32,16.5){\pos{cb}{$V$}}
\put(25,5){\line(1,0){12.5}}      \put(32,6.5){\pos{cb}{$W$}}
\end{picture}
\caption{\label{fig:DecompMatrixVars}%
Matrix with row index $X$ and columns indexed by the pair $(V,W)$.
(E.g., $X$ takes values in $\{ 0,1,2,3\}$ while $V$ and $W$ are both binary.)}
\end{figure}

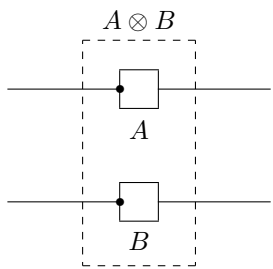
\begin{figure}[p]
\centering
\begin{picture}(35,34)(0,-1)
\put(0,22.5){\line(1,0){15}}
\put(15,20){\framebox(5,5){}}   \put(17.5,18.25){\pos{ct}{$A$}}
 \put(15,22.5){\markerDot}
\put(20,22.5){\line(1,0){15}}
\put(0,7.5){\line(1,0){15}}
\put(15,5){\framebox(5,5){}}    \put(17.5,3.5){\pos{ct}{$B$}}
 \put(15,7.5){\markerDot}
\put(20,7.5){\line(1,0){15}}
\put(10,-1){\dashbox(15,30){}}  \put(17.5,30.25){\pos{cb}{$A \otimes B$}}
\end{picture}
\caption{\label{fig:TensorProduct}%
Tensor product of matrices $A$ and $B$.}
\end{figure}

\begin{figure}[p]
\centering
\setlength{\unitlength}{0.85mm}
\begin{picture}(70,50)(0,0)
\put(10,0){\dashbox(50,50){}}
\put(27.5,12.5){\dashbox(15,25){}}
\put(0,40){\line(1,0){15}}
 \put(15,40){\markerDot}
\put(0,35){\line(1,0){15}}
 \put(15,35){\markerDot}
\put(15,32.5){\framebox(10,10){}}
\put(25,40){\line(1,0){45}}
\put(25,35){\line(1,0){5}}
\put(30,35){\line(1,-2){10}}
\put(40,15){\line(1,0){5}}
 \put(45,15){\markerDot}
\put(0,27.5){\line(1,0){70}}
\put(0,22.5){\line(1,0){70}}
\put(0,15){\line(1,0){30}}
 \put(30,15){\line(1,2){10}}
 \put(40,35){\line(1,0){30}}
\put(0,10){\line(1,0){45}}
 \put(45,10){\markerDot}
\put(45,7.5){\framebox(10,10){}}
\put(55,12.5){\line(1,0){15}}
\end{picture}
\caption{\label{fig:DecompUnitary}%
Decomposition of a unitary matrix into smaller unitary matrices.
Line switching as in the inner dashed box is itself a unitary matrix,
cf.\ \Fig{fig:SwapGate}.}
\end{figure}

\begin{figure}[p]
\centering
\begin{picture}(65,58)(2.5,1)
\put(2.5,55){\line(1,0){65}}
\put(2.5,45){\line(1,0){17.5}}
\put(20,42.5){\framebox(5,5){}}      \put(22.5,49){\pos{cb}{$B_k^\H$}}
 \put(25,45){\markerDot}
\put(25,45){\line(1,0){7.5}}
\put(32.5,42.5){\framebox(5,5){$=$}}
\put(37.5,45){\line(1,0){7.5}}
\put(45,42.5){\framebox(5,5){}}      \put(47.5,49){\pos{cb}{$B_k$}}
 \put(50,45){\markerDot}
\put(50,45){\line(1,0){17.5}}
\put(15,38){\dashbox(40,21){}}        \put(15,36.5){\pos{tl}{$A_k$}}
\put(35,42.5){\line(0,-1){10}}
\put(32.5,27.5){\framebox(5,5){$=$}}
 \put(37.5,30){\line(1,0){27.5}}      \put(62,31.25){\pos{cb}{$Y_k$}}
\put(35,27.5){\line(0,-1){10}}
\put(2.5,15){\line(1,0){17.5}}
\put(20,12.5){\framebox(5,5){}}      \put(22.5,11){\pos{ct}{$B_k$}}
 \put(20,15){\markerDot}
\put(25,15){\line(1,0){7.5}}
\put(32.5,12.5){\framebox(5,5){$=$}}
\put(37.5,15){\line(1,0){7.5}}
\put(45,12.5){\framebox(5,5){}}      \put(47.5,11){\pos{ct}{$B_k^\H$}}
 \put(45,15){\markerDot}
\put(50,15){\line(1,0){17.5}}
\put(2.5,5){\line(1,0){65}}
\put(15,1){\dashbox(40,21){}}        \put(15,23.5){\pos{bl}{$A_k^\H$}}
\end{picture}
\caption{\label{fig:PartialAsGeneralMeasurement}%
Measurements in \Fig{fig:PartialMeasurement} as a special case of 
\Fig{fig:GenMeasurement}.}
\end{figure}
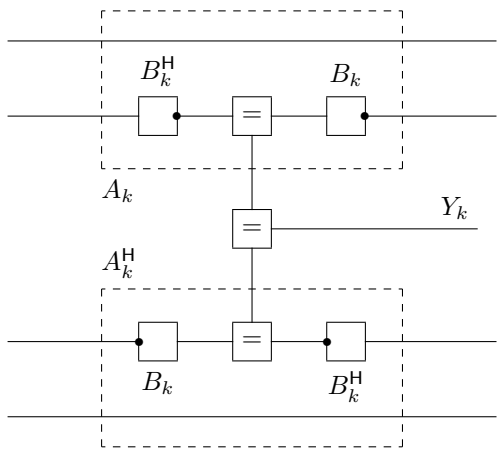

\begin{figure}[p]
\centering
\begin{picture}(77.5,64)(2.5,0)
\put(2.5,55){\line(1,0){70}}         \put(7.5,56){\pos{cb}{$X_{k,1}$}}
\put(2.5,45){\line(1,0){17.5}}       \put(7.5,46){\pos{cb}{$X_{k,2}$}}
\put(20,42.5){\framebox(5,5){}}      \put(22.5,49){\pos{cb}{$B_k^\H$}}
 \put(25,45){\markerDot}
\put(25,45){\line(1,0){7.5}}
\put(32.5,42.5){\framebox(5,5){$=$}}
\put(37.5,45){\line(1,0){7.5}}
\put(45,42.5){\framebox(5,5){}}      \put(47.5,49){\pos{cb}{$B_k$}}
 \put(50,45){\markerDot}
\put(50,45){\line(1,0){12.5}}
\put(35,42.5){\line(0,-1){10}}
\put(32.5,27.5){\framebox(5,5){$=$}}
 \put(37.5,30){\line(1,0){10}}       \put(45,31.25){\pos{cb}{$Y_k$}}
\put(35,27.5){\line(0,-1){10}}
\put(62.5,45){\line(0,-1){12.5}}
\put(60,27.5){\framebox(5,5){$=$}}
\put(62.5,27.5){\line(0,-1){12.5}}
\put(72.5,55){\line(0,-1){22.5}}
\put(70,27.5){\framebox(5,5){$=$}}
\put(72.5,27.5){\line(0,-1){22.5}}
\put(2.5,15){\line(1,0){17.5}}
\put(20,12.5){\framebox(5,5){}}      \put(22.5,11){\pos{ct}{$B_k$}}
 \put(20,15){\markerDot}
\put(25,15){\line(1,0){7.5}}
\put(32.5,12.5){\framebox(5,5){$=$}}
\put(37.5,15){\line(1,0){7.5}}
\put(45,12.5){\framebox(5,5){}}      \put(47.5,11){\pos{ct}{$B_k^\H$}}
 \put(45,15){\markerDot}
\put(50,15){\line(1,0){12.5}}         \put(7.5,16){\pos{cb}{$X_{k,2}'$}}
\put(2.5,5){\line(1,0){70}}           \put(7.5,6){\pos{cb}{$X_{k,1}'$}}
\put(15,0){\dashbox(65,60){}}         \put(45,61.5){\pos{cb}{$f_=$}}
\end{picture}
\caption{\label{fig:PartialMeasurementReduction}%
The exterior function of the dashed box is 
$f_=\big( (x_{k,1},x_{k,2}), (x_{k,1}',x_{k,2}') \big) 
= f_=(x_{k,1},x_{k,1}') f_=(x_{k,2},x_{k,2}')$.}
\end{figure}
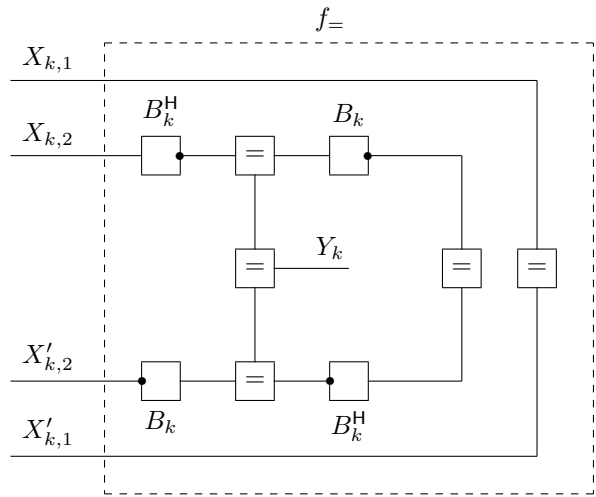

\section{Decompositions and Quantum Circuits, 
and Non-Unitary Operators from Unitary Interactions}
\label{sec:Decompositions}

Figs.\ \ref{fig:GenQFG} and~\ref{fig:GenMeasurement},
while fully general, do not do justice to 
the richness of quantum-mechanical probabilities and their factor-graph
representation, which we are now going to address.

\subsection{Decompositions and Partial Measurements}
\label{sec:Partial}

Consider the factor graph of \Fig{fig:PartialMeasurement}. 
The following points are noteworthy.
First, we note that the unitary matrices $U_0$, $U_1$, $U_2$
in \Fig{fig:PartialMeasurement} have more than two incident edges. 
This is to be understood as illustrated in \Fig{fig:DecompMatrixVars},
where the rows of some matrix are indexed by $X$ while its columns
are indexed by the pair $(V,W)$. 
More generally, rows (marked by a dot) and columns may both be indexed 
by several variables. 
Note that, in this way, bundling two unconnected matrices as in \Fig{fig:TensorProduct} 
represents the tensor product $A \otimes B$.
In \Fig{fig:PartialMeasurement}, all matrices are square, which implies that 
the product of the alphabet sizes 
of the row-indexing variables must equal the product of the alphabet sizes 
of the column-indexing variables.

Second, each edge in the factor graph of \Fig{fig:PartialMeasurement}
may actually represent several (finite-alphabet) variables, 
bundled into a single compound variable.

Third, each of the unitary matrices $U_0$, $U_1$, $U_2$, \ldots\
may itself be a product, either of smaller unitary matrices
as illustrated in \Fig{fig:DecompUnitary}, 
or of more general factors as exemplified by \Fig{fig:CNOT}; 
see also Section~\ref{sec:QuantumCircuits} below.

Forth, it is obvious from \Fig{fig:PartialMeasurement}
that each measurement involves only some of the variables 
while some other variables are left alone.
The actual measurements shown in \Fig{fig:PartialMeasurement}
are as in \Fig{fig:BasicQMFG}
(with unitary matrices $B_1,B_2\ldots$), 
but more general measurements could be used.

The measurements in \Fig{fig:PartialMeasurement} (including the uninvolved variables) 
are indeed a special case of measurements as in \Fig{fig:GenMeasurement}, 
as is obvious from \Fig{fig:PartialAsGeneralMeasurement}, 
from where we may also read 
$A_k(y_k) = I \otimes (B_k(y_k) B_k(y_k)^\H)$.
In order to verify (\ref{eqn:GenMeasurementCond}), we first recall 
its factor-graphic interpretation in \Fig{fig:QMBackwardBoxesGenMeas},
which, in this case, amounts to the obvious reduction of 
\Fig{fig:PartialMeasurementReduction} to an equality constraint.

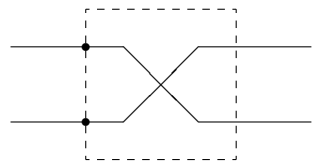
\begin{figure}
\centering
\begin{picture}(40,20)(0,0)
\put(0,15){\line(1,0){15}}
 \put(10,15){\markerDot}
\put(15,15){\line(1,-1){10}}
\put(25,5){\line(1,0){15}}
\put(0,5){\line(1,0){15}}
 \put(10,5){\markerDot}
\put(15,5){\line(1,1){10}}
\put(25,15){\line(1,0){15}}
\put(10,0){\dashbox(20,20){}}
\end{picture}
\caption{\label{fig:SwapGate}%
Swap gate.}
\end{figure}

\begin{figure}
\centering
\begin{picture}(35,27.5)(2.5,0)
\put(2.5,20){\line(1,0){15}}
 \put(12.5,20){\markerDot}
\put(17.5,17.5){\framebox(5,5){$=$}}
 \put(20,17.5){\line(0,-1){7.5}}
\put(22.5,20){\line(1,0){15}}
\put(2.5,7.5){\line(1,0){15}}
 \put(12.5,7.5){\markerDot}
\put(17.5,5){\framebox(5,5){$\oplus$}}
\put(22.5,7.5){\line(1,0){15}}
\put(12.5,0){\dashbox(15,27.5){}}
\end{picture}
\caption{\label{fig:CNOT}%
Controlled-NOT gate.}
\vspace{\floatsep}

\setlength{\unitlength}{0.9mm}
\begin{picture}(45,27.5)(5,0)

\put(5,20){\line(1,0){12.5}}
 \put(12.5,20){\markerDot}
\put(17.5,17.5){\framebox(5,5){$=$}}
 \put(20,17.5){\line(0,-1){7.5}}
\put(22.5,20){\line(1,0){10}}
\put(32.5,17.5){\framebox(5,5){$=$}}
 \put(35,17.5){\line(0,-1){7.5}}
\put(37.5,20){\line(1,0){12.5}}
\put(5,7.5){\line(1,0){12.5}}
 \put(12.5,7.5){\markerDot}
\put(17.5,5){\framebox(5,5){$\oplus$}}
\put(22.5,7.5){\line(1,0){10}}
\put(32.5,5){\framebox(5,5){$\oplus$}}
\put(37.5,7.5){\line(1,0){12.5}}
\put(12.5,0){\dashbox(30,27.5){}}
\end{picture}
\hspace{3mm}
\begin{picture}(5,27.5)(0,0)
\put(2.5,13.75){\cent{$=$}}
\end{picture}
\hspace{3mm}
\begin{picture}(30,27.5)(5,0)
\put(5,20){\line(1,0){12.5}}
 \put(12.5,20){\markerDot}
\put(17.5,17.5){\framebox(5,5){$=$}}
\put(22.5,20){\line(1,0){12.5}}
\put(5,7.5){\line(1,0){12.5}}
 \put(12.5,7.5){\markerDot}
\put(17.5,5){\framebox(5,5){$=$}}
\put(22.5,7.5){\line(1,0){12.5}}
\put(12.5,0){\dashbox(15,27.5){}}
\end{picture}
\caption{\label{fig:controlledNotUnitary}%
Proof that \Fig{fig:CNOT} is unitary: 
the exterior functions left and right are equal.}
\end{figure}

\afterpage{\clearpage}
\afterpage{\clearpage}

\subsection{Quantum Circuits}
\label{sec:QuantumCircuits}

Quantum gates \cite[Chap.~4]{NiChuang:QCI}
are unitary matrices used in quantum computation. 
(In Figs.\ \ref{fig:GenQFG} or \ref{fig:PartialMeasurement},
such quantum gates would appear as, or inside, $U_0, U_1, U_2,$ \@\ldots)
For example, \Fig{fig:SwapGate} shows a swap gate 
and \Fig{fig:CNOT} shows a controlled-NOT gate in factor-graph notation.
All variables in these two examples are $\{ 0,1\}$-valued
(rather than $\{ 1,2\}$-valued), 
both rows and columns are indexed by pairs of bits (cf.\ \Fig{fig:DecompMatrixVars}), 
and the factor $f_\oplus$ in \Fig{fig:CNOT} is defined 
as 
\begin{IEEEeqnarray}{rCl}
f_\oplus: & \hspace{0.5em} & \{ 0,1 \}^3 \rightarrow \{ 0, 1\}: \nonumber\\
&& f_\oplus(\xi_1,\xi_2,\xi_3) \eqdef 
\left\{ \begin{array}{ll}
 1, & \text{if $\xi_1+\xi_2+\xi_3$ is even\hspace{1em}} \\
 0, & \text{otherwise.}
 \IEEEeqnarraynumspace\label{eqn:EXOR}
\end{array}\right.
\end{IEEEeqnarray}
That \Fig{fig:CNOT} is a unitary matrix may be seen from 
\Fig{fig:controlledNotUnitary}.

Quantum circuits as in \cite[Chap.~4]{NiChuang:QCI} 
may then be viewed as, or are easily translated to, 
the upper half of factor graphs as in \Fig{fig:PartialMeasurement}. 
(However, this upper half cannot, by itself, properly represent the 
joint probability distribution of several measurements.)

\subsection{Non-unitary Operators from Unitary Interactions}
\label{sec:NonUnitaryFromInteraction}

Up to now, we have considered systems composed 
from only two elements: unitary evolution and measurement. 
(The role and meaning of the latter continues to be debated, 
see also Section~\ref{sec:MeasRecons}.)
However, a natural additional element is 
shown in \Fig{fig:MargSubsystem}, where a primary quantum system 
interacts once with a secondary quantum system. 

(The secondary quantum system might be a stray particle 
that arrives from ``somewhere'', interacts with the primary system, 
and travels off to somewhere else. 
Or, with exchanged roles, the secondary system might be a measurement apparatus
that interacts once with a particle of interest.)

Closing the dashed box in \Fig{fig:MargSubsystem} 
does not, in general, result in a unitary operator. 
Clearly, the exterior function of the dashed box in \Fig{fig:MargSubsystem}
can be represented as in \Fig{fig:Kraus},
which may be viewed as 
a measurement as in \Fig{fig:GenMeasurement} with unknown result $Y$.
Conversely, it is a well-known result 
that any operation as in \Fig{fig:Kraus},
subject only to the condition
\begin{equation}
\sum_y E(y)^\H E(y) = I
\end{equation}
(corresponding to (\ref{eqn:GenMeasurementCond}) and \Fig{fig:QMBackwardBoxesGenMeas}),
can be represented as a marginalized unitary interaction as in \Fig{fig:MargSubsystem}, 
cf.\ \cite[Box~8.1]{NiChuang:QCI}).

It seems natural to conjecture that classicality emerges 
out of such marginalized unitary interactions,
as has been proposed by Zurek \cite{Zu:einsel2003,Zu:qd2009} and others.

Finally, we mention some standard terminology associated with \Fig{fig:Kraus}. 
For fixed \mbox{$Y=y$}, $E(y)$ is a matrix, and these matrices in \Fig{fig:Kraus}
are called Kraus operators
(cf.\ the operator-sum representation in \cite[Sec.~8.2.3]{NiChuang:QCI}).
The exterior function of the dashed box in \Fig{fig:Kraus}, when viewed as a matrix 
with rows indexed by $(\tilde X,\tilde X')$ and columns indexed by $(X,X')$,
is called Liouville superoperator;
when viewed as a matrix with rows indexed by $(X,\tilde X)$ and columns indexed by $(X',\tilde X')$,
it is called Choi matrix (see, e.g., \cite{WBC:tngc2015}).

\begin{figure}[t]
\centering
\setlength{\unitlength}{0.94mm}
\begin{picture}(92.5,60)(-17.5,-5)

\put(-17.5,45){\line(1,0){47.5}}  \put(-13.5,46.5){\pos{cb}{$X$}}
\put(12.5,35){\line(1,0){5}}
\put(17.5,32.5){\framebox(5,5){}}
 \put(22.5,35){\markerDot}
\put(22.5,35){\line(1,0){7.5}}
\put(30,30){\framebox(10,20){}}
\put(40,45){\line(1,0){35}}       \put(71,46.5){\pos{cb}{$\tilde X$}}
 \put(40,45){\markerDot}
\put(40,35){\line(1,0){7.5}}
 \put(40,35){\markerDot}
\put(47.5,32.5){\framebox(5,5){}}
 \put(52.5,35){\markerDot}
\put(52.5,35){\line(1,0){5}}
\put(-2.5,22.5){\framebox(5,5){}}  \put(0,21.5){\pos{ct}{$p(\xi)$}}
\put(2.5,25){\line(1,0){7.5}}      \put(6.25,26.5){\pos{cb}{$\xi$}}
\put(10,22.5){\framebox(5,5){$=$}}
\put(12.5,27.5){\line(0,1){7.5}}
\put(12.5,22.5){\line(0,-1){7.5}}
\put(12.5,15){\line(1,0){5}}
\put(17.5,12.5){\framebox(5,5){}}
 \put(17.5,15){\markerDot}
\put(22.5,15){\line(1,0){7.5}}
 \put(30,15){\markerDot}
\put(-17.5,5){\line(1,0){47.5}}    \put(-13.5,6.5){\pos{cb}{$X'$}}
 \put(30,5){\markerDot}
\put(30,0){\framebox(10,20){}}
\put(40,15){\line(1,0){7.5}}
\put(47.5,12.5){\framebox(5,5){}}
 \put(47.5,15){\markerDot}
\put(52.5,15){\line(1,0){5}}
\put(40,5){\line(1,0){35}}        \put(71,6.5){\pos{cb}{$\tilde X'$}}
\put(57.5,35){\line(0,-1){7.5}}   
\put(55,22.5){\framebox(5,5){$=$}}
\put(57.5,22.5){\line(0,-1){7.5}}

\put(-7.5,-5){\dashbox(72.5,60){}}

\end{picture}
\caption{\label{fig:MargSubsystem}%
Two quantum systems interact unitarily. 
(All unlabeled boxes are unitary matrices.)
The resulting exterior function of the dashed box 
can be represented as in \Fig{fig:Kraus}.}
\vspace{\floatsep}

\setlength{\unitlength}{1mm}
\begin{picture}(35,40)(-10,0)
\put(0,0){\dashbox(15,40)}
\put(-10,30){\line(1,0){15}}   \put(-6,31.5){\pos{cb}{$X$}}
\put(5,27.5){\framebox(5,5){}} \put(7.5,34){\pos{cb}{$E$}}
 \put(10,30){\markerDot}
\put(10,30){\line(1,0){15}}    \put(21,31.5){\pos{cb}{$\tilde X$}}
\put(7.5,27.5){\line(0,-1){15}}  \put(8.75,20){\pos{cl}{$Y$}}
\put(-10,10){\line(1,0){15}}   \put(-6,11.5){\pos{cb}{$X'$}}
\put(5,7.5){\framebox(5,5){}}  \put(7.5,6){\pos{ct}{$E^\H$}}
 \put(5,10){\markerDot}
\put(10,10){\line(1,0){15}}    \put(21,11.5){\pos{cb}{$\tilde X'$}}
\end{picture}
\caption{\label{fig:Kraus}%
Factor graph of a general quantum operation 
using Kraus operators. 
Such an operation may be viewed as a measurement 
(as in \Fig{fig:GenMeasurement}) with unknown result $Y$.
}
\end{figure}

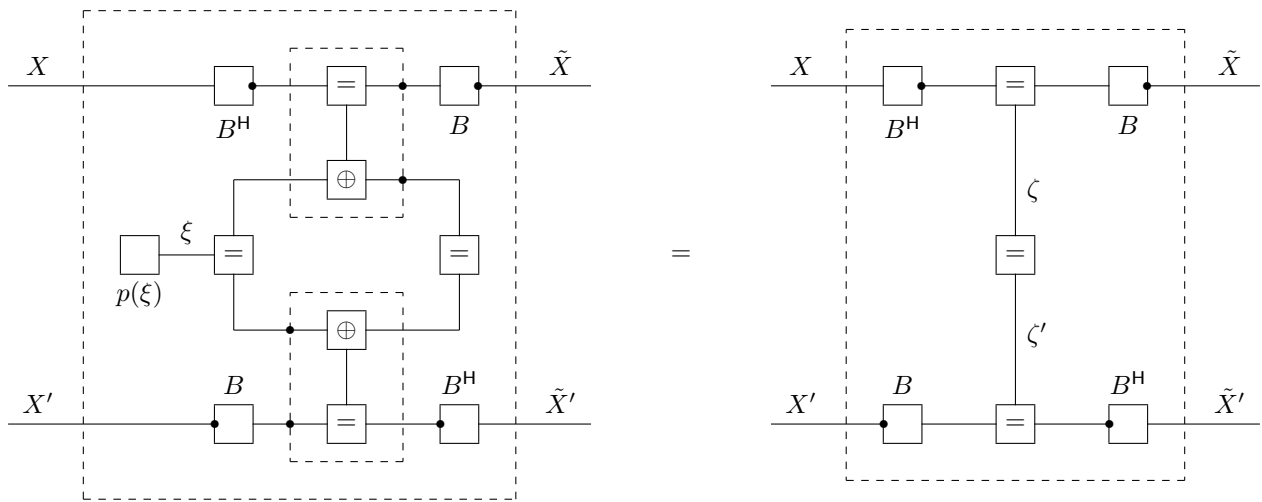
\begin{figure*}[t]
\centering
\begin{picture}(77.5,65)(-7.5,-7.5)

\put(-7.5,47.5){\line(1,0){27.5}}  \put(-3.5,49){\pos{cb}{$X$}}
\put(20,45){\framebox(5,5){}}     \put(22.5,43.5){\pos{ct}{$B^\H$}}
 \put(25,47.5){\connectionDot}
\put(25,47.5){\line(1,0){10}}
\put(35,45){\framebox(5,5){$=$}}
 \put(37.5,45){\line(0,-1){7.5}}
\put(40,47.5){\line(1,0){10}}
\put(50,45){\framebox(5,5){}}     \put(52.5,43.5){\pos{ct}{$B$}}
 \put(55,47.5){\connectionDot}
\put(55,47.5){\line(1,0){15}}     \put(66,49){\pos{cb}{$\tilde X$}}

\put(7.5,22.5){\framebox(5,5){}} \put(10,21.5){\pos{ct}{$p(\xi)$}}
\put(12.5,25){\line(1,0){7.5}}   \put(16.25,26.5){\pos{cb}{$\xi$}}
\put(20,22.5){\framebox(5,5){$=$}}
\put(22.5,27.5){\line(0,1){7.5}}
\put(22.5,22.5){\line(0,-1){7.5}}

\put(22.5,35){\line(1,0){12.5}}
\put(35,32.5){\framebox(5,5){$\oplus$}}
\put(40,35){\line(1,0){12.5}}

\put(22.5,15){\line(1,0){12.5}}
\put(35,12.5){\framebox(5,5){$\oplus$}}
\put(40,15){\line(1,0){12.5}}

\put(52.5,35){\line(0,-1){7.5}}
\put(52.5,15){\line(0,1){7.5}}
\put(50,22.5){\framebox(5,5){$=$}}

\put(-7.5,2.5){\line(1,0){27.5}}     \put(-3.5,4){\pos{cb}{$X'$}}
\put(20,0){\framebox(5,5){}}    \put(22.5,6.5){\pos{cb}{$B$}}
 \put(20,2.5){\connectionDot}
\put(25,2.5){\line(1,0){10}}
\put(35,0){\framebox(5,5){$=$}}
 \put(37.5,5){\line(0,1){7.5}}
\put(40,2.5){\line(1,0){10}}
\put(50,0){\framebox(5,5){}}    \put(52.5,6.5){\pos{cb}{$B^\H$}}
 \put(50,2.5){\connectionDot}
\put(55,2.5){\line(1,0){15}}    \put(66,4){\pos{cb}{$\tilde X'$}}

\put(30,30){\dashbox(15,22.5){}}
 \put(45,47.5){\connectionDot}
 \put(45,35){\connectionDot}

\put(30,-2.5){\dashbox(15,22.5){}}
 \put(30,15){\connectionDot}
 \put(30,2.5){\connectionDot}

\put(2.5,-7.5){\dashbox(57.5,65){}}
\end{picture}
\hspace{7mm}
\begin{picture}(5,65)(0,-7.5)
\put(2.5,25){\cent{$=$}}
\end{picture}
\hspace{7mm}
\begin{picture}(65,65)(0,-7.5)
\put(0,47.5){\line(1,0){15}}     \put(4,49){\pos{cb}{$X$}}
\put(15,45){\framebox(5,5){}}    \put(17.5,43.5){\pos{ct}{$B^\H$}}
 \put(20,47.5){\connectionDot}
\put(20,47.5){\line(1,0){10}}
\put(30,45){\framebox(5,5){$=$}}
\put(35,47.5){\line(1,0){10}}
\put(45,45){\framebox(5,5){}}    \put(47.5,43.5){\pos{ct}{$B$}}
 \put(50,47.5){\connectionDot}
\put(50,47.5){\line(1,0){15}}    \put(61,49){\pos{cb}{$\tilde X$}}

\put(32.5,27.5){\line(0,1){17.5}}  \put(34,32){\pos{bl}{$\zeta$}}
\put(30,22.5){\framebox(5,5){$=$}}
\put(32.5,22.5){\line(0,-1){17.5}} \put(34,16){\pos{tl}{$\zeta'$}}

\put(0,2.5){\line(1,0){15}}      \put(4,4){\pos{cb}{$X'$}}
\put(15,0){\framebox(5,5){}}     \put(17.5,6.5){\pos{cb}{$B$}}
 \put(15,2.5){\connectionDot}
\put(20,2.5){\line(1,0){10}}
\put(30,0){\framebox(5,5){$=$}}
\put(35,2.5){\line(1,0){10}}
\put(45,0){\framebox(5,5){}}     \put(47.5,6.5){\pos{cb}{$B^\H$}}
 \put(45,2.5){\connectionDot}
\put(50,2.5){\line(1,0){15}}     \put(61,4){\pos{cb}{$\tilde X'$}}

\put(10,-5){\dashbox(45,60){}}
\end{picture}
\caption{\label{fig:ProjMeasurementInteraction}%
Projection measurement (with unitary matrix $B$) as marginalized unitary interaction. 
Left: unitary interaction as in \Fig{fig:MargSubsystem};
the inner dashed boxes are unitary (cf.\ \Fig{fig:CNOT}).
Right: resulting projection measurement (with unknown result $\zeta$).
The exterior functions left and right are equal.}
\end{figure*}

\begin{figure}
\centering
\begin{picture}(30,35)(0,0)
\put(7.5,22.5){\line(1,0){5}}
\put(12.5,20){\framebox(5,5){$\oplus$}}
 \put(15,25){\line(0,1){10}}        \put(16.5,32){\pos{cl}{$\zeta$}}
\put(17.5,22.5){\line(1,0){5}}
\put(7.5,22.5){\line(0,-1){10}}
 \put(7.5,17.5){\knownBox}          \put(5.5,17.5){\pos{cr}{$\xi$}}
\put(22.5,22.5){\line(0,-1){10}}    \put(24,17.5){\pos{cl}{$\tilde\xi$}}
\put(7.5,12.5){\line(1,0){5}}
\put(12.5,10){\framebox(5,5){$\oplus$}}
 \put(15,10){\line(0,-1){10}}       \put(16.5,3){\pos{cl}{$\zeta'$}}
\put(17.5,12.5){\line(1,0){5}}
\put(0,7.5){\dashbox(30,20){}}
\end{picture}
\hspace{5mm}
\begin{picture}(5,35)(0,0)
\put(2.5,17.5){\cent{$=$}}
\end{picture}
\hspace{5mm}
\begin{picture}(5,35)(0,0)
\put(2.5,20){\line(0,1){10}}     \put(4,27){\pos{cl}{$\zeta$}}
\put(0,15){\framebox(5,5){$=$}}
\put(2.5,15){\line(0,-1){10}}    \put(4,8){\pos{cl}{$\zeta'$}}
\end{picture}
\caption{\label{fig:ProofProjMeasurementInteraction}%
Proof of the reduction in \Fig{fig:ProjMeasurementInteraction}.}
\end{figure}
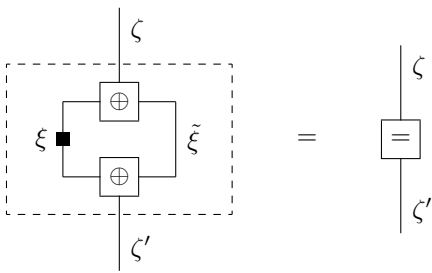

\begin{figure}[t]
\centering
\begin{picture}(65,65)(0,-7.5)
\put(0,47.5){\line(1,0){15}}     \put(4,49){\pos{cb}{$X$}}
\put(15,45){\framebox(5,5){}}    \put(17.5,43.5){\pos{ct}{$B^\H$}}
 \put(20,47.5){\connectionDot}
\put(20,47.5){\line(1,0){10}}
\put(30,45){\framebox(5,5){$=$}}
\put(35,47.5){\line(1,0){10}}
\put(45,45){\framebox(5,5){}}    \put(47.5,43.5){\pos{ct}{$B$}}
 \put(50,47.5){\connectionDot}
\put(50,47.5){\line(1,0){15}}    \put(61,49){\pos{cb}{$\tilde X$}}

\put(32.5,27.5){\line(0,1){17.5}}
\put(30,22.5){\framebox(5,5){$=$}}
 \put(35,25){\line(1,0){10}}     \put(40,26.5){\pos{cb}{$\zeta$}}
 \put(45,22.5){\framebox(5,5){}} \put(47.5,21){\pos{ct}{$p(y \cond \zeta)$}}
 \put(50,25){\line(1,0){15}}     \put(61,26.5){\pos{cb}{$Y$}}
\put(32.5,22.5){\line(0,-1){17.5}}

\put(0,2.5){\line(1,0){15}}      \put(4,4){\pos{cb}{$X'$}}
\put(15,0){\framebox(5,5){}}     \put(17.5,6.5){\pos{cb}{$B$}}
 \put(15,2.5){\connectionDot}
\put(20,2.5){\line(1,0){10}}
\put(30,0){\framebox(5,5){$=$}}
\put(35,2.5){\line(1,0){10}}
\put(45,0){\framebox(5,5){}}     \put(47.5,6.5){\pos{cb}{$B^\H$}}
 \put(45,2.5){\connectionDot}
\put(50,2.5){\line(1,0){15}}     \put(61,4){\pos{cb}{$\tilde X'$}}

\put(10,-5){\dashbox(45,60){}}
\end{picture}
\caption{\label{fig:MeasuringQuantumStateDMC}%
Observing the post-measurement variable $\zeta$ 
in \Fig{fig:ProjMeasurementInteraction} (right)
via a (classical) ``channel''
$p(y\cond \zeta)$.}
\end{figure}
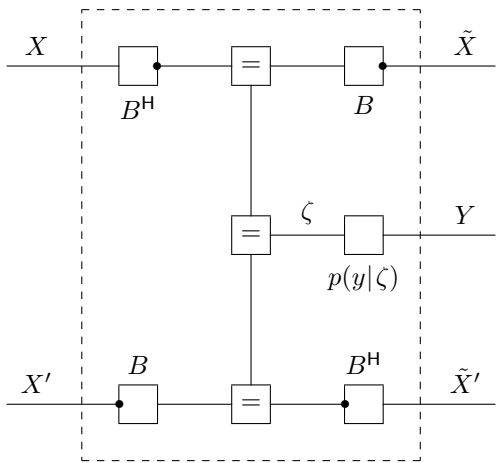

\section{Measurements Reconsidered}
\label{sec:MeasRecons}

Our tour through quantum-mechanical concepts 
followed the traditional route where ``measurement'' 
is an unexplained primitive. 
However, based on the mentioned correspondence between 
\Fig{fig:Kraus} and \Fig{fig:MargSubsystem},
progress has been made in understanding measurement 
as interaction \cite{BrPe:oqs,Schl:dmp2004}. 

There thus emerges a view of quantum mechanics
fundamentally involving only unitary transforms and marginalization.
This view is still imperfectly developed (cf.\ \cite{Schl:dmp2004}), 
but the basic idea can be explained quite easily.

\subsection{Projection Measurements}

The realization of a projection measurement 
by a unitary interaction is exemplified 
in \Fig{fig:ProjMeasurementInteraction}.
As will be detailed below, \Fig{fig:ProjMeasurementInteraction} (left)
is a unitary interaction as in \Fig{fig:MargSubsystem}
while \Fig{fig:ProjMeasurementInteraction} (right)
is a projection measurement (with unknown result $\zeta$).
We will see that the exterior functions 
of \Fig{fig:ProjMeasurementInteraction} (left) and 
\Fig{fig:ProjMeasurementInteraction} (right) are equal.

All variables in \Fig{fig:ProjMeasurementInteraction} (left)
take values in the set $\{ 0,\ldots, M{-}1 \}$ 
(rather than in $\{ 1,\ldots, M \}$) 
and the box labeled ``$\oplus$'' generalizes (\ref{eqn:EXOR}) to 
\begin{IEEEeqnarray}{rCl}
f_\oplus: & \hspace{0.5em} & \{ 0,\ldots, M{-}1 \}^3 \rightarrow \{ 0, 1\}: \nonumber\\
&& \hspace{-1.5em}
 f_\oplus(\xi_1,\xi_2,\xi_3) \eqdef 
 \left\{ \begin{array}{ll}
 1, & \text{if $(\xi_1+\xi_2+\xi_3) \bmod M = 0$\hspace{1.5em}} \\
 0, & \text{otherwise.}
 \IEEEeqnarraynumspace
\end{array}\right.
\end{IEEEeqnarray}
We first note that the two inner dashed boxes 
in \Fig{fig:ProjMeasurementInteraction} (left)
are unitary matrices, as is easily verified from \Fig{fig:controlledNotUnitary}.
Therefore, \Fig{fig:ProjMeasurementInteraction} (left)
is indeed a special case of \Fig{fig:MargSubsystem}.

The key step in the reduction of \Fig{fig:ProjMeasurementInteraction}
(left) to \Fig{fig:ProjMeasurementInteraction} (right)
is shown in \Fig{fig:ProofProjMeasurementInteraction},
which in turn can be verified as follows:
the product of the two factors in the box 
in \Fig{fig:ProofProjMeasurementInteraction} (left)
is zero unless both
\begin{equation} \label{eqn:ProofUnintaryProjMeas1}
\xi+\zeta+\tilde\xi = 0  \mod M
\end{equation}
and
\begin{equation} \label{eqn:ProofUnintaryProjMeas2}
\xi+\zeta'+\tilde\xi = 0  \mod M,
\end{equation}
which is equivalent to $\zeta = \zeta'$ and (\ref{eqn:ProofUnintaryProjMeas1}).
For fixed $\xi$ 
and $\zeta$, (\ref{eqn:ProofUnintaryProjMeas1}) allows only 
one value for $\tilde\xi$, which proves the reduction in \Fig{fig:ProofProjMeasurementInteraction}.

The generalization from fixed $\xi$ to 
arbitrary $p(\xi)$ is straightforward.

We have thus established that 
the (marginalized) unitary interaction in \Fig{fig:ProjMeasurementInteraction} (left)
acts like the projection measurement in \Fig{fig:ProjMeasurementInteraction} (right)
and thereby creates the random variable $\zeta$.

Moreover, projection measurements are repeatable, i.e., 
repeating the same measurement (immediately after the first measurement)
leaves the measured quantum system unchanged.
(In fact, this property characterizes projection measurements.)
Therefore, the random variable $\zeta$ 
is an objective 
property of the quantum system after the measurement/interaction;
it can be cloned, and it can, in principle, be observed, 
either directly or via some ``channel'' $p(y \cond \zeta)$, as illustrated in  
\Fig{fig:MeasuringQuantumStateDMC}.
The conditional-probability factor $p(y\cond \zeta)$ allows, in particular,
that $\zeta$ is not fully observable,
i.e., different values of $\zeta$ may lead to the same observation $Y=y$.

\begin{figure*}
\centering
\begin{picture}(117.5,72.5)(-17.5,-17.5)

\put(-17.5,45){\line(1,0){47.5}}  \put(-13.5,46.5){\pos{cb}{$X$}}
\put(12.5,35){\line(1,0){5}}
\put(17.5,32.5){\framebox(5,5){}}
 \put(22.5,35){\markerDot}
\put(22.5,35){\line(1,0){7.5}}
\put(30,30){\framebox(10,20){}}
\put(40,45){\line(1,0){60}}        \put(96,46.5){\pos{cb}{$\tilde X$}}
 \put(40,45){\markerDot}
\put(-2.5,22.5){\framebox(5,5){}}  \put(0,21.5){\pos{ct}{$p(\xi)$}}
\put(2.5,25){\line(1,0){7.5}}      \put(5.75,26.5){\pos{cb}{$\xi$}}
\put(10,22.5){\framebox(5,5){$=$}}
\put(12.5,27.5){\line(0,1){7.5}}
\put(12.5,22.5){\line(0,-1){7.5}}
\put(12.5,15){\line(1,0){5}}
\put(17.5,12.5){\framebox(5,5){}}
 \put(17.5,15){\markerDot}
\put(22.5,15){\line(1,0){7.5}}
 \put(30,15){\markerDot}
\put(-17.5,5){\line(1,0){47.5}}    \put(-13.5,6.5){\pos{cb}{$X'$}}
 \put(30,5){\markerDot}
\put(30,0){\framebox(10,20){}}
\put(40,5){\line(1,0){60}}         \put(96,6.5){\pos{cb}{$\tilde X'$}}

\put(40,35){\line(1,0){7.5}}
 \put(40,35){\markerDot}
\put(47.5,32.5){\framebox(5,5){}}  \put(50,31.25){\pos{ct}{$B^\H$}}
 \put(52.5,35){\markerDot}
\put(52.5,35){\line(1,0){7.5}}
\put(60,32.5){\framebox(5,5){$=$}}

\put(40,15){\line(1,0){7.5}}
\put(47.5,12.5){\framebox(5,5){}}  \put(50,18.75){\pos{cb}{$B$}}
 \put(47.5,15){\markerDot}
\put(52.5,15){\line(1,0){7.5}}
\put(60,12.5){\framebox(5,5){$=$}}

{\gray
\multiput(72.5,35)(-1.5,0){5}{\line(-1,0){1}}
\put(72.5,32.5){\framebox(5,5){}}    \put(75,31.25){\pos{ct}{$B$}}
 \put(77.5,35){\connectionDot}
\multiput(82.5,35)(-1.5,0){3}{\line(-1,0){1}}
\multiput(82.5,35)(0,-1.5){5}{\line(0,-1){1}}
\put(80,22.5){\framebox(5,5){$=$}}
\multiput(82.5,15)(0,1.5){5}{\line(0,1){1}}
\multiput(65,15)(1.5,0){5}{\line(1,0){1}}
\put(72.5,12.5){\framebox(5,5){}}    \put(75,18.75){\pos{cb}{$B^\H$}}
 \put(72.5,15){\connectionDot}
\multiput(77.5,15)(1.5,0){3}{\line(1,0){1}}
} 

\put(62.5,32.5){\line(0,-1){15}}
\put(62.5,12.5){\line(0,-1){12.5}}  \put(63.5,8.5){\pos{cl}{$\zeta$}}
\put(60,-5){\framebox(5,5){}}       \put(66.5,-2.5){\pos{cl}{$p(y\cond \zeta)$}}
\put(62.5,-5){\line(0,-1){12.5}}    \put(63.5,-17.25){\pos{bl}{$Y$}}

\put(-7.5,-10){\dashbox(97.5,65){}}
\end{picture}
\caption{\label{fig:MeasurementInteraction}%
General measurement as unitary interaction and marginalization.
The matrix $B$ and the unlabeled solid boxes are unitary matrices.
The part with the dashed edges is redundant.}
\vspace{\dblfloatsep}

\centering
\begin{picture}(77.5,55)(-7.5,-5)
\put(-2.5,17.5){\framebox(5,5){}}  \put(0,16.5){\pos{ct}{$p(x_0)$}}
\put(2.5,20){\line(1,0){10}}       \put(7.5,21){\pos{cb}{$X_0$}}
\put(12.5,17.5){\framebox(5,5){$=$}}
\put(15,22.5){\line(0,1){10}}
\put(15,32.5){\line(1,0){10}}
\put(25,25){\framebox(7.5,15){}}
\put(32.5,35){\connectionDot}
\put(32.5,35){\line(1,0){25}}     \put(45,36.5){\pos{cb}{$X$}}
\put(32.5,30){\connectionDot}
\put(15,17.5){\line(0,-1){10}}
\put(15,7.5){\line(1,0){10}}
 \put(25,7.5){\connectionDot}
\put(25,0){\framebox(7.5,15){}}
\put(32.5,5){\line(1,0){25}}     \put(45,3.75){\pos{ct}{$X'$}}
\put(32.5,30){\line(1,0){10}}
\put(42.5,30){\line(0,-1){7.5}}   \put(44,29){\pos{tl}{$\xi$}}
\put(40,17.5){\framebox(5,5){$=$}}
\put(42.5,17.5){\line(0,-1){7.5}} \put(44,11){\pos{bl}{$\xi'$}}
\put(32.5,10){\line(1,0){10}}
\put(-7.5,-5){\dashbox(57.5,50){}}   \put(23,46){\pos{cb}{$\rho(x,x')$}}
\put(57.5,32.5){\framebox(5,5){}}
\put(62.5,35){\connectionDot}
\put(62.5,35){\line(1,0){5}}
\put(67.5,35){\line(0,-1){12.5}}
\put(65,17.5){\framebox(5,5){$=$}}
\put(67.5,17.5){\line(0,-1){12.5}}
\put(62.5,5){\line(1,0){5}}
\put(57.5,2.5){\framebox(5,5){}}
\put(57.5,5){\connectionDot}
\end{picture}
\caption{\label{fig:MargSep}%
Marginalization over $\xi$ turns $\xi$ into a random variable.
(The unlabeled boxes are unitary matrices.)}
\vspace{\dblfloatsep}

\centering
\newcommand{\shiftbox}{\begin{picture}(15,25)%
 \put(0,0){\framebox(15,25){}}
 \put(7.5,20.5){\cent{$\vdots$}}
 \put(0,15){\line(1,0){5}}
 \put(5,15){\line(1,-1){5}}
 \put(10,10){\line(1,0){5}}
 \put(0,10){\line(1,0){5}}
 \put(5,10){\line(1,-1){5}}
 \put(10,5){\line(1,0){5}}
 \end{picture}}
\newcommand{\shiftboxflipped}{\begin{picture}(15,25)%
 \put(0,0){\framebox(15,25){}}
 \put(0,15){\line(1,0){5}}
 \put(5,15){\line(1,1){5}}
 \put(10,20){\line(1,0){5}}
 \put(0,10){\line(1,0){5}}
 \put(5,10){\line(1,1){5}}
 \put(10,15){\line(1,0){5}}
 \put(7.5,6){\cent{$\vdots$}}
 \end{picture}}
\begin{picture}(147.5,65)(-10,0)
\put(-10,30){\framebox(5,5){}}    \put(-7.5,29){\pos{ct}{$p(x_0)$}}
\put(-5,32.5){\line(1,0){10}}     \put(0,33.5){\pos{cb}{$X_0$}}
\put(5,30){\framebox(5,5){$=$}}
\put(7.5,35){\line(0,1){17.5}}
\put(7.5,52.5){\line(1,0){7.5}}
\put(15,40){\framebox(10,25){}}
\put(37.5,61){\cent{$\vdots$}}
\put(25,55){\line(1,0){25}}
 \put(25,55){\connectionDot}
\put(25,50){\line(1,0){25}}
 \put(25,50){\connectionDot}
\put(7.5,30){\line(0,-1){17.5}}
\put(7.5,12.5){\line(1,0){7.5}}
 \put(15,12.5){\connectionDot}
\put(15,0){\framebox(10,25){}}
\put(25,15){\line(1,0){25}}
\put(25,10){\line(1,0){25}}
\put(37.5,5){\cent{$\vdots$}}
\put(25,45){\line(1,0){12.5}}
 \put(25,45){\connectionDot}
\put(37.5,45){\line(0,-1){10}}   \put(38.75,40){\pos{cl}{$\xi_1$}}
\put(35,30){\framebox(5,5){$=$}}
\put(25,20){\line(1,0){12.5}}
\put(37.5,20){\line(0,1){10}}
\put(50,40){\shiftbox}
\put(50,0){\shiftboxflipped}
\put(65,45){\line(1,0){12.5}}
\put(77.5,45){\line(0,-1){10}}  \put(78.75,40){\pos{cl}{$\xi_2$}}
\put(75,30){\framebox(5,5){$=$}}
\put(65,20){\line(1,0){12.5}}
\put(77.5,20){\line(0,1){10}}
\put(77.5,61){\cent{$\vdots$}}
\put(65,55){\line(1,0){25}}
\put(65,50){\line(1,0){25}}
\put(90,40){\shiftbox}
\put(90,0){\shiftboxflipped}
\put(65,15){\line(1,0){25}}
\put(65,10){\line(1,0){25}}
\put(77.5,5){\cent{$\vdots$}}
\put(105,45){\line(1,0){12.5}}
\put(117.5,45){\line(0,-1){10}}  \put(118.75,40){\pos{cl}{$\xi_3$}}
\put(115,30){\framebox(5,5){$=$}}
\put(105,20){\line(1,0){12.5}}
\put(117.5,20){\line(0,1){10}}
\put(117.5,61){\cent{$\vdots$}}
\put(105,55){\line(1,0){25}}
\put(105,50){\line(1,0){25}}
\put(105,15){\line(1,0){25}}
\put(105,10){\line(1,0){25}}
\put(117.5,5){\cent{$\vdots$}}
\put(135,32.5){\cent{$\ldots$}}
\end{picture}
\caption{\label{fig:ShiftProc}%
Stochastic process without measurement.
The rectangular boxes are unitary operators.}
\end{figure*}

\subsection{General Measurements}

A very general form of (indirect) measurement is shown 
in \Fig{fig:MeasurementInteraction}, 
which is identical to \Fig{fig:MargSubsystem} except 
for the observable variable $Y$.
The figure is meant to be interpreted as follows.
Some primary quantum system (with variables $X,X',\tilde X,\tilde X'$)
interacts once with a secondary quantum system, which in 
turn is measured by a projection measurement 
as in \Fig{fig:MeasuringQuantumStateDMC}.
It is not difficult to verify 
(e.g., by adapting the procedure in \cite[Box~8.1]{NiChuang:QCI})
that an interaction as in \Fig{fig:MeasurementInteraction}
can realize any measurement as in \Fig{fig:GenMeasurement}.

\section{Random Variables Reconsidered}
\label{sec:RVRev}

Up to Section~\ref{sec:QuantumCircuits},
all random variables were either part of the initial conditions
(such as $X_0$ in \Fig{fig:PartialMeasurement})
or else created by measurements 
(such as $Y_1$ and $Y_2$ in \Fig{fig:PartialMeasurement}).
In Section~\ref{sec:MeasRecons}, we have outlined 
an emerging view
of quantum mechanics where measurements are no longer undefined primitives,
but explained as unitary interactions. 

We now re-examine the creation of random variables in this setting. 
We find that, fundamentally, random variables are not created 
by interaction, but by the end of it. 
The mechanism is illustrated in \Fig{fig:MargSep}:
a quantum system with potentially entangled variables 
$(X,X')$ and $(\xi,\xi')$
splits 
such that $(X,X')$ and $(\xi,\xi')$ do not interact in the future.
In this case, $(\xi,\xi')$ can be marginalized away 
by closing the dashed box in \Fig{fig:MargSep},
which amounts to forming the density 
matrix $\rho(x,x')$ as a partial trace of 
$\rho\big((x,\xi),(x',\xi')\big)$.
In this reduced model, 
$\xi$ is a random variable 
(inside the representation of the density matrix $\rho(x,x')$),
as is obvious in \Fig{fig:MargSep}).

In other words, random variables are created 
as a byproduct of separation: if a quantum system
splits into two parts that do not interact in the future,
then focussing on one subsystem (by marginalizating 
the other subsystem away) turns the state variable(s) of the other subsystem 
into random variables. 

The number of random variables that can be created in this way 
is limited by the initial state: 
the product of the alphabet sizes of $X$ and $\xi$ 
must equal the alphabet size of $X_0$ in \Fig{fig:MargSep}.

In particular, a stochastic process $\xi_1$, $\xi_2$, \ldots, cannot be created
in this way (i.e., without measurements or additional quantum variables) 
if the alphabet of $X_0$ is finite.

If we drop the restriction to finite alphabets, 
then stochastic processes are possible. 
For example, for $k=1,2,3,\ldots$, let 
\begin{equation}
X_k = (X_{k,1}, X_{k,2},\ldots)
\end{equation}
with $X_{k,\ell} \in \{ 1,\ldots, M \}$, let
$\xi_k = X_{k,1}$,
and let
\begin{equation}
X_{k+1} = (X_{k,2}, X_{k,3},\ldots),
\end{equation}
as illustrated in \Fig{fig:ShiftProc}.
Clearly, $\xi_1,\xi_2,\ldots$ is a discrete-time stochastic process 
generated by a quantum system without measurement.

\section{On Quantum Codes and Channels}
\label{sec:CodesChannels}

In this final section, we briefly outline 
the basic concepts of quantum coding \cite{NiChuang:QCI}
in terms of the factor-graph representation.

A quantum channel is an operator that maps 
a density matrix into another density matrix,
as will be discussed below.
The purpose of quantum coding is to 
create an overall quantum system, around the channel, 
that is insensitive (within some limits) to the action of the channel.

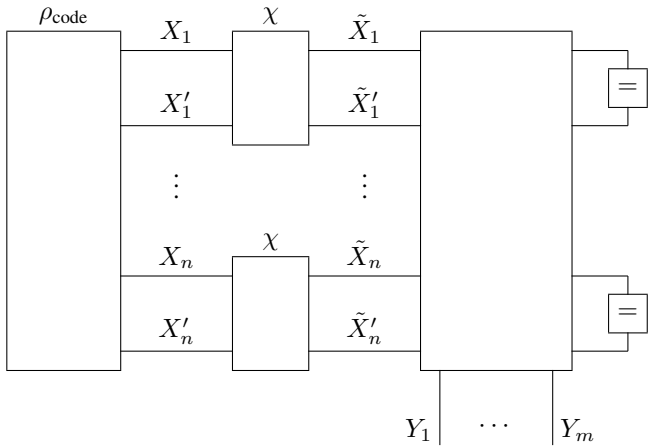
\begin{figure}
\centering
\begin{picture}(85,59)(0,-10)
\put(0,0){\framebox(15,45){}}   \put(7.5,46){\pos{cb}{$\rho_\text{code}$}}
\put(15,42.5){\line(1,0){15}}   \put(22.5,43.5){\pos{cb}{$X_1$}}
\put(15,32.5){\line(1,0){15}}   \put(22.5,33.5){\pos{cb}{$X_1'$}}
\put(30,30){\framebox(10,15){}} \put(35,46){\pos{cb}{$\chi$}}
\put(22.5,25.5){\cent{$\vdots$}}
\put(15,12.5){\line(1,0){15}}   \put(22.5,13.5){\pos{cb}{$X_n$}}
\put(15,2.5){\line(1,0){15}}    \put(22.5,3.5){\pos{cb}{$X_n'$}}
\put(30,0){\framebox(10,15){}}  \put(35,16){\pos{cb}{$\chi$}}
\put(40,42.5){\line(1,0){15}}   \put(47.5,43.5){\pos{cb}{$\tilde X_1$}}
\put(40,32.5){\line(1,0){15}}   \put(47.5,33.5){\pos{cb}{$\tilde X_1'$}}
\put(47.5,25.5){\cent{$\vdots$}}
\put(40,12.5){\line(1,0){15}}   \put(47.5,13.5){\pos{cb}{$\tilde X_n$}}
\put(40,2.5){\line(1,0){15}}    \put(47.5,3.5){\pos{cb}{$\tilde X_n'$}}
\put(55,0){\framebox(20,45){}}
\put(57.5,0){\line(0,-1){10}}  \put(56.5,-9.5){\pos{br}{$Y_1$}}
\put(65,-7.5){\cent{$\cdots$}}
\put(72.5,0){\line(0,-1){10}}  \put(73.5,-9.5){\pos{bl}{$Y_m$}}
\put(75,42.5){\line(1,0){7.5}}
\put(82.5,42.5){\line(0,-1){2.5}}
\put(80,35){\framebox(5,5){$=$}}
\put(82.5,35){\line(0,-1){2.5}}
\put(75,32.5){\line(1,0){7.5}}
\put(75,12.5){\line(1,0){7.5}}
\put(82.5,12.5){\line(0,-1){2.5}}
\put(80,5){\framebox(5,5){$=$}}
\put(82.5,5){\line(0,-1){2.5}}
\put(75,2.5){\line(1,0){7.5}}
\end{picture}
\caption{\label{fig:CodeChannelModel}%
Factor graph of length-$n$ quantum code, memoryless quantum channel, and detector. 
Note the visual arrangement of the variables 
into pairs $(X_1,X_1')$, \ldots, $(X_n,X_n')$,
which differs from most other figures in this paper.}
\end{figure}

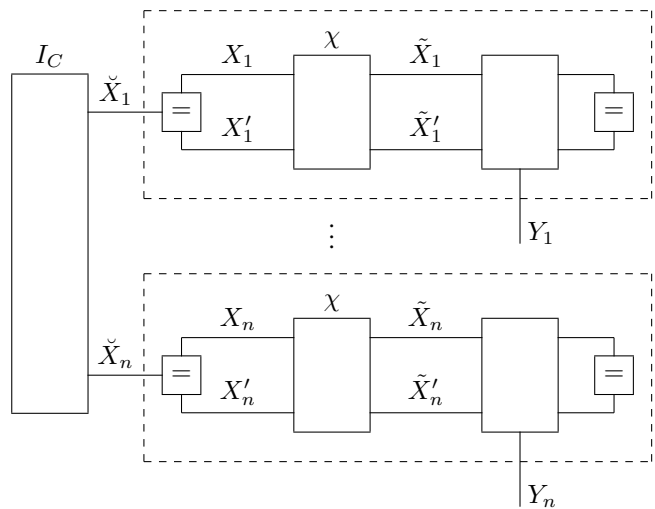
\begin{figure}
\centering
\begin{picture}(85,66)(-7.5,-10)
\put(-7.5,2.5){\framebox(10,45){}}  \put(-2.5,48.5){\pos{cb}{$I_C$}}
\put(2.5,42.5){\line(1,0){10}}    \put(6.25,43.5){\pos{cb}{$\breve X_1$}}
\put(2.5,7.5){\line(1,0){10}}     \put(6.25,8.5){\pos{cb}{$\breve X_n$}}
\put(10,31){\dashbox(67.5,25){}}
\put(35,27){\cent{$\vdots$}}
\put(10,-4){\dashbox(67.5,25){}}
\put(12.5,40){\framebox(5,5){$=$}}
\put(15,45){\line(0,1){2.5}}
\put(15,47.5){\line(1,0){15}}   \put(22.5,48.5){\pos{cb}{$X_1$}}
\put(15,40){\line(0,-1){2.5}}
\put(15,37.5){\line(1,0){15}}   \put(22.5,38.5){\pos{cb}{$X_1'$}}
\put(30,35){\framebox(10,15){}} \put(35,51){\pos{cb}{$\chi$}}
\put(15,12.5){\line(1,0){15}}   \put(22.5,13.5){\pos{cb}{$X_n$}}
\put(15,2.5){\line(1,0){15}}    \put(22.5,3.5){\pos{cb}{$X_n'$}}
\put(30,0){\framebox(10,15){}}  \put(35,16){\pos{cb}{$\chi$}}
\put(12.5,5){\framebox(5,5){$=$}}
\put(15,10){\line(0,1){2.5}}
\put(40,47.5){\line(1,0){15}}   \put(47.5,48.5){\pos{cb}{$\tilde X_1$}}
\put(15,5){\line(0,-1){2.5}}
\put(40,37.5){\line(1,0){15}}   \put(47.5,38.5){\pos{cb}{$\tilde X_1'$}}
\put(55,35){\framebox(10,15){}}
\put(60,35){\line(0,-1){10}}   \put(61,25){\pos{bl}{$Y_1$}}
\put(40,12.5){\line(1,0){15}}   \put(47.5,13.5){\pos{cb}{$\tilde X_n$}}
\put(40,2.5){\line(1,0){15}}    \put(47.5,3.5){\pos{cb}{$\tilde X_n'$}}
\put(55,0){\framebox(10,15){}}
\put(60,0){\line(0,-1){10}}    \put(61,-10){\pos{bl}{$Y_n$}}
\put(65,47.5){\line(1,0){7.5}}
\put(72.5,47.5){\line(0,-1){2.5}}
\put(70,40){\framebox(5,5){$=$}}
\put(72.5,40){\line(0,-1){2.5}}
\put(65,37.5){\line(1,0){7.5}}
\put(65,12.5){\line(1,0){7.5}}
\put(72.5,12.5){\line(0,-1){2.5}}
\put(70,5){\framebox(5,5){$=$}}
\put(72.5,5){\line(0,-1){2.5}}
\put(65,2.5){\line(1,0){7.5}}
\end{picture}
\caption{\label{fig:CodeChannelClassical}%
Quantum channel turned into classical channel and used with classical code
with indicator function $I_C$.
(Note the visual arrangement of the variables 
into pairs $(X_1,X_1')$, \ldots, $(X_n,X_n')$ 
as in \Fig{fig:CodeChannelModel}.)
}
\end{figure}

A quantum system with error correction 
comprises four parts: an encoder, a channel, 
a detector, and a reconstruction device. 
The encoder of a quantum code maps some given (classical or quantum) information
into a quantum system with state variables 
$(X_1,X_1'),\ldots,(X_n,X_n')$,
which is fed as input to the quantum channel. 
The output of the quantum channel is processed by the detector,
which involves measurements with results $Y_1,\ldots,Y_m$. 
From these results, the reconstruction device attempts 
to recover either the pre-channel quantum state or 
the (classical or quantum) information that was encoded. 

\Fig{fig:CodeChannelModel} shows the factor graph of such a system. 
More precisely, the figure shows the factor graph 
of a general code with density matrix $\rho_\text{code}$, 
a memoryless channel, and a general detector. 
A channel is called memoryless if it operates separately on $X_1, X_2, \ldots$,
as shown in \Fig{fig:CodeChannelModel}. 
The reconstruction device is not shown in \Fig{fig:CodeChannelModel}. 

In the special case where the code and the detector 
can be represented as in \Fig{fig:CodeChannelClassical},
the quantum channel is effectively transformed into 
a classical memoryless channel with $m=n$ and 
\begin{equation}
p(y_1,\ldots,y_n \cond \breve x_1,\ldots, \breve x_n)
=  \prod_{\ell=1}^n p(y_\ell\cond \breve x_\ell),
\end{equation}
and $\rho_\text{code}$ effectively reduces to the indicator function
\begin{equation}
I_C(\breve x_1,\ldots, \breve x_n) \eqdef \left\{ \begin{array}{ll}
     1, & (\breve x_1,\ldots, \breve x_n) \in C \\
     0, & \text{otherwise}
   \end{array} \right.
\end{equation}
of a classical code $C$ (up to a scale factor).
In this case, standard classical decoding algorithms can 
be used. For example, if $C$ is a low-density parity-check code,
it can be decoded by iterative sum-product message passing 
in the factor graph of $C$ \cite{Lg:ifg2004,RiUr:mct} .

By contrast, in genuine quantum coding, the detector 
does not split as in \Fig{fig:CodeChannelClassical}. 

\begin{figure}[tp]
\centering
\begin{picture}(55,49)(0,0)
\put(0,35){\line(1,0){27.5}}      \put(4,36){\pos{cb}{$X_\ell$}}
\put(27.5,32.5){\framebox(5,5){}} \put(30,39){\pos{cb}{$A_\ell(\xi_\ell)$}}
 \put(32.5,35){\connectionDot}
\put(32.5,35){\line(1,0){22.5}}   \put(51,36){\pos{cb}{$\tilde X_\ell$}}
\put(15,20){\framebox(5,5){}}     \put(17.5,19){\pos{ct}{$p(\xi_\ell)$}}
\put(20,22.5){\line(1,0){7.5}}   \put(23.75,23.75){\pos{cb}{$\xi_\ell$}}
\put(27.5,20){\framebox(5,5){$=$}}
 \put(30,25){\line(0,1){7.5}}
\put(30,20){\line(0,-1){7.5}}
\put(0,10){\line(1,0){27.5}}      \put(4,11){\pos{cb}{$X'_\ell$}}
\put(27.5,7.5){\framebox(5,5){}}  \put(30,6){\pos{ct}{$A_\ell(\xi_\ell)^\H$}}
 \put(27.5,10){\connectionDot}
\put(32.5,10){\line(1,0){22.5}}   \put(51,11){\pos{cb}{$\tilde X'_\ell$}}
\put(10,0){\dashbox(35,45){}}   \put(27.5,46){\pos{cb}{$\chi$}}
\end{picture}
\caption{\label{fig:SelectedMatrixChannel}%
Factor graph of a general channel model (for use in \Fig{fig:CodeChannelModel}).
\hspace{3em} The node/factor $p(\xi_\ell)$ may be missing.}
\vspace{\floatsep}
\centering
\begin{picture}(50,41.5)(0,0)
\put(0,27.5){\line(1,0){22.5}}    \put(4,28.5){\pos{cb}{$X_\ell$}}
\put(22.5,25){\framebox(5,5){}}   \put(25,31.5){\pos{cb}{$A_\ell$}}
 \put(27.5,27.5){\connectionDot}
\put(27.5,27.5){\line(1,0){22.5}}  \put(46,28.5){\pos{cb}{$\tilde X_\ell$}}
\put(0,10){\line(1,0){22.5}}      \put(4,11){\pos{cb}{$X'_\ell$}}
\put(22.5,7.5){\framebox(5,5){}}  \put(25,6){\pos{ct}{$A_\ell^\H$}}
 \put(22.5,10){\connectionDot}
\put(27.5,10){\line(1,0){22.5}}   \put(46,11){\pos{cb}{$\tilde X'_\ell$}}
\put(10,0){\dashbox(30,37.5){}}   \put(27.5,38.5){\pos{cb}{$\chi$}}
\end{picture}
\caption{\label{fig:FixedMatrixChannel}%
Simplified version of \Fig{fig:SelectedMatrixChannel} for fixed $\xi_\ell$.}
\end{figure}

\subsection{On Channels}

A factor graph of a quite general class of memoryless channel models 
is shown in \Fig{fig:SelectedMatrixChannel},
which may be interpreted in several different ways.
For example, the matrix $A_\ell(\xi_\ell)$ might be an unknown unitary matrix
that is selected by the random variable $\xi_\ell$ 
with probability density function $p(\xi_\ell)$.
Or, in an other interpretation, \Fig{fig:SelectedMatrixChannel}
without the node/factor $p(\xi_\ell)$
is a general operation as in \Fig{fig:Kraus}.

Many quantum coding schemes distinguish only between 
``no error'' in position $\ell$ (i.e., $A_\ell(\xi_\ell)=I$) 
and ``perhaps some error'' 
(where $A_\ell(\xi_\ell)$ is arbitrary, but nonzero);
no other distinction is made and no prior $p(\xi_\ell)$ is assumed.
For the analysis of such schemes, 
\Fig{fig:SelectedMatrixChannel} can often be replaced by
the simpler \Fig{fig:FixedMatrixChannel}. 
In such an analysis,
it may be helpful to express the (fixed, but unknown) matrix 
$A_\ell$ in \Fig{fig:FixedMatrixChannel} in some pertinent basis.
For example, any matrix $A\in \C^{2\times 2}$ 
can be written as
\begin{equation} \label{eqn:ErrADecompPauli}
A = \sum_{k=0}^3 w_k \sigma_k
\end{equation}
with $w_0,\ldots,w_3 \in \C$ and 
where $\sigma_0,\ldots,\sigma_3$ are the Pauli matrices
\begin{equation} \label{eqn:Pauli0}
\sigma_0 \eqdef \left( \begin{array}{cc}
   1 & 0 \\
   0 & 1
   \end{array} \right),
\end{equation}
\begin{equation} \label{eqn:Pauli1}
\sigma_1 \eqdef \left( \begin{array}{cc}
   0 & 1 \\
   1 & 0
   \end{array} \right),
\end{equation}
\begin{equation} \label{eqn:Pauli2}
\sigma_2 \eqdef \left( \begin{array}{cc}
   0 & -i \\
   i & 0
   \end{array} \right),
\end{equation}
and
\begin{equation} \label{eqn:Pauli3}
\sigma_3 \eqdef \left( \begin{array}{cc}
   1 & 0 \\
   0 & -1
   \end{array} \right).
\end{equation}
The matrices $\sigma_0,\ldots,\sigma_3$ are unitary and Hermitian, 
and they form a basis of $\C^{2\times 2}$.

\subsection{Repetition Codes of Length 2 and~3}
\label{sec:RepeatCode}

\begin{figure*}
\centering
\begin{picture}(60,64)(0,-5)
\put(0,0){\framebox(7.5,55){}}    \put(3.75,56.5){\pos{cb}{$\rho_0$}}
\put(7.5,50){\line(1,0){17.5}}
\put(25,47.5){\framebox(5,5){$=$}}
 \put(27.5,47.5){\line(0,-1){2.5}}
\put(30,50){\line(1,0){10}}
\put(40,47.5){\framebox(5,5){$=$}}
 \put(42.5,47.5){\line(0,-1){10}}
\put(45,50){\line(1,0){15}}       \put(55,51){\pos{cb}{$X_1$}}
\put(17.5,42.5){\knownBox}        \put(15.5,42.5){\pos{cr}{0}}
\put(17.5,42.5){\line(1,0){7.5}}
\put(25,40){\framebox(5,5){$\oplus$}}
\put(30,42.5){\line(1,0){30}}     \put(55,43.5){\pos{cb}{$X_2$}}
\put(17.5,35){\knownBox}          \put(15.5,35){\pos{cr}{0}}
\put(17.5,35){\line(1,0){22.5}}
\put(40,32.5){\framebox(5,5){$\oplus$}}
\put(45,35){\line(1,0){15}}       \put(55,36){\pos{cb}{$X_3$}}
\put(17.5,20){\knownBox}          \put(15.5,20){\pos{cr}{0}}
\put(17.5,20){\line(1,0){22.5}}
\put(40,17.5){\framebox(5,5){$\oplus$}}
\put(45,20){\line(1,0){15}}       \put(55,21){\pos{cb}{$X_3'$}}
\put(17.5,12.5){\knownBox}        \put(15.5,12.5){\pos{cr}{0}}
\put(17.5,12.5){\line(1,0){7.5}}
\put(25,10){\framebox(5,5){$\oplus$}}
\put(30,12.5){\line(1,0){30}}     \put(55,13.5){\pos{cb}{$X_2'$}}
\put(7.5,5){\line(1,0){17.5}}
\put(25,2.5){\framebox(5,5){$=$}}
 \put(27.5,7.5){\line(0,1){2.5}}
\put(30,5){\line(1,0){10}}
\put(40,2.5){\framebox(5,5){$=$}}
 \put(42.5,7.5){\line(0,1){10}}
\put(45,5){\line(1,0){15}}        \put(55,6){\pos{cb}{$X_1'$}}
\end{picture}
\hspace{2cm}
\begin{picture}(70,60)(0,-5)
\put(0,50){\line(1,0){15}}    \put(5,51){\pos{cb}{$\tilde X_1$}}
\put(15,47.5){\framebox(5,5){$=$}}
 \put(17.5,47.5){\line(0,-1){10}}
\put(20,50){\line(1,0){20}}
\put(40,47.5){\framebox(5,5){$=$}}
 \put(42.5,47.5){\line(0,-1){2.5}}
\put(45,50){\line(1,0){22.5}}   
\put(0,42.5){\line(1,0){40}}  \put(5,43.5){\pos{cb}{$\tilde X_2$}}
\put(40,40){\framebox(5,5){$\oplus$}}
\put(45,42.5){\line(1,0){5}}
\put(0,35){\line(1,0){15}}   \put(5,36){\pos{cb}{$\tilde X_3$}}
\put(15,32.5){\framebox(5,5){$\oplus$}}
\put(20,35){\line(1,0){5}}
\put(25,35){\line(0,-1){5}}
\put(22.5,25){\framebox(5,5){$=$}}
\put(25,25){\line(0,-1){5}}
\put(27.5,27.5){\line(1,0){2.5}}
\put(30,27.5){\line(0,-1){32.5}}    \put(29,-4.5){\pos{br}{$Y_2$}}
\put(50,30){\line(0,1){12.5}}
\put(47.5,25){\framebox(5,5){$=$}}
\put(50,25){\line(0,-1){12.5}}
\put(52.5,27.5){\line(1,0){2.5}}
\put(55,27.5){\line(0,-1){32.5}}    \put(54,-4.5){\pos{br}{$Y_1$}}
\put(67.5,30){\line(0,1){20}}
\put(65,25){\framebox(5,5){$=$}}
\put(67.5,25){\line(0,-1){20}}
\put(0,20){\line(1,0){15}}   \put(5,21){\pos{cb}{$\tilde X_3'$}}
\put(15,17.5){\framebox(5,5){$\oplus$}}
\put(20,20){\line(1,0){5}}
\put(0,12.5){\line(1,0){40}}  \put(5,13.5){\pos{cb}{$\tilde X_2'$}}
\put(40,10){\framebox(5,5){$\oplus$}}
\put(45,12.5){\line(1,0){5}}
\put(0,5){\line(1,0){15}}    \put(5,6){\pos{cb}{$\tilde X_1'$}}
\put(15,2.5){\framebox(5,5){$=$}}
 \put(17.5,7.5){\line(0,1){10}}
\put(20,5){\line(1,0){20}}
\put(40,2.5){\framebox(5,5){$=$}}
 \put(42.5,7.5){\line(0,1){2.5}}
\put(45,5){\line(1,0){22.5}}
\end{picture}
\caption{\label{fig:RepeatCodeDec}%
An encoder (left) and a detector (right) for a repetition code of length $n=3$.}
\end{figure*}
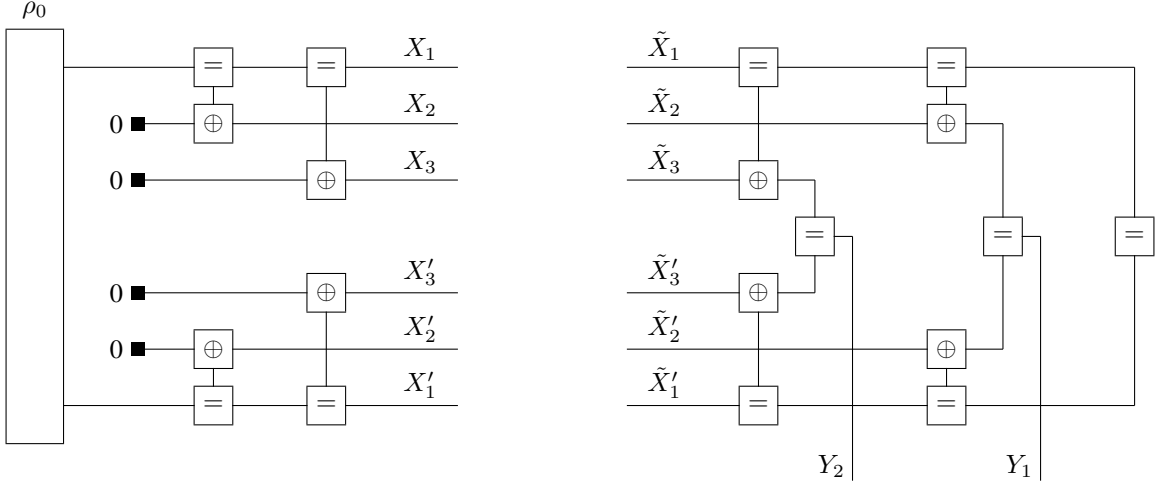

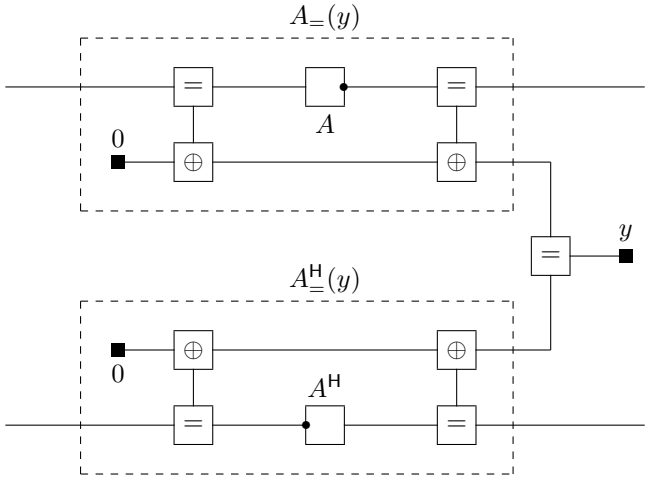
\begin{figure}
\centering
\begin{picture}(85,64)(0,1)
\put(0,52.5){\line(1,0){22.5}}
\put(22.5,50){\framebox(5,5){$=$}}
 \put(25,50){\line(0,-1){5}}
\put(27.5,52.5){\line(1,0){12.5}}     
\put(40,50){\framebox(5,5){}}         \put(42.5,48.75){\pos{ct}{$A$}}
 \put(45,52.5){\connectionDot}
\put(45,52.5){\line(1,0){12.5}}       
\put(57.5,50){\framebox(5,5){$=$}}
 \put(60,50){\line(0,-1){5}}
\put(62.5,52.5){\line(1,0){22.5}}
\put(15,42.5){\knownBox}           \put(15,44.5){\pos{cb}{$0$}}
\put(15,42.5){\line(1,0){7.5}}
\put(22.5,40){\framebox(5,5){$\oplus$}}
\put(27.5,42.5){\line(1,0){30}}
\put(57.5,40){\framebox(5,5){$\oplus$}}
\put(62.5,42.5){\line(1,0){10}}
\put(72.5,42.5){\line(0,-1){10}}
\put(10,36){\dashbox(57.5,23){}}    \put(42.5,60){\pos{cb}{$A_=(y)$}}
\put(70,27.5){\framebox(5,5){$=$}}
\put(75,30){\line(1,0){7.5}}         \put(82.5,32){\pos{cb}{$y$}}
\put(82.5,30){\knownBox}
\put(15,17.5){\knownBox}           \put(15,15.5){\pos{ct}{$0$}}
\put(15,17.5){\line(1,0){7.5}}
\put(22.5,15){\framebox(5,5){$\oplus$}}
\put(27.5,17.5){\line(1,0){30}}
\put(57.5,15){\framebox(5,5){$\oplus$}}
\put(62.5,17.5){\line(1,0){10}}
\put(72.5,17.5){\line(0,1){10}}
\put(0,7.5){\line(1,0){22.5}}
\put(22.5,5){\framebox(5,5){$=$}}
 \put(25,10){\line(0,1){5}}
\put(27.5,7.5){\line(1,0){12.5}}     
\put(40,5){\framebox(5,5){}}         \put(42.5,11.25){\pos{cb}{$A^\H$}}
 \put(40,7.5){\connectionDot}
\put(45,7.5){\line(1,0){12.5}}       
\put(57.5,5){\framebox(5,5){$=$}}
 \put(60,10){\line(0,1){5}}
\put(62.5,7.5){\line(1,0){22.5}}
\put(10,1){\dashbox(57.5,23){}}    \put(42.5,25){\pos{cb}{$A_=^\H(y)$}}
\end{picture}
\caption{\label{fig:SyndromeBitEquPath}%
Effective channel (created by encoder, channel, and detector) 
of repetition code of length $n=2$ with an error in the direct path.}
\end{figure}

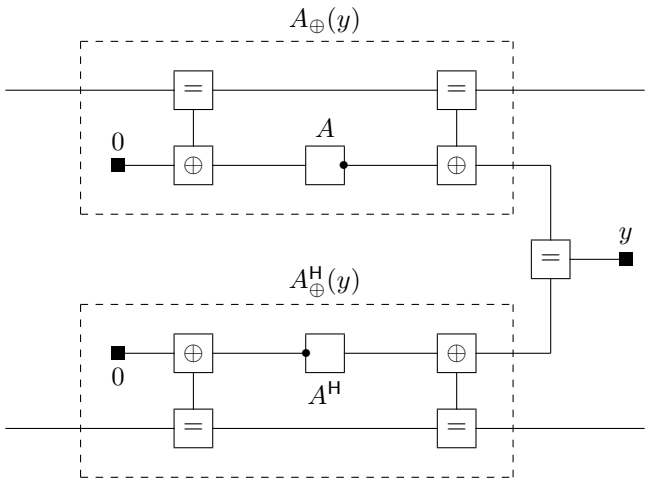
\begin{figure}
\centering
\begin{picture}(85,64)(0,1)
\put(0,52.5){\line(1,0){22.5}}
\put(22.5,50){\framebox(5,5){$=$}}
 \put(25,50){\line(0,-1){5}}
\put(27.5,52.5){\line(1,0){30}}
\put(57.5,50){\framebox(5,5){$=$}}
 \put(60,50){\line(0,-1){5}}
\put(62.5,52.5){\line(1,0){22.5}}
\put(15,42.5){\knownBox}           \put(15,44.5){\pos{cb}{$0$}}
\put(15,42.5){\line(1,0){7.5}}
\put(22.5,40){\framebox(5,5){$\oplus$}}
\put(27.5,42.5){\line(1,0){12.5}}
\put(40,40){\framebox(5,5){}}         \put(42.5,46.25){\pos{cb}{$A$}}
 \put(45,42.5){\connectionDot}
\put(45,42.5){\line(1,0){12.5}}
\put(57.5,40){\framebox(5,5){$\oplus$}}
\put(62.5,42.5){\line(1,0){10}}
\put(72.5,42.5){\line(0,-1){10}}
\put(10,36){\dashbox(57.5,23){}}    \put(42.5,60){\pos{cb}{$A_\oplus(y)$}}
\put(70,27.5){\framebox(5,5){$=$}}
\put(75,30){\line(1,0){7.5}}         \put(82.5,32){\pos{cb}{$y$}}
\put(82.5,30){\knownBox}
\put(15,17.5){\knownBox}           \put(15,15.5){\pos{ct}{$0$}}
\put(15,17.5){\line(1,0){7.5}}
\put(22.5,15){\framebox(5,5){$\oplus$}}
\put(27.5,17.5){\line(1,0){12.5}}
\put(40,15){\framebox(5,5){}}       \put(42.5,13.75){\pos{ct}{$A^\H$}}
 \put(40,17.5){\connectionDot}
\put(45,17.5){\line(1,0){12.5}}
\put(57.5,15){\framebox(5,5){$\oplus$}}
\put(62.5,17.5){\line(1,0){10}}
\put(72.5,17.5){\line(0,1){10}}
\put(0,7.5){\line(1,0){22.5}}
\put(22.5,5){\framebox(5,5){$=$}}
 \put(25,10){\line(0,1){5}}
\put(27.5,7.5){\line(1,0){30}}
\put(57.5,5){\framebox(5,5){$=$}}
 \put(60,10){\line(0,1){5}}
\put(62.5,7.5){\line(1,0){22.5}}
\put(10,1){\dashbox(57.5,23){}}    \put(42.5,25){\pos{cb}{$A_\oplus^\H(y)$}}
\end{picture}
\caption{\label{fig:SyndromeBitCheckPath}%
Effective channel (created by encoder, channel, and detector) 
of repetition code of length $n=2$ with an error in the check path.}
\end{figure}

\Fig{fig:RepeatCodeDec} (left) shows the factor graph of an encoder
of a simple code of length $n=3$. 
All variables in this factor graph are binary, 
and the initial density matrix $\rho_0$ is arbitrary.
Note that this encoder can be realized with two controlled-not gates
(cf.\ \Fig{fig:CNOT})
and two ancillary qubits with fixed initial state zero.

A detector for this code is shown in \Fig{fig:RepeatCodeDec} (right).
This detector can be realized with two controlled-not gates
and two qubit measurements. 
The unitary part 
of this detector 
inverts the unitary part of the encoder,
and the measured bits $Y_1$ and $Y_2$ (henceforth called syndrome bits)
correspond to the ancillary qubits in the encoder.

The code of \Fig{fig:RepeatCodeDec} is not very useful in itself,
but it suffices to demonstrate some basic ideas of quantum coding 
and it further illustrates the use of factor graphs.
Moreover, once this simple code is understood, 
it is easy to proceed to
the Shor code \cite{NiChuang:QCI}, 
which can correct an arbitrary single-qubit error. 

The encoder-detector pair of \Fig{fig:RepeatCodeDec} may be viewed as two nested 
encoder-detector pairs for a repetition code of length $n=2$: 
the inner encoder-detector pair produces 
the syndrome bit $Y_2$, and the outer encoder-detector pair produces 
the syndrome bit $Y_1$.

Therefore, we now consider the net effect of 
the encoder, the channel, and the detector 
of a repetition code of length $n=2$ as shown in 
Figs.~\ref{fig:SyndromeBitEquPath} and~\ref{fig:SyndromeBitCheckPath}. 
We assume that at most one qubit error occurs, 
either in the direct path (as in \Fig{fig:SyndromeBitEquPath})
or in the check path (as in \Fig{fig:SyndromeBitCheckPath}).
This single potential error is a general nonzero matrix $A \in \C^{2\times 2}$ 
(as in \Fig{fig:FixedMatrixChannel}) with row and column indices in $\{0, 1\}$.

For fixed \mbox{$Y=y$}, the net effect of the encoder, the channel, and the detector 
amounts to a matrix $A_=(y)$ or $A_\oplus(y)$ 
corresponding to the dashed boxes in 
Figs.~\ref{fig:SyndromeBitEquPath} and~\ref{fig:SyndromeBitCheckPath},
respectively.

If $A=I$ (i.e., if there is no error), we necessarily have $Y=0$
and $A_=(0) = A_\oplus(0) = I$.
For general nonzero $A$, parameterized as in (\ref{eqn:ErrADecompPauli}), 
we have
\begin{IEEEeqnarray}{rCl}
A_=(0) & = & 
   \left( \begin{array}{cc}
           A(0,0)  &  0 \\
             0     &  A(1,1)
          \end{array} \right) \IEEEeqnarraynumspace\\
 & = & w_0 \sigma_0 + w_3 \sigma_3,
       \label{eqn:SyndromeEffectEq0}
\end{IEEEeqnarray}
i.e., the projection of $A$ onto the space spanned by $\sigma_0$ and $\sigma_3$,
and
\begin{IEEEeqnarray}{rCl}
A_=(1) & = & 
   \left( \begin{array}{cc}
             0  &  A(0,1) \\
         A(1,0) &  0
          \end{array} \right) \\
 & = & w_1 \sigma_1 + w_2 \sigma_2,
       \label{eqn:SyndromeEffectEq1}
\end{IEEEeqnarray}
i.e., the projection of $A$ onto the space spanned by $\sigma_1$ and $\sigma_2$.
Moreover,
\begin{IEEEeqnarray}{rCl}
A_\oplus(0) & = & A_=(0),  \IEEEeqnarraynumspace
       \label{eqn:SyndromeEffectCheck0}
\end{IEEEeqnarray}
and
\begin{IEEEeqnarray}{rCl}
A_\oplus(1) & = & 
   \left( \begin{array}{cc}
          A(1,0) &  0 \\
            0    &  A(0,1)
          \end{array} \right) \\
 & = &  \sigma_1 A_=(1) \\
 & = & w_1 \sigma_0 + w_2 i \sigma_3.
       \label{eqn:SyndromeEffectCheck1}
\end{IEEEeqnarray}

We now return to \Fig{fig:RepeatCodeDec},
which we consider as two nested encoder-detector pairs 
as in Figs.~\ref{fig:SyndromeBitEquPath} and~\ref{fig:SyndromeBitCheckPath}. 
We assume that at most 
one qubit error occurs, or, equivalently, $A_\ell=I$ 
except for a single index $\ell \in \{1,2,3\}$.
For the inner encoder-detector pair, 
the above analysis of Figs.~\ref{fig:SyndromeBitEquPath} and~\ref{fig:SyndromeBitCheckPath} 
applies immediately. For the outer encoder-detector pair,
the same analysis can be reused, with the error matrix $A$ 
replaced by $A_=(y_2)$ or $A_\oplus(y_2)$ from the inner code.
The resulting effective channel 
from the encoder input to the detector output in \Fig{fig:RepeatCodeDec},
as a function of $Y_1$ and $Y_2$,
is tabulated in Table~\ref{table:Length3CodeEffChanPos}.

From Table~\ref{table:Length3CodeEffChanPos},
we observe that the syndrome bits $Y_1$ and $Y_2$ uniquely
determine the resulting effective channel,
which allows us to compress Table~\ref{table:Length3CodeEffChanPos}
into Table~\ref{table:Length3CodeEffChan}.
Note that the four unknown parameters $w_0,w_1,w_2,w_3$ 
of the error matrix (\ref{eqn:ErrADecompPauli})
are thus converted into only two unknown parameters 
(either $w_0$ and $w_3$ or $w_1$ and $w_2$, depending on $Y_1,Y_2$).

\begin{table}[tp]
\caption{\label{table:Length3CodeEffChanPos}%
Net effect of encoder and detector of \Fig{fig:RepeatCodeDec}
if at most one qubit error occurs.
Both the single-qubit error and the resulting effective channel
are parameterized as in (\ref{eqn:ErrADecompPauli}).}
\vspace{-3ex}
\[
\begin{array}{cc|c|c|c}
    &     &  \multicolumn{3}{c}{\text{error location}} \\
Y_2 & Y_1 &  1  &  2  &  3 \\
\hline
 0  &  0  & w_0 \sigma_0 + w_3 \sigma_3  
          & w_0 \sigma_0 + w_3 \sigma_3  
          & w_0 \sigma_0 + w_3 \sigma_3 \\
 0  &  1  & \text{impossible}  
          & w_1 \sigma_0 + w_2 i \sigma_3
          & \text{impossible} \\
 1  &  0  &  \text{impossible} 
          & \text{impossible} 
          & w_1 \sigma_0 + w_2 i \sigma_3\\
 1  &  1  & w_1 \sigma_1 + w_2 \sigma_2 
          & \text{impossible} 
          & \text{impossible}
\end{array}
\]
\vspace{\floatsep}
\caption{\label{table:Length3CodeEffChan}%
Compressed version of Table~\ref{table:Length3CodeEffChanPos}.}
\vspace{-3ex}
\[
\begin{array}{cc|c}
Y_2 & Y_1 &  \text{effective channel} \\
\hline
 0  &  0  & w_0 \sigma_0 + w_3 \sigma_3 \\
 0  &  1  & w_1 \sigma_0 + w_2 i \sigma_3 \\
 1  &  0  & w_1 \sigma_0 + w_2 i \sigma_3\\
 1  &  1  & w_1 \sigma_1 + w_2 \sigma_2 
\end{array}
\]
\end{table}

In the special case where we consider only bit flips, 
i.e., if we assume $w_2=w_3=0$, then it is obvious from 
Table~\ref{table:Length3CodeEffChan} 
that the code of \Fig{fig:RepeatCodeDec}
can correct a single bit flip in any position.
In fact, from Table~\ref{table:Length3CodeEffChanPos},
we see that a bit flip in qubit~1 is manifested 
in the syndrome $Y_1=Y_2=1$ while a bit flip in qubit~2 or in qubit~3 
has no effect on the resulting effective channel,
except for an irrelevant scale factor. 
However, we wish to be able to deal with more general errors.

\begin{figure}
\centering
\setlength{\unitlength}{0.9mm}
\begin{picture}(92.5,77)(0,7.5)
\put(0,80){\line(1,0){15}}
\put(15,77.5){\framebox(5,5){$=$}}
\put(20,80){\line(1,0){10}}
\put(30,77.5){\framebox(5,5){$=$}}
\put(35,80){\line(1,0){10}}
\put(45,77.5){\framebox(5,5){$H$}}
\put(50,80){\line(1,0){10}}
\put(60,77.5){\framebox(5,5){$=$}}
\put(65,80){\line(1,0){10}}
\put(75,77.5){\framebox(5,5){$=$}}
\put(80,80){\line(1,0){12.5}}       \put(87.5,81){\pos{cb}{$X_1$}}
\put(62.5,77.5){\line(0,-1){2.5}}
\put(60,70){\framebox(5,5){$\oplus$}}
 \put(55,72.5){\knownBox}  \put(53,72.5){\pos{cr}{$0$}}
 \put(55,72.5){\line(1,0){5}}
 \put(65,72.5){\line(1,0){27.5}}   \put(87.5,73.5){\pos{cb}{$X_2$}}
\put(77.5,77.5){\line(0,-1){10}}
\put(75,62.5){\framebox(5,5){$\oplus$}}
 \put(70,65){\knownBox}    \put(68,65){\pos{cr}{$0$}}
 \put(70,65){\line(1,0){5}}
 \put(80,65){\line(1,0){12.5}}     \put(87.5,66){\pos{cb}{$X_3$}}
\put(10,52.5){\knownBox}   \put(8,52.5){\pos{cr}{$0$}}
\put(10,52.5){\line(1,0){5}}
\put(17.5,55){\line(0,1){22.5}}
\put(15,50){\framebox(5,5){$\oplus$}}
\put(20,52.5){\line(1,0){25}}
\put(45,50){\framebox(5,5){$H$}}
\put(50,52.5){\line(1,0){10}}
\put(60,50){\framebox(5,5){$=$}}
\put(65,52.5){\line(1,0){10}}
\put(75,50){\framebox(5,5){$=$}}
\put(80,52.5){\line(1,0){12.5}}    \put(87.5,53.5){\pos{cb}{$X_4$}}
\put(62.5,47.5){\line(0,1){2.5}}
\put(60,42.5){\framebox(5,5){$\oplus$}}
 \put(55,45){\knownBox}  \put(53,45){\pos{cr}{$0$}}
 \put(55,45){\line(1,0){5}}
 \put(65,45){\line(1,0){27.5}}     \put(87.5,46){\pos{cb}{$X_5$}}
\put(77.5,40){\line(0,1){10}}
\put(75,35){\framebox(5,5){$\oplus$}}
 \put(70,37.5){\knownBox}    \put(68,37.5){\pos{cr}{$0$}}
 \put(70,37.5){\line(1,0){5}}
 \put(80,37.5){\line(1,0){12.5}}   \put(87.5,38.5){\pos{cb}{$X_6$}}
\put(25,25){\knownBox}
\put(25,25){\line(1,0){5}}
\put(32.5,27.5){\line(0,1){50}}
\put(30,22.5){\framebox(5,5){$\oplus$}}
\put(35,25){\line(1,0){10}}
\put(45,22.5){\framebox(5,5){$H$}}
\put(50,25){\line(1,0){10}}
\put(60,22.5){\framebox(5,5){$=$}}
\put(65,25){\line(1,0){10}}
\put(75,22.5){\framebox(5,5){$=$}}
\put(80,25){\line(1,0){12.5}}     \put(87.5,26){\pos{cb}{$X_7$}}
\put(62.5,20){\line(0,1){2.5}}
\put(60,15){\framebox(5,5){$\oplus$}}
 \put(55,17.5){\knownBox}  \put(53,17.5){\pos{cr}{$0$}}
 \put(55,17.5){\line(1,0){5}}
 \put(65,17.5){\line(1,0){27.5}}   \put(87.5,18.5){\pos{cb}{$X_8$}}
\put(77.5,12.5){\line(0,1){10}}
\put(75,7.5){\framebox(5,5){$\oplus$}}
 \put(70,10){\knownBox}    \put(68,10){\pos{cr}{$0$}}
 \put(70,10){\line(1,0){5}}
 \put(80,10){\line(1,0){12.5}}    \put(87.5,11){\pos{cb}{$X_9$}}
\end{picture}
\caption{\label{fig:ShorEncoder}%
Encoder of the Shor code. 
The figure shows only the upper half of the factor graph.}
\visual{\vspace{3ex}}
\end{figure}

\subsection{Correcting a Single Error: The Shor Code}

\Fig{fig:ShorEncoder} shows an encoder of the Shor code \cite{NiChuang:QCI}. 
The figure shows only the upper half of the factor graph
(i.e., the quantum circuit). 
The nodes labeled ``$H$'' represent the normalized Hadamard matrix
\begin{equation} \label{eqn:DefHadamardMatrix}
H \eqdef \frac{1}{\sqrt{2}}
  \left( \begin{array}{cc}
   1 & 1 \\
   1 & -1
  \end{array} \right),
\end{equation}
which is symmetric and unitary and satisfies 
$H\sigma_1 = \sigma_3 H$, and $H\sigma_2 = - \sigma_2 H$.
Note that this encoder uses four copies of the encoder 
in \Fig{fig:RepeatCodeDec}: three independent inner encoders 
are glued together with an outer encoder.

As a detector, we use the obvious generalization of \Fig{fig:RepeatCodeDec} (right),
i.e., the mirror image of the encoder.

This encoder-detector pair is easily analyzed 
using the results of Section~\ref{sec:RepeatCode}. 
For this analysis, we assume that at most a single qubit error occurs
(i.e., $A_\ell \neq I$ for at most one index $\ell \in \{1,\ldots,9\}$).
In consequence, two of the three inner encoder-detector pairs are error-free 
and reduce to an identity matrix.
The remaining inner encoder-detector pair 
is described by Table~\ref{table:Length3CodeEffChan}.
The multiplication by $H$ both in the encoder and in the detector 
changes Table~\ref{table:Length3CodeEffChan} to 
Table~\ref{table:Length3CodeHadamardEffChan}.
Note that the resulting effective channel is either 
of the form $a\sigma_0 + b\sigma_1$ or $c\sigma_2 + d\sigma_3$,
and the detector knows which case applies.

The outer encoder-detector pair thus sees an error in at most 
one position, and the potential error is described 
by Table~\ref{table:Length3CodeEffChan}, 
except that the underlying channel is not (\ref{eqn:ErrADecompPauli}),
but as in Table~\ref{table:Length3CodeHadamardEffChan}. 
Revisiting Table~\ref{table:Length3CodeEffChan} accordingly 
yields Table~\ref{table:ShorTypeCodeEffChan},
which describes the net effect of the outer encoder-detector pair.
In any case, the resulting effective channel is of the form 
$\alpha \sigma_k$ for some nonzero $\alpha\in \C$ and some (known) $k\in \{0,1,2,3\}$. 
In other words, the effective channel (from encoder input to detector output) 
is fully determined by the 8 syndrome bits,
up to an irrelevant scale factor.
In consequence, the (arbitrary) original quantum state can exactly be restored.

\begin{table}[tp]
\newcommand{\llinewidth}{\linewidth}
\caption{\label{table:Length3CodeHadamardEffChan}%
\mbox{\hspace{2.5cm}Effective channel of Table~\ref{table:Length3CodeEffChan}\hspace{2.5cm}}
with pre- and post-multiplication by $H$.
}
\vspace{-3ex}
\[
\begin{array}{cc|c}
Y_2 & Y_1 &  \text{effective channel} \\
\hline
 0  &  0  & w_0 \sigma_0 + w_3 \sigma_1 \\
 0  &  1  & w_1 \sigma_0 + w_2 i \sigma_1 \\
 1  &  0  & w_1 \sigma_0 + w_2 i \sigma_1\\
 1  &  1  & w_1 \sigma_3 - w_2 \sigma_2 
\end{array}
\]
\vspace{\floatsep}
\caption{\label{table:ShorTypeCodeEffChan}%
Net effect of encoder (as in \Fig{fig:ShorEncoder}) 
and mirror-image detector of Shor code, 
assuming that at most one qubit error occurs.
``Inner code'' refers to the inner encoder-detector pair
with the potential error.}
\vspace{-3ex}
\[
\begin{array}{cc|c|c}
\multicolumn{2}{c|}{\text{outer detector}} 
 & \multicolumn{2}{c}{\text{effect of inner code}} \\
Y_2 & Y_1 &  a \sigma_0 + b \sigma_1 & c \sigma_2 + d\sigma_3 \\
\hline
 0  &  0  & a \sigma_0 & d \sigma_3 \\
 0  &  1  & b \sigma_0 & c i \sigma_3 \\
 1  &  0  & b \sigma_0 & c i \sigma_3 \\
 1  &  1  & b \sigma_1 & c \sigma_2
\end{array}
\]
\visual{\vspace{1ex}}
\end{table}

\section{Conclusion}
\label{sec:Concl}

We have proposed factor graphs for 
quantum-mechanical probabilities involving any number of measurements,
both for basic projection measurements and for general measurements. 
Our factor graphs represent factorizations of 
complex-valued functions $q$ as in (\ref{eqn:pfromq})
such that the joint probability distribution of all random variables 
(in a given quantum system) is a marginal of $q$.
Therefore (and in contrast to other graphical representations of quantum mechanics),
our factor graphs are fully compatible with standard statistical models.
We have also interpreted a variety of concepts and quantities of quantum mechanics 
in terms of factorizations and marginals of such functions $q$.
We have further illustrated the use of factor graphs 
by an elementary introduction to quantum coding. 

In Appendix~\ref{sec:Related},
we offer some additional remarks on the prior literature.
In Appendix~\ref{sec:WignerWeyl}, 
we derive factor graphs for the Wigner--Weyl representation. 
In Appendix~\ref{sec:MonteCarlo}, we point out that the factor graphs of this paper 
are amenable (at least in principle) to Monte Carlo algorithms. 

We hope that our approach makes quantum-mechanical probabilities 
more accessible to non-physicists and 
further promotes the exchange of concepts and algorithms between 
physics, statistical inference, and error correcting codes 
in the spirit of \cite{MeMo:ipc,VoLg:FGEN2002,Cw:spe2015}.

Finally, we mention that the factor graphs of this paper have been used in \cite{CaoVo:ISIT2017} 
for estimating the information rate of certain quantum channels,
and iterative sum-product message passing in such factor graphs 
is considered in \cite{CaoVo:defg2017c}.

\appendices

\section{Additional Remarks About Related Work}
\label{sec:Related}

\subsection{Tensor Networks}
\label{sec:MoreAboutTensorNetworks}

With hindsight, the factor graphs of this paper 
are quite similar to tensor networks 
\cite{CiVe:rtp2009,Coe:qp2010,WBC:tngc2015,ChiAP:pp2010}, 
which have recently moved into the heart of theoretical physics 
\cite{Cw:spe2015}.

Tensor networks (and related graphical notation) 
have been used to represent the wave function 
$| \Psi \rangle$ of several entangled spins at a
given time. In general, the resulting states are called tensor network states
(TNS), but depending on the structure of the tensor network, more specialized
names like matrix product states (MPS), tree tensor states (TTS), etc., are
used. A very nice overview of this line of work is given in the survey paper
by Cirac and Verstraete \cite{CiVe:rtp2009}, 
which also explains the connection of TNS to techniques like
the density matrix renormalization group (DMRG), the multiscale entanglement
renormalization ansatz (MERA), and projected entangled pair states
(PEPS). 

If such tensor diagrams are used to represent quantities like 
$\langle \Psi | \Psi \rangle$ or $\langle \Psi | \sigma_2 \sigma_4 | \Psi \rangle$
(see, e.g., Fig.~2 in \cite{CiVe:rtp2009}),
they have two conjugate parts, like the factor graphs in the present paper
(\Fig{fig:BasicExampleQFG}, etc.).

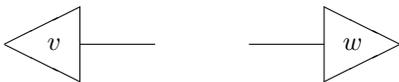
\begin{figure}
\begin{center}
\begin{picture}(20,10)(0,0)
\put(0,5){\line(2,1){10}}
\put(0,5){\line(2,-1){10}}
\put(10,0){\line(0,1){10}}   \put(7.5,5){\pos{cr}{$v$}}
\put(10,5){\line(1,0){10}} 
\end{picture}
\hspace{10mm}
\begin{picture}(20,10)(0,0)
\put(0,5){\line(1,0){10}}
\put(10,0){\line(0,1){10}}   \put(12.5,5){\pos{cl}{$w$}}
\put(10,0){\line(2,1){10}}
\put(10,10){\line(2,-1){10}}
\end{picture}
\caption{\label{fig:TensorNetworkKetBra}%
Tensor network notation. Left: bra (row vector); right: ket (column vector).
Note that the meaning of the symbol depends on its orientation on the page.}
\end{center}
\end{figure}

It should be noted, however, that the graphical conventions of tensor networks 
differ from factor graphs in this point: 
the meaning of a tensor network diagram frequently
depends on its orientation on the page (see, e.g., \cite{WBC:tngc2015}), 
and exchanging left and right amounts to a Hermitian transposition, 
as illustrated in \Fig{fig:TensorNetworkKetBra}.

\subsection{Quantum Bayesian Networks and Quantum Belief Propagation}

Whereas the present paper uses conventional Forney factor graphs
(with standard semantics and algorithms),
various authors have proposed modified graphical models 
or specific ``quantum algorithms'' for quantum mechanical quantities 
\cite{Tu:qi,Pa:qfg2001,LePo:qgm2008}. 
Such graphical models (or algorithms) are not compatible 
with standard statistical models; they are not based on (\ref{eqn:pfromq}) 
and they lack Proposition~\ref{prop:FGMain}.

\subsection{Keldysh Formalism}

There are some high-level similarities between the graphical models 
in the present paper and some diagrams that appear in the context of
the Keldysh formalism (see, e.g., \cite{DSAB:Keld2006}); 
in particular, both have ``two branches along the time axis.''

However, there are also substantial dissimilarities:
first, the diagrams in the Keldysh formalism also have a third branch along
the imaginary axis; second, our factor graphs are arguably more
explicit than the diagrams in the Keldysh formalism.

\subsection{Normal Factor Graphs, Classical Analytical Mechanics, and Feynman
  Path Integrals}

In \cite{Vo:fgLH2011c}, it is shown how Forney factor graphs 
(= normal factor graphs) can be used 
for computations in classical ana\-lytical mechanics. 
In particular, it is shown how to represent the action $S(x)$ of a trajectory $x$
and how to use the stationary-sum algorithm for finding the path 
where the action is stationary.

It is straightforward to modify the factor graphs in \cite{Vo:fgLH2011c}
in order to compute, at least in principle, Feynman path integrals,
where 
$\exp\bigl(\frac{i}{\hbar} S(x) \bigr)$ 
is integrated over a suitable domain of paths $x$:
essentially by replacing the function nodes $f(\,\cdot\,)$ in \cite{Vo:fgLH2011c}
by $\exp\bigl( \frac{i}{\hbar} f(\,\cdot\,) \bigr)$,
and by replacing the stationary-sum algorithm by standard sum-product message passing
\cite{Lg:ifg2004}.

\section{Wigner--Weyl Representation}
\label{sec:WignerWeyl}

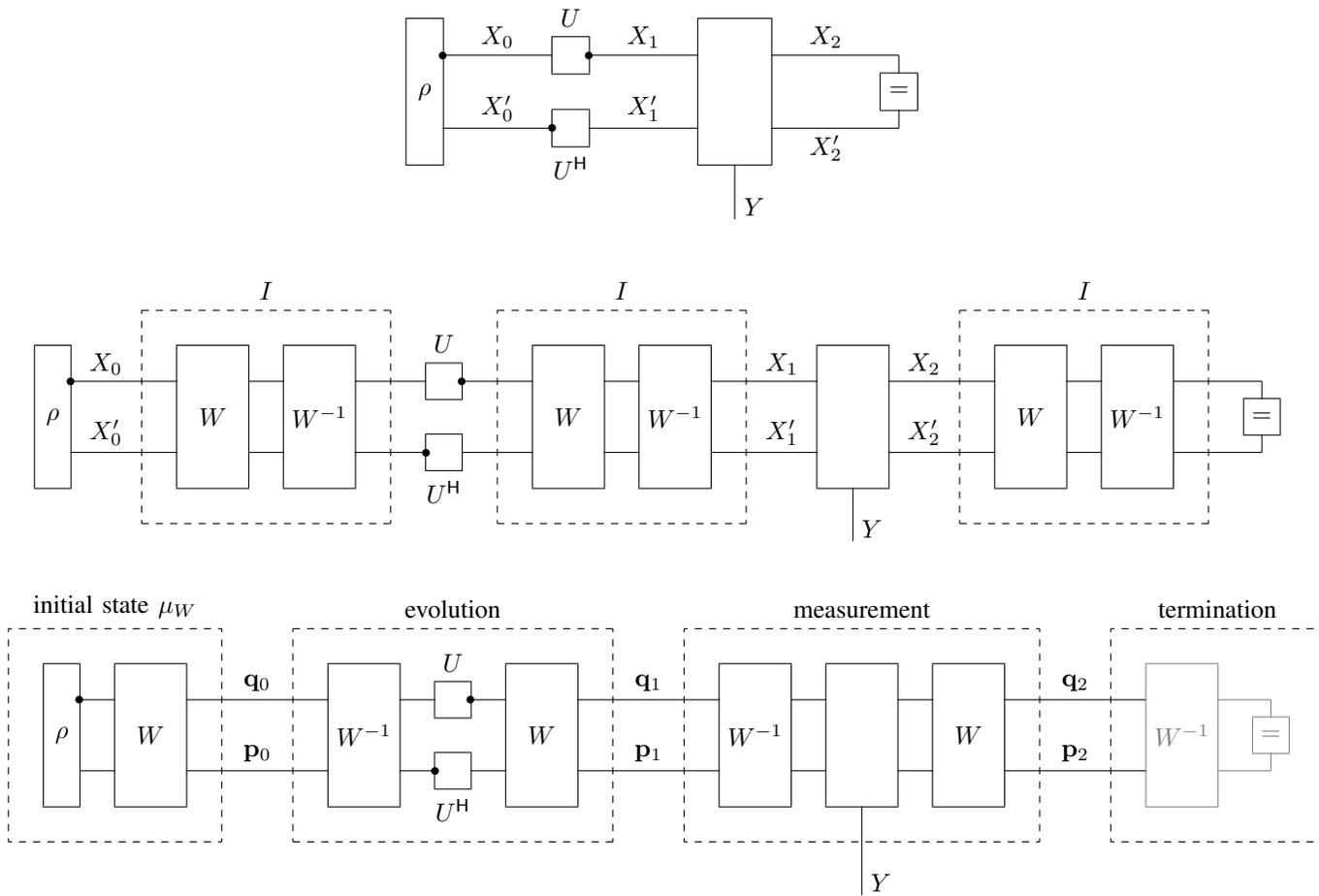
\begin{figure*}
\centering
\begin{picture}(70,29)(0,-7.5)

\put(0,0){\framebox(5,20){$\rho$}}
 \put(5,15){\markerDot}
\put(5,15){\line(1,0){15}}        \put(12.5,16){\pos{cb}{$X_0$}}
\put(5,5){\line(1,0){15}}         \put(12.5,6){\pos{cb}{$X_0'$}}
\put(20,12.5){\framebox(5,5){}}   \put(22.5,19){\pos{cb}{$U$}}
 \put(25,15){\markerDot}
\put(20,2.5){\framebox(5,5){}}    \put(22.5,1){\pos{ct}{$U^\H$}}
 \put(20,5){\markerDot}
\put(25,15){\line(1,0){15}}       \put(32.5,16){\pos{cb}{$X_1$}}
\put(25,5){\line(1,0){15}}        \put(32.5,6){\pos{cb}{$X_1'$}}
\put(40,0){\framebox(10,20){}}
 \put(45,0){\line(0,-1){7.5}}     \put(46.25,-7){\pos{bl}{$Y$}}
\put(50,15){\line(1,0){17.5}}     \put(57.5,16){\pos{cb}{$X_2$}}
\put(50,5){\line(1,0){17.5}}      \put(57.5,4){\pos{ct}{$X_2'$}}
\put(67.5,12.5){\line(0,1){2.5}}
\put(65,7.5){\framebox(5,5){$=$}}
\put(67.5,7.5){\line(0,-1){2.5}}
\end{picture}
\vspace{\dblfloatsep}

{\setlength{\unitlength}{0.975mm}
\begin{picture}(175,36.5)(0,-7.5)

\put(0,0){\framebox(5,20){$\rho$}}
 \put(5,15){\markerDot}
\put(5,15){\line(1,0){15}}       \put(10,16){\pos{cb}{$X_0$}}
\put(5,5){\line(1,0){15}}        \put(10,6){\pos{cb}{$X_0'$}}

\put(20,0){\framebox(10,20){$W$}}
\put(30,15){\line(1,0){5}}
\put(30,5){\line(1,0){5}}
\put(35,0){\framebox(10,20){$W^{-1}$}}
\put(15,-5){\dashbox(35,30){}}   \put(32.5,26.5){\pos{cb}{$I$}}

\put(45,15){\line(1,0){10}}
\put(45,5){\line(1,0){10}}
\put(55,12.5){\framebox(5,5){}}  \put(57.5,19){\pos{cb}{$U$}}
 \put(60,15){\markerDot}
\put(55,2.5){\framebox(5,5){}}   \put(57.5,1){\pos{ct}{$U^\H$}}
 \put(55,5){\markerDot}
\put(60,15){\line(1,0){10}}
\put(60,5){\line(1,0){10}}

\put(70,0){\framebox(10,20){$W$}}
\put(80,15){\line(1,0){5}}
\put(80,5){\line(1,0){5}}
\put(85,0){\framebox(10,20){$W^{-1}$}}
\put(65,-5){\dashbox(35,30){}}   \put(82.5,26.5){\pos{cb}{$I$}}

\put(95,15){\line(1,0){15}}      \put(105,16){\pos{cb}{$X_1$}}
\put(95,5){\line(1,0){15}}       \put(105,6){\pos{cb}{$X_1'$}}
\put(110,0){\framebox(10,20){}}
 \put(115,0){\line(0,-1){7.5}}   \put(116.25,-7){\pos{bl}{$Y$}}
\put(120,15){\line(1,0){15}}     \put(125,16){\pos{cb}{$X_2$}}
\put(120,5){\line(1,0){15}}      \put(125,6){\pos{cb}{$X_2'$}}

\put(135,0){\framebox(10,20){$W$}}
\put(145,15){\line(1,0){5}}
\put(145,5){\line(1,0){5}}
\put(150,0){\framebox(10,20){$W^{-1}$}}
\put(130,-5){\dashbox(35,30){}}   \put(147.5,26.5){\pos{cb}{$I$}}

\put(160,15){\line(1,0){12.5}}
\put(160,5){\line(1,0){12.5}}
\put(172.5,12.5){\line(0,1){2.5}}
\put(170,7.5){\framebox(5,5){$=$}}
\put(172.5,7.5){\line(0,-1){2.5}}
\end{picture}
}
\vspace{\dblfloatsep}

{\setlength{\unitlength}{0.975mm}
\begin{picture}(185,41)(-5,-12.5)

\put(0,0){\framebox(5,20){$\rho$}}
 \put(5,15){\markerDot}
\put(5,15){\line(1,0){5}}
\put(5,5){\line(1,0){5}}
\put(10,0){\framebox(10,20){$W$}}
\put(-5,-5){\dashbox(30,30){}}   \put(10,26.5){\pos{cb}{initial state $\mu_W$}}

\put(20,15){\line(1,0){20}}      \put(30,16.25){\pos{cb}{$\wig{q}_0$}}
\put(20,5){\line(1,0){20}}       \put(30,6.25){\pos{cb}{$\wig{p}_0$}}
\put(40,0){\framebox(10,20){$W^{-1}$}}
\put(50,15){\line(1,0){5}}
\put(50,5){\line(1,0){5}}
\put(55,12.5){\framebox(5,5){}}  \put(57.5,19){\pos{cb}{$U$}}
 \put(60,15){\markerDot}
\put(55,2.5){\framebox(5,5){}}   \put(57.5,1){\pos{ct}{$U^\H$}}
 \put(55,5){\markerDot}
\put(60,15){\line(1,0){5}}
\put(60,5){\line(1,0){5}}
\put(65,0){\framebox(10,20){$W$}}
\put(35,-5){\dashbox(45,30){}}   \put(57.5,26.5){\pos{cb}{evolution}}

\put(75,15){\line(1,0){20}}       \put(85,16.25){\pos{cb}{$\wig{q}_1$}}
\put(75,5){\line(1,0){20}}        \put(85,6.25){\pos{cb}{$\wig{p}_1$}}
\put(95,0){\framebox(10,20){$W^{-1}$}}
\put(105,15){\line(1,0){5}}
\put(105,5){\line(1,0){5}}
\put(110,0){\framebox(10,20){}}
 \put(115,0){\line(0,-1){12.5}}   \put(116.25,-12){\pos{bl}{$Y$}}
\put(120,15){\line(1,0){5}}
\put(120,5){\line(1,0){5}}
\put(125,0){\framebox(10,20){$W$}}
\put(90,-5){\dashbox(50,30){}}    \put(115,26.5){\pos{cb}{measurement}}

\put(135,15){\line(1,0){20}}      \put(145,16.25){\pos{cb}{$\wig{q}_2$}}
\put(135,5){\line(1,0){20}}       \put(145,6.25){\pos{cb}{$\wig{p}_2$}}
{\gray
\put(155,0){\framebox(10,20){$W^{-1}$}}
\put(165,15){\line(1,0){7.5}}
\put(165,5){\line(1,0){7.5}}
\put(172.5,12.5){\line(0,1){2.5}}
\put(170,7.5){\framebox(5,5){$=$}}
\put(172.5,7.5){\line(0,-1){2.5}}
}
\put(150,-5){\dashbox(30,30){}}   \put(165,26.5){\pos{cb}{termination}}
\end{picture}
}
\caption{\label{fig:HolWignerTransform}%
Wigner--Weyl transform of a quantum factor graph
with $W$ as defined in \Fig{fig:WignerBoxDirac}. 
Top: quantum system with a single measurement yielding $Y$.
Middle: inserting neutral factors (identity operators) $I=WW^{-1}$
does not change the exterior function $p(y)$.
Bottom: closing the dashed boxes yields the 
factor graph of the Wigner--Weyl representation.
The termination box reduces to an empty box.
}
\end{figure*}

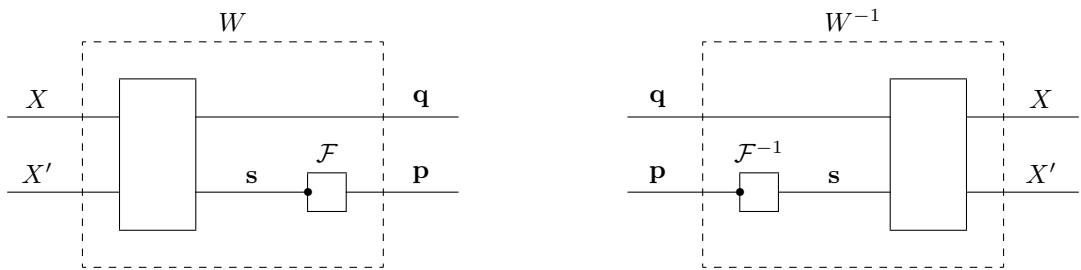
\begin{figure*}
\centering
\begin{picture}(60,34)(0,-5)
\put(0,15){\line(1,0){15}}     \put(4,16.25){\pos{cb}{$X$}}
\put(0,5){\line(1,0){15}}      \put(4,6.25){\pos{cb}{$X'$}}
\put(15,0){\framebox(10,20){}}
%
\put(25,15){\line(1,0){35}}     \put(55,16.25){\pos{cb}{$\wig{q}$}}
\put(25,5){\line(1,0){15}}      \put(32.5,6.25){\pos{cb}{$\wig{s}$}}
\put(40,2.5){\framebox(5,5){}}   \put(42.5,9){\pos{cb}{$\FT$}}
 \put(40,5){\markerDot}
\put(45,5){\line(1,0){15}}      \put(55,6.25){\pos{cb}{$\wig{p}$}}
\put(10,-5){\dashbox(40,30){}}  \put(30,26.5){\pos{cb}{$W$}}
\end{picture}
\hspace{20mm}
\begin{picture}(60,34)(0,-5)
\put(0,15){\line(1,0){35}}     \put(4,16.25){\pos{cb}{$\wig{q}$}}
\put(0,5){\line(1,0){15}}      \put(4,6.25){\pos{cb}{$\wig{p}$}}
\put(15,2.5){\framebox(5,5){}}   \put(17.5,9){\pos{cb}{$\FT^{-1}$}}
 \put(15,5){\markerDot}
\put(20,5){\line(1,0){15}}      \put(27.5,6.25){\pos{cb}{$\wig{s}$}}
\put(35,0){\framebox(10,20){}}
\put(45,15){\line(1,0){15}}     \put(55,16.25){\pos{cb}{$X$}}
\put(45,5){\line(1,0){15}}      \put(55,6.25){\pos{cb}{$X'$}}
\put(10,-5){\dashbox(40,30){}}  \put(30,26.5){\pos{cb}{$W^{-1}$}}
\end{picture}
\caption{\label{fig:WignerBoxDirac}%
Factor graphs of Wigner--Weyl transformation operator $W$ (left) and its inverse (right). 
The unlabeled box inside $W$ represents the factor (\ref{eqn:WignerDeltaConstraints});
the unlabeled box inside $W^{-1}$ represents 
the factor (\ref{eqn:WignerDeltaConstraintsInv}).}
\end{figure*}

The Wigner--Weyl representation of quantum mechanics 
expresses the latter in terms of the ``phase-space'' coordinates 
$\wig{q}$ and $\wig{p}$
(corresponding to the position and the momentum, respectively, of classical mechanics).
When transformed into this representation, the density matrix turns into 
a real-valued function.

So far in this paper, all variables were assumed to take values 
in some finite set without any structure. 
However, the Wigner--Weyl representation requires that 
both the original coordinates $X$ and $X'$ and the new coordinates $\wig{p}$ and $\wig{q}$
can be added and subtracted and admit a Fourier transform 
as in (\ref{eqn:WignerFourierKernel}) and (\ref{eqn:WignerTransform}) below. 
In the following, we assume 
$X_k, X_k', \wig{p}_k, \wig{q}_k \in \R^N$ for all $k$.

In a factor graph with continuous variables, 
the exterior function of a box is defined by integrating over the internal variables, 
i.e., the sum in (\ref{eqn:InnerBoxExt}) and (\ref{eqn:OuterBoxExt}) 
is replaced by an integral. 
Moreover, the equality constraint function (\ref{eqn:EqualityConstraint}) 
becomes 
\begin{equation}
f_=(x_1,\ldots,x_n) = \delta(x_1-x_2)\cdots \delta(x_{n-1}-x_n),
\end{equation}
where $\delta$ is the Dirac delta.
Finally, matrices (cf.\ Section~\ref{sec:FGMatrices})
are generalized to operators, i.e., the sums in (\ref{eqn:MatrixMult}) 
and (\ref{eqn:Trace}) are replaced by integrals.

The transformation to the Wigner--Weyl representation uses 
an operator $W$ that will be described below. 
Factor graphs for the Wigner--Weyl representation 
may then be obtained from the factor graphs in 
Sections \ref{sec:ElementaryQM}--\ref{sec:Decompositions}
by a transformation as in \Fig{fig:HolWignerTransform}. 
The example in this figure is a factor graph 
as in \Fig{fig:GenQFG} with a single measurement, 
but the generalization to any number of measurements is obvious. 
Starting from the original factor graph (top in \Fig{fig:HolWignerTransform}),
we first insert neutral factors (identity operators) factored as $I=WW^{-1}$ 
as shown in \Fig{fig:HolWignerTransform} (middle); clearly, 
this manipulation does not change $p(y)$. 
We then regroup the factors as in \Fig{fig:HolWignerTransform} (bottom),
which again leaves $p(y)$ unchanged. 
The Wigner--Weyl factor graph is then obtained by closing 
the dashed boxes in \Fig{fig:HolWignerTransform} (bottom).
(The Wigner--Weyl representation has thus been obtained as 
a ``holographic'' factor graph transform as in \cite{BaMa:nfght2011,FoVo:pfnfg2011c}.)

The operator $W$ encodes the relations 
\begin{IEEEeqnarray}{rCl}
X  & = & \wig{q} - \wig{s} \label{eqn:WignerTransfX}\\
X' & = & \wig{q} + \wig{s} \label{eqn:WignerTransfXprime}
   \IEEEeqnarraynumspace
\end{IEEEeqnarray}
and the Fourier transform with kernel
\begin{equation} \label{eqn:WignerFourierKernel}
\FT(\wig{s},\wig{p}) = \left(\frac{1}{\pi \hbar}\right)^N\! e^{(i/\hbar)2\wig{p}^\T \wig{s}}.
\IEEEeqnarraynumspace
\end{equation}
For the purpose of this paper, $\hbar$ (the reduced Planck constant) 
is an arbitrary positive scale factor.

The factor-graph representation of the operator $W$
(shown left in \Fig{fig:WignerBoxDirac}) consists of two factors:
the first factor is 
\begin{equation} \label{eqn:WignerDeltaConstraints}
\delta\big(x-(\wig{q}-\wig{s})\big) \delta\big(x'-(\wig{q}+\wig{s})\big),
\end{equation}
which encodes the contraints (\ref{eqn:WignerTransfX})
and (\ref{eqn:WignerTransfXprime});
the second factor is the Fourier kernel (\ref{eqn:WignerFourierKernel}).

The factor-graph representation of $W^{-1}$
(right in \Fig{fig:WignerBoxDirac}) consists of 
the inverse Fourier transform kernel
\begin{equation} \label{eqn:WignerFourierKernelInv}
\FT^{-1}(\wig{s},\wig{p}) = e^{(-i/\hbar)2\wig{p}^\T \wig{s}}
\end{equation}
and the factor
\begin{equation} \label{eqn:WignerDeltaConstraintsInv}
\delta\big( \wig{q} - \frac{1}{2}(x+x') \big)
\delta\big( \wig{s} - \frac{1}{2}(-x+x') \big).
\end{equation}

Closing the ``initial state'' box in \Fig{fig:HolWignerTransform}
yields the function%
\begin{equation} \label{eqn:WignerTransform}
\mu_W(\wig{q}, \wig{p})
= \int_{-\boldsymbol\infty}^{\boldsymbol\infty} 
  \left(\frac{1}{\pi \hbar}\right)^N\! 
  e^{(i/\hbar)2\wig{p}^\T \wig{s}}
  \rho(\wig{q}-\wig{s}, \wig{q}+\wig{s})\, d\wig{s}
\end{equation}
for $\wig{q}=\wig{q_0}$ and $\wig{p}=\wig{p_0}$,
which is easily seen to be real 
(since $\rho(x,x') = \ccj{\rho(x',x)}$).

Closing the ``termination'' box in \Fig{fig:HolWignerTransform} 
yields the function 
\begin{IEEEeqnarray}{rCl}
\IEEEeqnarraymulticol{3}{l}{
\int_{-\boldsymbol\infty}^{\boldsymbol\infty}
\int_{-\boldsymbol\infty}^{\boldsymbol\infty}
\int_{-\boldsymbol\infty}^{\boldsymbol\infty}
   e^{(-i/\hbar)2\wig{p}^\T \wig{s}} 
   \,
   \delta\big( \wig{q} - \frac{1}{2}(x+x') \big)
   \hspace{7em}
  }\nonumber\\
\IEEEeqnarraymulticol{3}{l}{
   \hspace{7em}
   \delta\big( \wig{s} - \frac{1}{2}(-x+x') \big)
   \delta(x-x')\, dx'\, dx\, d\wig{s}
  }\nonumber\\\quad
& = & 
\int_{-\boldsymbol\infty}^{\boldsymbol\infty}
 e^{(-i/\hbar)2\wig{p}^\T \wig{s}} \delta(\wig{s})\, d\wig{s}
\IEEEeqnarraynumspace \\
& = & 1.
\end{IEEEeqnarray}
The termination box thus reduces to an empty box 
and can be omitted.

\section{Monte Carlo Methods}
\label{sec:MonteCarlo}

Let $f(x_1,\ldots,x_n)$ be a nonnegative real 
function of finite-alphabet variables $x_1,\ldots,x_n$.
Many quantities of interest in statistical physics, 
information theory, and machine learning
can be expressed as a partition sum
\begin{equation}
Z_f \eqdef \sum_{x_1,\ldots,x_n} f(x_1,\ldots,x_n)
\end{equation}
of such a function $f$.
The numerical computation of such quantities is often hard. 
When other methods fail, good results can sometimes be obtained 
by Monte Carlo methods \cite{MK:mct1998,Neal:proinf1993r,LgMo:2DMC}.
A key quantity in such Monte Carlo methods is the 
probability mass function
\begin{equation} \label{eqn:pf}
p_f(x_1,\ldots,x_n) \eqdef f(x_1,\ldots,x_n) / Z_f.
\end{equation}

An extension of such Monte Carlo methods 
to functions $f$ that can be negative or complex 
was outlined in \cite{LgMo:ITW2012c}. 
However, only the real case (where $f$ can be negative) 
was addressed in some detail in \cite{LgMo:ITW2012c}. 
We now substantiate the claim from \cite{LgMo:ITW2012c} 
that complex functions $q$ 
as represented by the factor graphs of this paper 
can be handled as in the real case.

We will use the abbreviation $x \eqdef (x_1,\ldots,x_n)$,
and, following \cite{LgMo:ITW2012c}, we define
\begin{equation}
Z_{|f|} \eqdef \sum_{x} |f(x)|
\end{equation}
and the probability mass function 
\begin{equation} \label{eqn:pfabs}
p_{|f|}(x) \eqdef \frac{|f(x)|}{Z_{|f|}}
\end{equation}
Note that $p_{|f|}$ inherits factorizations (and thus factor graphs) 
from $f$. 
This also applies to more general distributions of the form 
\begin{equation}  \label{eqn:pfabsdamped}
p(x ; \rho) \propto |f(x)|^\rho
\end{equation}
for $0<\rho<1$.

For the real case, 
the gist of the Monte Carlo methods of \cite{LgMo:ITW2012c}
is as follows: 
\begin{enumerate}
\item
Generate a list of samples $\smpl{x}{1},\ldots,\smpl{x}{K}$
either from $p_{|f|}(x)$, or from a uniform distribution over $x$, 
or from an auxiliary distribution $p(x; \rho)$ as in (\ref{eqn:pfabsdamped}).
\item
Estimate $Z$ (and various related quantities) 
from sums such as 
\begin{equation} \label{eqn:StatSumfPlus}
\sum_{k: f(\smpl{x}{k})>0} f(\smpl{x}{k})
\end{equation}
and
\begin{equation} \label{eqn:StatSumfMinus}
\sum_{k: f(\smpl{x}{k})<0} f(\smpl{x}{k}),
\end{equation}
or
\begin{equation} \label{eqn:StatSumInvfPlus}
\sum_{k: f(\smpl{x}{k})>0} \frac{1}{f(\smpl{x}{k})}
\end{equation}
and
\begin{equation} \label{eqn:StatSumInvfMinus}
\sum_{k: f(\smpl{x}{k})<0} \frac{1}{f(\smpl{x}{k})}\, ,
\end{equation}
or, more generally,
\begin{equation} \label{eqn:MultiLevelStatSumPlus}
\sum_{k: f(\smpl{x}{k})>0} \frac{f(\smpl{x}{k})^{\rho_1}}{f(\smpl{x}{k})^{\rho_2}}
\end{equation}
and
\begin{equation} \label{eqn:MultiLevelStatSumMinus}
\sum_{k: f(\smpl{x}{k})<0} \frac{f(\smpl{x}{k})^{\rho_1}}{f(\smpl{x}{k})^{\rho_2}}
\end{equation}
\end{enumerate}
The idea is always that the sampling probability 
equals the denominator (up to a scale factor), 
which results in simple expectations for these sums.
(The quantities (\ref{eqn:MultiLevelStatSumPlus}) 
and (\ref{eqn:MultiLevelStatSumMinus}) 
are not actually mentioned in \cite{LgMo:ITW2012c},
but they arise from translating multi-temperature Monte Carlo methods 
(cf.\ \cite{Neal:proinf1993r,LgMo:2DMC})
to the setting of \cite{LgMo:ITW2012c}.)

Note that Step~1 above (the generation of samples) 
generalizes immediately to the complex case;
our issue here is Step~2, where the generalization is less obvious.

Recall now that all factor graphs 
in Sections \ref{sec:ElementaryQM}--\ref{sec:Decompositions} 
represent functions with the structure
\begin{equation}
q(x,x',y) = g(x,y) \ccj{g(x',y)}
\end{equation}
as in \Fig{fig:BasicExampleQFG}. 
But any such function satisfies
\begin{equation}
q(x,x',y) = \ccj{q(x',x,y)}.
\end{equation}
Under any of the probability distributions in Step~1 above, 
a configuration $(x,x',y)$ then has the same probability as the 
conjugate configuration $(x',x,y)$
(i.e., $(x,x',y)$ and $(x',x,y)$ are so-called antithetic variates). 
We can thus double the list of samples in Step~1 
by adding all the conjugate configurations.
For the augmented list of samples, 
the sum (\ref{eqn:StatSumfPlus}) becomes
\begin{equation} \label{eqn:StatSumfPlusCcj}
\sum_{k: f(\smpl{x}{k}) + \ccj{f(\smpl{x}{k})}>0} f(\smpl{x}{k}) + \ccj{f(\smpl{x}{k})},
\end{equation}
and the sums (\ref{eqn:StatSumfMinus})--(\ref{eqn:MultiLevelStatSumMinus})
can be handled analogously.

\section*{Acknowledgment}

The material of this paper has been developing over a long time. 
Its beginnings have benefitted from discussions, mostly long ago,  
with 
Niclas Wiberg, Justin Dauwels, Frank Kschischang, and Nikolai Nefedov. 
More recently, we have profited from discussions with 
Ali Al-Bashabsheh, G.~David Forney, Jr., and Yongyi Mao.
We also thank the reviewers of \cite{LVo:ISIT2012c}
and Alexey Kovalev for pointing out pertinent work in the physics literature.

\newcommand{\IT}{IEEE Trans.\ Inf.\ Theory}
\newcommand{\COM}{IEEE Trans.\ Communications}
\newcommand{\ComLett}{IEEE Communications Letters}
\newcommand{\JSAC}{IEEE J.\ on Selected Areas in Communications}
\newcommand{\WirelessComm}{IEEE Trans.\ Wireless Communications}
\newcommand{\SP}{IEEE Trans.\ Signal Processing}
\newcommand{\SPMag}{IEEE Sig.\ Proc.\ Mag.}
\newcommand{\ProcIEEE}{Proceedings of the IEEE}

\end{document}